# Table of Content

# Earth-affecting Solar Transients: A Review of Progresses in Solar Cycle 24

Jie Zhang[1]
Corresponding author
Email: jzhang7@gmu.edu

Manuela Temmer[2]
Email: manuela.temmer@uni-graz.at

Nat Gopalswamy[3]
Email: natchimuthuk.gopalswamy-1@nasa.gov

Olga Malandraki[4]
Email: omaland@noa.gr

Nariaki V. Nitta[5]
Email: nitta@lmsal.com

Spiros Patsourakos[6]
Email: spatsour@uoi.gr

Fang Shen[7]
Email: fshen@spaceweather.ac.cn

Bojan Vršnak[8]
Email: bvrsnak@gmail.com

Yuming Wang[9]
Email: ymwang@ustc.edu.cn

David Webb[10]
Email: david.webb@bc.edu

Mihir I. Desai[11,15]
Email: mihir.desai@swri.org

Karin Dissauer[2]
Email: karin.dissauer@uni-graz.at

Nina Dresing[12]
Email: dresing@physik.uni-kiel.de

Mateja Dumbović[8]
Email: mateja.dumbovic@geof.unizg.hr

Xueshang Feng[7]
Email: fengx@spaceweather.ac.cn

Stephan G. Heinemann[2]
Email: Stephan.heinemann@hmail.at

Monica Laurenza[13]
Email: monica.laurenza@inaf.it

Noé Lugaz[14]
Email: noe.lugaz@unh.edu

Bin Zhuang[14]
Email: Bin.Zhuang@unh.edu

[1] Department of Physics and Astronomy, George Mason University, 4400 University Dr., MSN 3F3, Fairfax, Virginia 22030, USA

[2] Institute of Physics, University of Graz, Graz, Austria

[3] Goddard Space Flight Center, Greenbelt, Maryland, USA

[4] National Observatory of Athens, Institute for Astronomy, Astrophysics, Space Applications and Remote Sensing, Penteli, Athens, Greece

[5] Lockheed Martin Solar and Astrophysics Laboratory, Palo Alto, California, USA

[6] Department of Physics, University of Ioannina, Ioannina, Greece

[7] SIGMA Weather Group, State Key Laboratory of Space Weather, National Space Science Center, Chinese Academy of Sciences, Beijing, 100190, China

[8] Hvar Observatory, Faculty of Geodesy, University of Zagreb, Kaciceva 26, HR-10000 Zagreb, Croatia

[9] CAS Key Laboratory of Geospace Environment, Department of Geophysics and Planetary Sciences, University of Science and Technology of China, Hefei, Anhui 230026, P.R., China

[10] ISR, Boston College, 140 Commonwealth Ave., Chestnut Hill, MA 02467, USA

[11] Southwest Research Institute, 6220 Culebra Road, San Antonia, TX 78023, USA

[12] Institute fuer Experimentelle und Angewandte Physik, University of Kiel, Germany

[13]INAF-Istituto di Astrofisica e Planetologia Spaziali, Via del Fosso del Cavaliere, 100, I-00133 Roma, Italy

[14]Space Science Center and Department of Physics, University of New Hampshire, Durham, New Hampshire, USA

[15]Department of Physics and Astronomy, University of Texas at San Antonio, San Antonio, Texas 78249, USA

# Abstract

This review article summarizes the advancement in the studies of Earth-affecting solar transients in the last decade that encompasses most of solar cycle 24. It is a part of the effort of the International Study of Earth-affecting Solar Transients (ISEST) project, sponsored by the SCOSTEP/VarSITI program (2014-2018). The Sun-Earth is an integrated physical system in which the space environment of the Earth sustains continuous influence from mass, magnetic field and radiation energy output of the Sun in varying time scales from minutes to millennium. This article addresses short time-scale events, from minutes to days that directly cause transient disturbances in the Earth's space environment and generate intense adverse effects on advanced technological systems of human society. Such transient events largely fall into the following four types: (1) solar flares, (2) coronal mass ejections (CMEs) including their interplanetary counterparts ICMEs, (3) solar energetic particle (SEP) events, and (4) stream interaction regions (SIRs) including corotating interaction regions (CIRs). In the last decade, the unprecedented multi-viewpoint observations of the Sun from space, enabled by STEREO Ahead/Behind spacecraft in combination with a suite of observatories along the Sun-Earth lines, have provided much more accurate and global measurements of the size, speed, propagation direction and morphology of CMEs in both 3-D and over a large volume in the heliosphere. Many CMEs, fast ones in particular, can be clearly characterized as a two-front (shock front plus ejecta front) and three-part (bright ejecta front, dark cavity and bright core) structure. Drag-based kinematic models of CMEs are developed to interpret CME propagation in the heliosphere and are applied to predict their arrival times at 1 AU in an efficient manner. Several advanced MHD models have been developed to simulate realistic CME events from the initiation on the Sun until their arrival at 1 AU. Much progress has been made on detailed kinematic and dynamic behaviors of CMEs, including non-radial motion, rotation and deformation of CMEs, CME-CME interaction, and stealth CMEs and problematic ICMEs. The knowledge about SEPs has also been significantly improved. An outlook of how to address critical issues related to Earth-affecting solar transients concludes this article.

# Keywords

# 1. Introduction

Earth-affecting solar transients refer to a broad range of energetic and/or eruptive events occurring on the Sun that have direct effects on the space environment near the Earth and cause adverse space weather impact on advanced technological systems of human society. They occur near the Sun on time scales of minutes to hours, and the resulting effects on the Earth can take place in minutes to days. These transient events are commonly categorized in four different types: (1) solar flares, (2) coronal mass ejections (CMEs) and their interplanetary counterparts, Interplanetary CMEs (ICMEs), (3) solar energetic particle (SEP) events, and (4) stream interaction regions (SIRS) including corotating interaction regions (CIRs). These four types of Earth-affecting transient events differ in their observational appearances, physical origin or processes, as well as the geoeffectiveness in their own unique ways (Table 1.1). Other energetic events on the Sun, such as filament eruptions, coronal dimmings, waves, etc, can be usually treated as an associated phenomenon with solar flares and/or CMEs. In the following, we briefly introduce the definition of these phenomena and their possible geoeffectiveness, along with selected review articles that discuss in depth of these phenomena. The detailed review of these phenomena, including theoretical interpretations and numerical modellings, are given in the subsequent sections of this article.

*Table 1-1. Four types of Earth-affecting solar transients and their key physical processes and geoeffectiveness*

| Earth-affecting Solar Transients | Key Physical Processes | Effects on Near-Earth Space Environment | Effects on Technological System and Life |
|---|---|---|---|
| Solar Flares | Magnetic reconnection; Particle acceleration; Plasma heating | Disturbances in the ionosphere; Heating and expansion of upper atmosphere | High frequency radio communication; Satellite drag (Earth climate from long term variation of solar irradiance) |
| CMEs and ICMEs | Ideal MHD instability; Flux rope formation; Shock formation; Particle acceleration; aerodynamic drag; CME-CME interaction; Magnetic reconnection | Geomagnetic storms; Substorms; Disturbances in the ionosphere; Ionosphere scintillations; Radiation belt storms; | GPS systems and navigation; Satellite communication; High frequency radio communication; Electric power transmission; Satellite degradation and failure (Single event upset; Dielectric material charging and discharging; Surface charging); Radiation hazards to astronauts; Radiation hazards to aircraft crew and passengers |
| SEPs | Particle acceleration; Injection; Propagation; Turbulence | Particle Radiation Storms | Satellite degradation and failure; High frequency radio communication; Radiation hazards to astronauts; Atmospheric chemistry |

| SIRs/CIRs | Stream interaction; Particle acceleration | Substorms; Geomagnetic storms | Similar to CMEs to a lesser extent |
| --- | --- | --- | --- |

## 1.1 Solar flares

Solar flares are probably the oldest transient phenomenon ever observed on the Sun. They were first discovered as a flash in white light by Carrington (1859) and Hodgson (1859) when observing sunspots. In the modern era, solar flares are observed as sudden enhancement in electromagnetic radiation over a broad range of wavelengths including radio, visible light, EUV, X-rays and gamma rays. The radiation energy released during a flare is about $10^{28}$ to $10^{32}$ ergs during a time scale of minutes to hours. It is well accepted that magnetic reconnection in a configuration of current sheet is the central mechanism that converts free magnetic energy in the corona into particle acceleration and plasma heating, producing a solar flare. Shibata and Magara (2011) provided a review on solar flares with focus on theoretical magnetohydrodynamic process. Fletcher et al. (2011) made a review on solar flares from observational point of view. Hudson (2011) discussed flares from the perspective of their global properties. Note that it is now well known that the process of flares is strongly coupled with that of CMEs (Harrison 1995; J. Zhang et al. 2001b; M. Temmer et al. 2008). Therefore, any further discussion on the origin of flares will be included in the discussions on the origin of CMEs, which will be extensively reviewed in this article. Very often, the term of solar eruptions is used to refer to transient and large-scale energy release on the Sun, and a solar eruption contains both flare and CME, along with other associated phenomena, such as coronal dimmings and global coronal waves etc.

While the total electromagnetic radiation, or irradiance from the Sun, is nearly a constant with an amplitude of approximately 0.1% over the 11-year solar cycle that affects the long term climate of the Earth (Lean 1991), the EUV and X-ray irradiances during solar flares can increase by many fold to orders of magnitude. One space weather effect of solar flares is from EUV radiation in particular the Lyman-alpha radiation at 121.6 nm wavelength absorbed in the Earth's upper atmosphere causing its instantaneous heating and expansion, which results in a sudden drag and lowering of low-orbiting satellites (Schwenn 2006). Enhanced X-ray emissions from solar flares can penetrate to the bottom of the ionosphere and create an enhancement in the electron content, which may affect high-frequency radio communication.

## 1.2. CMEs and ICMEs

Since the discovery of solar flares by Carrington in 1859, it had long been conceived that there was a cause and effect relation between solar flares and geomagnetic activities on the Earth. Only starting from 1980s, it became clear that the only type of solar transients that has a clear cause-effect relation to geomagnetic activity lies in CMEs, not in flares (e.g., Schwenn 1983; Sheeley et al. 1985; Gosling 1993; Reames 1999; Zhang et al. 2007). It is now well accepted that CMEs are the solar transient that have the most profound effect on space environment and inflict the most adverse space weather effect (N. Gopalswamy 2016).

CMEs are transient and energetic expulsion of mass and magnetic flux in a large scale from the low corona into interplanetary space. While the basic configuration of shock-driven magnetic structures from the Sun have been proposed to explain geomagnetic storm sudden commencement (Gold 1962) and various types of non-thermal solar radio bursts (Fokker 1963), CMEs were first directly imaged in white light from space by the OSO-7 coronagraph in the early 1970s (Tousey 1973). The speed of CMEs in the outer corona ranges from ~100 km/s to ~3000 km/s at maximum with an average speed from 300 km/s to 500 km/s depending on the phase of the solar cycle (S. Yashiro et al. 2004). Mass of CMEs is mostly in the range from $10^{13}$ g to $10^{16}$ g with a peak value at $3.4 \times 10^{14}$ g, and their kinetic energy is mostly between $10^{27}$ to $10^{32}$ ergs with a peak value at $8.5 \times 10^{29}$ ergs (Vourlidas et al. 2010). CMEs reach their peak speed in the time range from minutes to hours with a median value at ~54 min (Zhang and Dere 2006).

The following is a list of review articles on CMEs, or more generally in solar eruptions, in the last decade. Schrijver (2009) reviewed the drivers of major solar flares and eruptions with a focus on flux emergence and its interaction with the ambient field. Chen (2011) provided a comprehensive overview of theoretical models and their observational basis of CMEs. The review of Webb and Howard (2012) focused  the observational aspects of CMEs. Schmieder, Aulanier, and Vršnak (2015) further reviewed on observational perspectives of flare-CME models. Gopalswamy (2016) reviewed major discoveries on CMEs observed by spaceborne coronagraphs that show the growing significance of CMEs as the primary source of severe space weather.  More recently, reviews were made focusing on magnetic structures of solar eruptions, from the perspective of magnetic flux rope (Xin Cheng, Guo, and Ding 2017a) and modeling of magnetic field (Guo, Cheng, and Ding 2017). Chen (2017) made a review on the physics of erupting solar flux ropes in the aspects of both theory and observation. The origin, early evolution and predictability of solar eruptions were recently reviewed by Green et al. (2018). The recent review of Patsourakos et al. (2020) discusses the formation and the nature of the pre-eruptive magnetic configuration of CMEs. We would also like to point out the following earlier reviews on CME models (Forbes 2000; Klimchuk 2001; Lin, Soon, and Baliunas 2003). Most recently, Lamy et al. (2019) made an extensive review on statistical properties of CMEs covering two complete solar cycles 23 and 24. Gopalswamy et al. (2020) also reviewed how CME properties varied with solar cycle, taking advantage of the availability of uniform and extensive observations made over two complete solar cycles.

Following the ejection from the corona, a CME largely maintains its magnetic configuration or topology that is well organized by a twisted magnetic flux rope, thus is able to continuously propagate outward through the heliosphere to a large distance, interacting with the ambient solar wind and impacting planets along its path. Its counterpart in the heliosphere is called interplanetary CME (ICME). Howard and Tappin (2009) reviewed the theory of ICMEs observed in the heliosphere. Rouillard (2011) provided a short review relating white light CMEs near the Sun and in-situ ICMEs. Zhao and Dryer (2014) summarized the status of CME/shock arrival time prediction to that date.  The physical processes of CME/ICME evolution are reviewed in Manchester et al. (2017). Lugaz et al. (2017) review focused on the interaction of successive CMEs. Shen et al. (2017) also reviewed on CME interaction with a focus on analyzing the physical nature of the interaction. More recently, Vourlidas, Patsourakos, and Savani (2019) made an overview of predicting the geoeffectiveness properties of CMEs, including current status, open issues and a path forward. The review by Kilpua et al. (2019) focused on the forecasting of magnetic structure and orientation of CMEs. The most recent review on ICMEs was by Luhmann et al. (2020).

Besides the ejecta or magnetic flux rope component, a fast CME is capable of driving a wider shock ahead and forming a thick sheath region between the eject front and the shock front. While the shock is the main source of solar energetic particle (SEPs), the sheath, like the magnetic flux rope, is also an important transient structure for causing geomagnetic storms. The properties and importance of shock, sheath regions, as well as CME ejecta, are viewed in (E. K. J. Kilpua et al. 2017; E. Kilpua, Koskinen, and Pulkkinen 2017a)

ICMEs passing the Earth can significantly distort and energize the Earth's magnetosphere and generate a cascade of effects in the different layers of the Earth's space environment, including in the magnetosphere, radiation belt, ionosphere and upper atmosphere, and even in the lithosphere. These are collectively defined as space weather. The term space weather refers to conditions on the Sun and in the solar wind, magnetosphere, ionosphere, and thermosphere that can influence the performance and reliability of space-borne and ground-based technological systems and that can affect human life and health (Schwenn 2006). The terrestrial perspective and impacts of space weather are summarized in Pulkkinen (2007). Space weather effects in the Earth's radiation belts were recently reviewed in Baker et al. (2018). CMEs can cause extensive ionospheric anomalies (e.g., Wang et al. 2016), and disturbances in the atmosphere-ionosphere coupling system (Yiğit et al. 2016). The cradle to grave process of some extreme space weather events are outlined in Riley et al. (2018).

The impacts due to severe space weather storms caused by CMEs/ICMEs on technological systems are profound (Lanzerotti 2017). These impacts include hazards to astronauts, satellite degradation and failure through single event upset. dielectric material charging and discharging and surface charging, error and failure in GPS navigation, effects on satellite communication, effects on high frequency communication, effects on power grids and aviation etc. The potential catastrophic societal effect of the May 1967 great storm is revisited by Knipp et al. (2016). The economic impact of space weather is reviewed and analyzed in Eastwood et al. (2017).

## 1.3 SEPs

Solar energetic particle events are enhancements of electrons, protons and heavy ion fluxes observed in the heliosphere related to both solar flares and CMEs,. SEP events present energy spectra that span more than six orders of magnitude, from a few keV superthermal to GeV relativistic energies.

High energy particles from the Sun were first observed as a sudden increase in intensity in ground-based ion chambers and neutron monitors during large solar flares (Forbush 1946). Such ground-level enhancements (GLEs) consist of the strongest set of SEPs events that are mostly detected from space. For half a century following its discovery, it was generally assumed that energetic particles originated from solar flares, i.e., the point-like source in time and space. However, in 1990s it became clear that there are two types of SEPs: impulsive type and gradual type, whose source is of impulsive flares on the Sun and of large-scale-long-lasting shocks driven by CMEs, respectively (Reames 1999b). The gradual SEPs typically last for several days, while the impulsive events only last for a few hours. It is now believed that CME-driven coronal and interplanetary shocks are the most prolific producers of SEPs that pose radiation hazards for our environment and our assets on Earth and in space. Particle enhancements accompanying CME-driven interplanetary shocks that are passing near the Earth are known as energetic storm particles or ESP events, because they are often associated with "Sudden Storm Commencements".

Besides the seminal review paper that summarized the paradigm shift on the origin of SEPs in (Reames 1999b), a series of review papers were also published in the last decade (Reames 2013; 2015; 2018; 2020). Reames (2013) provided a comprehensive account on the two sources of solar energetic particles, which highlighted the early evidence from fast-drifting type III and slow-drifting type II solar radio emissions (Wild, Smerd, and Weiss 1963). Reames (2015) focused on element abundances and source plasma temperatures of SEPs. Reames (2018) extended the topics including abundances, ionization states, temperatures and FIP (First Ionization Potential) in SEPs. Most recently, Reames (2020) categorized SEPs into four basic populations and discussed the four distinct pathways, to account for the mixture of SEPs from pure impulsive and pure gradual events. Desai and Giacalone (2016) provided a comprehensive review on large gradual SEP events to date. Klein and Dalla (2017) also made a review on the acceleration and propagation of SEPs. More recently, a set of 10 review papers on SEPs are collected in a published book (O. E. Malandraki and Crosby 2018b), which built upon the 2-year HESPERIA (High Energy Solar Particle Events Forecasting and Analysis) project of the EU HORIZON 2020 program.

Like flares and CMEs, it is apparent that SEPs pose a threat to modern technology strongly relying on spacecraft and are a serious radiation hazard to humans in space (Jiggens et al. 2014; O. E. Malandraki and Crosby 2018a). High energy charged particles have been found to have damaging impacts on various components of spacecraft, including instruments, electronic components, solar arrays etc. SEPs also effect signal propagation between Earth and satellites due to Polar Cap Absorption (PCA) which results from intense ionization of the D-layer of the polar ionosphere. In the instances when SEP events reach aviation latitudes, they are also a concern for human health as the radiation dose received can increase. This applies specifically to high latitude flights and polar routes for commercial aviation. It can be a risk for frequent flyers and for aircrew. SEP forecasting is relied upon to mitigate against the effects SEP events.

## 1.4. SIRs/CIRs

The solar wind reveals long-term, and most often periodic, variations in terms of high speed solar wind streams that may lead to geomagnetic storms. The solar wind also transports short-term disturbances, such as CMEs. Investigating the solar wind is therefore of crucial interest for Space Weather forecasting. The interplay between fast and slow solar wind causes stream interaction regions (SIRs). SIRs are related to coronal holes (CHs), long-lived regions on the Sun with predominantly open magnetic field. Due to the quasi-stationary location of low-latitude CHs, the interaction of high and slow speed solar wind streams results in a compression of plasma and magnetic field that occurs at certain distance from the Sun. As the Sun rotates, recurring SIRs are referred to as corotating interaction regions (CIRs).

Recent reviews on solar wind high speed streams are given by Cranmer, Gibson, and Riley (2017) ; Living review by Richardson (2018) on SIRs and corona and solar wind by Cranmer and Winebarger (2019). Cranmer et al. (2017) gives an overview of the community's recent progress and understanding of two major problems associated with HSSs: the coronal heating and the acceleration of the fast and slow solar wind. They discuss recent observational, theoretical, model and forecasting techniques with a positive forecast to the future. The review by Richardson (2018) focuses entirely on the interaction of slow and fast solar wind leading to the formation of SIRs and discusses the acceleration processes of energetic particles in stream-stream interaction regions, modulation of the galactic cosmic ray count, resulting geomagnetic disturbances as well as MHD modeling results. The very recent review by Cranmer and Winebarger (2019) gives detailed insight into high resolution observations of the solar corona as well as 3D numerical simulations which hint towards small scale entangled, twisted and braided magnetic fields. These processes, that may lead to reconnection, are, despite their limitations, thought to be a main source of heating the solar corona.

The geoeffectiveness and Space Weather impact of SIRs/CIRs can be observed as variations in the Earth's magnetosphere, ionosphere, and even neutral density in the thermosphere. The solar wind delivers significant energy that causes various Space Weather phenomena. CIR-related storms are more hazardous to space-based assets, particularly at geosynchronous orbit compared to CMEs, because CIRs are of longer duration and have hotter plasma sheets causing a stronger spacecraft charging (Borovsky and Denton 2006). Details on the geoeffectiveness of SIRs/CIRs can be found in recent reviews and statistical papers (E. K. J. Kilpua et al. 2017; Bojan Vršnak et al. 2017; Yu I. Yermolaev et al. 2018).

## 1.5. ISEST project

This review article is part of the collective effort made by the International Study of Earth-affecting Solar Transients (ISEST) project, which is one of the four research projects of the Variability of the Sun and Its Terrestrial Impact (VarSITI)) program, sponsored by the Scientific Committee on Solar-Terrestrial Physics (SCOSTEP) for the period of 2014 – 2018. The VarSITI program is summarized in a companion article (Shiokawa et al. 2020 to be added). The stated overarching goal of the ISEST project is to understand the origin, propagation, and evolution of solar transients through the space between the Sun and the Earth, and develop the prediction capability of space weather. Toward this goal, the ISEST project has organized four dedicated workshops in three different geographic locations across the globe: 17 – 20 June 2013 in Hvar, Croatia, 26 – 30 October 2015 in Mexico City, Mexico, 18 – 22 September 2017 in Jeju, South Korea, and 24 – 28 September 2018 again in Hvar, Croatia. The ISEST project maintains a standing website for hosting event catalogs, data, and presentations, and offers a forum for discussion at http://solar.gmu.edu/heliophysics/index.php/Main_Page.
The ISEST project has resulted in a Topic Issue in the journal of Solar Physics with a collection of 32 articles (Zhang et al. 2018a); this collection is then converted to a published book (Zhang et al. 2018b). A similar but earlier project in the CAWSES-II era, Climate and Weather of the Sun-Earth System of SCOSTEP (2010-2014), is the project of "Short-term variability of the Sun-Earth system"; the summary of the activity from 2010-2014 is in Gopalswamy, Tsurutani, and Yan (2015).

The implementation of the ISEST project is centered around several working groups, which are (1) data, (2) theory, (3) simulation, (4) campaign study, (5) SEP, (6) Bs challenge, and (7) MiniMax24 campaign. The sections of this article largely contain the contribution of these seven working groups, respectively. We organize the articles as follows. Section 2 is on observational progress on CMEs and ICMEs. Section 3 is on theoretical progress on CMEs and ICMEs, while Section 4 summarizes the progress in simulation studies of CMEs and ICMEs. Campaign-style studies are reviewed in Section 5. SEP studies are reviewed in Section 6. Section 7 reviews stream interaction regions. Forecasting CMEs are reviewed in Section 8. Section 9 summarizes the activity of MiniMax24 campaign. The conclusion and outlook are in Section 10.

## 2. Progress in Observations of CMEs/ICMEs

## 2.1 Introduction

The capacity of observing and studying the solar-terrestrial system has increased dramatically during solar cycle 24. We can observe and track solar eruptions nearly continuously in time and space from Sun to Earth, thanks to a large set of sensitive remote-sensing and in-situ instruments onboard a fleet of spacecraft. These include the Solar Terrestrial Relations Observatory Ahead/Behind (STEREO A/B) (twin spacecraft launched in 2006; drifting along the Earth orbit) (Kaiser et al. 2008), the Solar Dynamics Observatory (SDO) (geosynchronous orbit; launched in 2010) (Pesnell, Thompson, and Chamberlin 2012), the Solar and Heliospheric Observatory (SOHO) (L1 point; launched in 1995) (Domingo, Fleck, and Poland 1995), Hinode mission (polar orbit of Earth; launched in 2006) (Kosugi et al. 2007), the Advance Composition Explorer (ACE) (L1 point; launched in 1997) (Smith et al. 1998), Wind spacecraft (L1 point; launched in 1994) (Ogilvie et al. 1995) , and other space-based spacecraft and ground-based observatories. In particular, the SECCHI (Sun Earth Connection Coronal and Heliospheric Investigation) suite onboard STEREO comprised five imaging telescopes, which together observe the solar corona from the solar disk to beyond 1 AU; these telescopes are: EUVI (Extreme Ultraviolet Imager: 1-1.7 Rs), COR-1 (Coronagraph 1: 1.5-4 Rs), COR-2 (Coronagraph 2: 2.5-15 Rs), HI-1 (Heliospheric Imager 1: 15-84 Rs, or 4°-24° in elongation angle) and HI-2 (Heliospheric Imager 2: 66-318 Rs, or 19°-89° in elongation angle ) (R. A. Howard et al. 2008). Two recent missions dedicated to solar and heliospheric physics are the Parker Solar Probe (PSP) (varying elliptical orbit around the Sun with perihelia < $10R_S$; launched in 2018) (Fox et al. 2016) and the Solar Orbiter (highly elliptical and inclined orbit around the Sun; launched in 2020) (Müller et al. 2020). These missions will provide observations of the unexplored territories of the Sun-heliosphere system, but results from these two spacecraft will not be included in this review.

The global and long-lasting nature of CMEs makes the so-called Sun-Earth connection truly meaningful. Erupting from the low corona of the Sun and propagating into the outer corona and interplanetary space, a typical fast CME largely contains two volumetric components that are persistent in time and space: the magnetic ejecta component and the shock sheath component. Each of the two volumetric components has its own unique front: the ejecta front and the shock front, respectively. This is evident in both remote-sensing imaging observations in white light as well as from in-situ one-point time-series sampling when the CME passes through the spacecraft, as illustrated in Figure 2-1. The ejecta contains the erupted magnetic field and plasma originating in the low corona, while the sheath region contains the magnetic field and plasma corresponding to the ambient solar wind that is disturbed and compressed by the forward shock (or forward compressing waves for slower CMEs). Through its propagation from the Sun to Earth, a CME ejecta is believed to maintain its curved flux-rope shape in a quasi-self-similar manner and keep its two legs remaining rooted on the surface of the Sun for days and even longer.

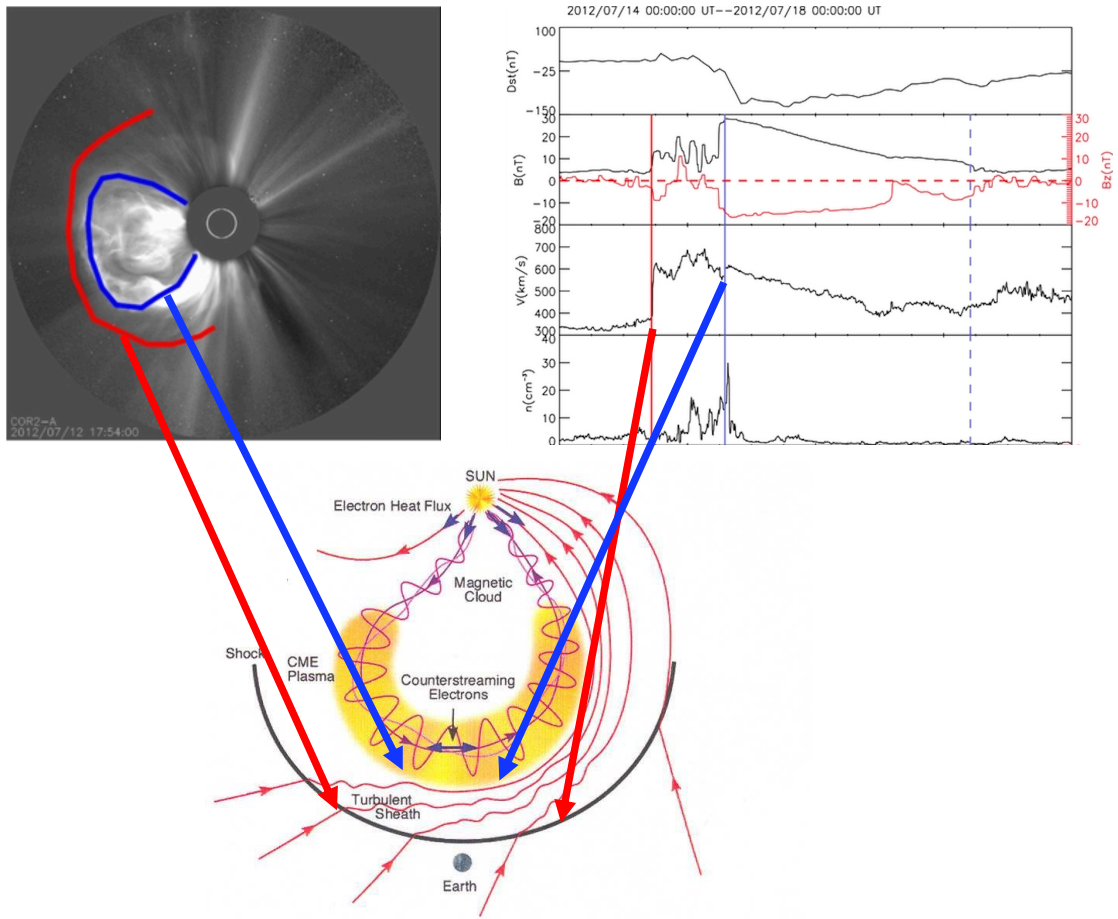

*Figure 2-1. A schematic of a CME and its interplanetary counterpart ICME. Top left: the CME image in white light near the Sun (an event on July 12, 2012, adopted from Hess and Zhang (2014). The red and blue curves outline the shock front and CME ejecta front, respectively. The Sun is indicated by the white circle in the center. Top right: the in-situ data of the resulting ICME near the Earth (adopted from Hess & Zhang 2014). From top to bottom, the five panels show the Dst index, solar wind magnetic field, velocity and density. The vertical red line indicates the arrival time of the shock, and two vertical blue lines indicate the beginning and ending time of the CME ejecta. Bottom: A schematic of CME/ICME illustrates its geometry and internal components including the shock front, turbulent sheath and draped ambient magnetic field, twisted magnetic field in the CME ejecta and electron heat flux along magnetic fields. (adopted from Zurbuchen & Richardson 2006).*

In light of the fact that the behavior of CMEs is dominated by different kinematics and dynamics at different distances within the vast space between the Sun and Earth, we loosely divide the whole Sun-Earth domain into three sub-domains: (1) in the corona where CME evolution is dominated by its internal magnetic force; this is also the region imaged by coronagraphic instruments, (2) farther from the Sun in the interplanetary space where CME evolution is mostly dominated by the aerodynamic drag force, i.e., momentum transfer between the CME and the ambient solar wind flow; practically, this is the area observed by Heliospheric Imagers onboard STEREO, and (3) near the Earth (or other locations near 1 AU) where most in-situ sampling data are taken. These in-situ data provide detailed diagnostics of plasma, magnetic and abundance properties of CME ejecta and driven shock, albeit limited at one particular point in space or a particular sampling line for a traveling ICME. Remote-sensing of the CME-driven shock has been enabled by tracking type II radio bursts with the radio instruments on board the Wind and STEREO missions. A CME in these three sub-domains can be conveniently called as CME in the traditional sense, an ICME in the interplanetary space and in-situ ICME, respectively. Note that there is certainly no boundary or barrier between the aforementioned corona and interplanetary space, which can be anywhere between 4 Rs and 30 Rs. For the sake of simplicity only, one could arbitrarily adopt a value of 20 Rs (roughly coinciding with the Alfvenic critical point) to separate domains 1 and 2.

In this Section, we review the basic morphology and geometry (Section 2.2) as well as kinematic behavior of CMEs (Section 2.3) in the corona and in the interplanetary space. The properties of source regions in the low corona where CMEs originate are reviewed in Section 2.4. Section 2.5 reviews statistical properties and solar cycle variation of CMEs and ICMEs. A summary is given in Section 2.6

## 2.2 CME Morphology, Geometry and Their Evolution

## 2.2.1 Basic morphology of CMEs

One of the fundamental properties of CMEs is its morphology near the Sun, as obtained from the direct interpretation of outer corona images made by white-light coronagraphs. Prior to the SOHO era, the morphology of CMEs had been characterized by the so-called three-part structure: a bright frontal shell, followed by a relatively dark cavity surrounding a bright core (Illing and Hundhausen 1986). The expected shock fronts missing in this traditional structure were later routinely identified, thanks to the improved sensitivity of coronagraphs on the SOHO and STEREO spacecraft. The shock front appears as an outline or boundary of a weakly brightened region that contains displaced or kinked coronal streamers and rays (Sheeley, Hakala, and Wang 2000; B. E. Wood and Howard 2009a; Ontiveros and Vourlidas 2009; Hess and Zhang 2014b; Ying D. Liu et al. 2017; N. Gopalswamy, Thompson, et al. 2009; Nat Gopalswamy and Yashiro 2011). A shock fronts is expected to form when the speed of the CME ejecta in the frame of the ambient solar wind is faster than the local Alfven speed. Figure 2-2 shows one example of the identification and geometrical fitting of the fronts of the ejecta (green wireframe) and the shock (red wireframe). Thus, the overall morphology of a typically large and fast CME can be characterized by two fronts: a large fuzzy shock/wave front followed by a bright loop-like ejecta front which can be interpreted as the plasma pileup at the boundary of the expanding magnetic flux rope, irrespective of whether a three-part structure can be identified following the loop-like front (A. Vourlidas et al. 2013).

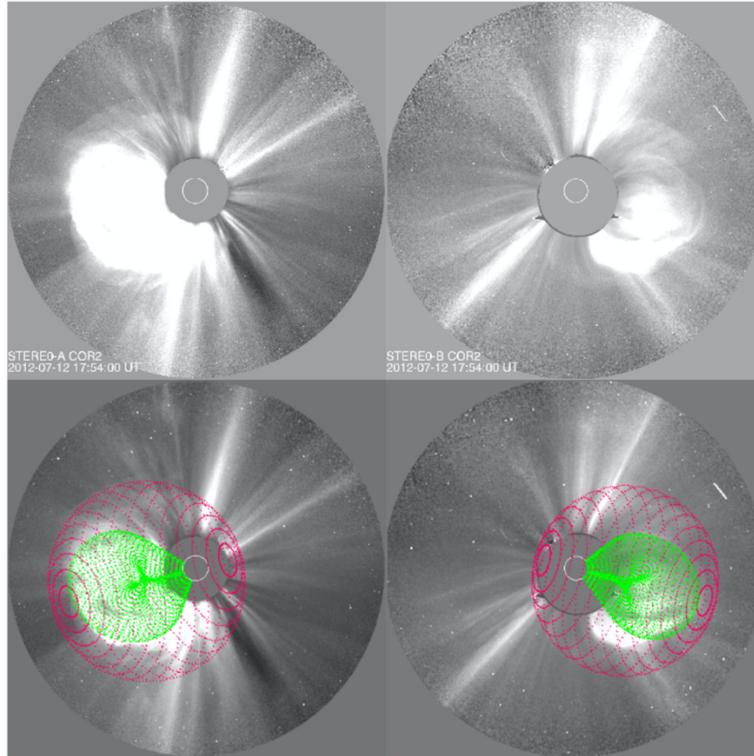

*Figure 2-2. Forward model fitting of CME ejecta front and CME-driven shock front of July 12, 2012 event based on STEREO-A COR2 (left) and STEREO-B COR (right) images, along with (bottom) and without the raytrace mesh. The green mesh shows the GCS fitting to the eject front, whereas the red mesh shows the spheroid fitting to the shock front (adapted from Hess and Zhang 2014)*

To quantitatively capture the 3D morphology (i.e., shape and size) and geometry (i.e., location and orientation) of a CME ejecta, the graduated cylindrical shell (GCS) model has been widely used (Thernisien, Howard, and Vourlidas 2006; Thernisien, Vourlidas, and Howard 2009; Thernisien 2011). This 3D geometric model, meant to reproduce flux-rope-like CMEs, consists of a tubular section forming the main body of the flux rope attached to two cones that correspond to the "legs" of the flux rope that are connected to the surface of the Sun. In this model, the bright frontal shell of the ejecta corresponds to the surface of the flux rope, and the cavity as the body of the flux rope,  being consonant with the common view (e.g., Chen et al. 1997; Cremades, Bothmer, and Tripathi 2006).  This model contains a central axis that threads through the center of the curved tube and the conical legs. The shell surface of the flux rope exhibits rotational symmetry around the central axis at each cross section perpendicular to the axis. This axis also defines the geometric plane of the flux rope. The GCS model, as the one implemented in Solar Software (IDL), has the following six free parameters, (1) propagation longitude, (2) propagation latitude, (3) tilt angle of the curved central axis (the plane of the flux rope), (4) height of the leading edge of the front, (5) half angle between the axes of the legs and (6) aspect ratio between the radius of the circular cross section of the tubular shell and the distance to the outer edge of the shell from the Sun center. The first three parameters define the geometry, while the later three parameters define the morphology or sizes of the CME. As will be discussed below, CME geometry changes significantly close to the Sun, but is assumed to remains largely constant in the interplanetary space. On the contrary, CME morphology remains self-similar in the corona, but distorts significantly in the interplanetary space.

To capture both the ejecta and shock fronts of CMEs, Kwon, Zhang, and Olmedo (2014) developed a compound model, in which the shock front is modeled as an ellipsoid, which can be spherical or ellipsoidal depending on events as well as on the evolution stage of the event of study; the ejecta front simply follows the GCS model. The ellipsoid model of the shock front also has six free parameters that define the geometry and morphology in 3D. Using this model, Kwon, Zhang, and Olmedo (2014) demonstrated that the footprints of expanding shock waves seen in the outer corona correspond well to the EUV wave front observed on the solar disk in the early development of CMEs. Similar results of reconciling CME-driven shock and EUV coronal waves are obtained in other studies (X. Cheng et al. 2012; Veronig et al. 2018), revealing the global behavior of CME-driven shocks.

The reconstruction of shock fronts in 3D also reveals the global properties of halo CMEs. It had been widely believed that the halo appearance of a CME is caused by the geometric projection effect, i.e., a CME moves along the Sun-Earth line and project in all directions on the plane of the sky surrounding the occulter. However, Kwon, Zhang, and Vourlidas (2015) found that 66% of halo CMEs from 2010 to 2012 are seen as halos in all three spacecraft, SOHO, STEREO-A and STREO-B when they are in quadrature configuration. They concluded that the halo structure largely represents the shock/wave that propagates in all directions, with a lesser dependence on the projection effect of the CME ejecta that has a limited size. Shen et al. (2013) also found that very fast (> 900 km/s) full-halo CMEs originating far from the vicinity of solar disk center have a small projection effect. This global reach of CME-driven shock, even having a component propagating in the opposite direction of the CME ejecta, helps explain that some SEP events have a wide range of helio-longitude distribution, even allowing particle intensity increase at poorly connected spacecraft (Lario et al. 2014; Ying D. Liu et al. 2017)

While the physical properties of CME-driven shocks were refined in the last decade, as discussed above, the physical nature of the core of CMEs has been recently questioned in several studies (T. A. Howard et al. 2017; Song et al. 2017; Veronig et al. 2018; Song et al. 2019). It has long been believed that the bright core inside the CME cavity originates from entrained erupting filament/prominence material which has a high density. However, through investigating source region of CMEs on the solar disk and tracking eruptions continuously into the coronagraph FOV from multiple viewpoints in space, unambiguous observational evidence shows that many "classical" three-part CMEs do not contain an erupting filament/prominence (T. A. Howard et al. 2017; Song et al. 2017). Howard et al. (2017) suggested that the core could be the result of a mathematical caustic produced by the geometric projection of a twisted/writhed flux rope, implying the same flux rope produces both the cavity and core; they also suggested another possible cause that could arise spontaneously from the eruption of a flux rope. Through investigating the well-observed highly-structured CME on 2017 September 10, Veronig et al. (2018) argued that the bright core rises from the hot plasma generated through magnetic reconnection but adds onto the rim of the rising flux rope, implying that the core is the flux rope. Song et al. (2019) suggested that the core might correspond to the entirety of the flux rope in the early phase, but expand continuously and fill-in the entire cavity at a later time. The physical nature of the observed CME core and cavity remains to be an open question.

## 2.2.2 CME morphological evolution in the corona: self-similar expansion

How does the morphology of CMEs evolve in the corona (as well as in the interplanetary space; see next subsection)? This issue is far from settled. One simple question is whether such evolution is self-similar or not. Since it is a structured 3D entity, a CME evolves along three principal directions in a 3D space. Thus, to properly answer this question, one has to define which direction the self-similarity refers to. For the clarity of discuss hereafter, we define three principal directions in the frame of the flux rope with the apex of the central axis of the flux rope at the origin: toroidal direction (T), poloidal direction (P) and radial direction (R). The radial direction is the vector line connecting the Sun center toward the apex of the central axis, the toroidal direction, which is on the plane of the flux rope, is along the direction of the central axis at the apex, while the poloidal direction is perpendicular to the plane of the flux rope. Both toroidal and poloidal directions are perpendicular to the radial direction. If the tilt angle of the flux rope is zero, the toroidal direction will be exactly along the heliographic longitude, while the poloidal direction will be along the heliographic latitude. The linear sizes of the flux ropes can be characterized by $L_T$, $L_P$ and $L_R$, respectively. Similarly, one can define three aspect ratios: $\kappa_T = L_T/d_A$, $\kappa_P = L_P/d_A$, $\kappa_R = L_R/d_A$, respectively, where $d_A$ is the distance of the apex of the flux rope central axis. A constant aspect ratio along one particular direction defines the self-similar evolution in that direction.

Since the advent of multi-viewpoint observations of CMEs, the aforementioned GCS model is widely used to determine CME morphology in 3D for a large number of CMEs (e.g., Poomvises, Zhang, and Olmedo 2010; Kilpua et al. 2012; Colaninno, Vourlidas, and Wu 2013; Subramanian et al. 2014; Hess and Zhang 2014; Veronig et al. 2018; Chi et al. 2018). These studies found good agreement between GCS-generated flux rope shells and the observed CME appearances. One particular interesting result, relevant to the morphology, is that the constant aspect ratio and angular width can be adopted for a particular CME observed in different times, implying a self-similar evolution of the morphology in all three principle directions. Note that the aspect ratio in the GCS model is the same as $\kappa_P$ and $\kappa_R$ defined above, and $\kappa_P = \kappa_R$ since the GCS flux rope has a circular cross section perpendicular to the central axis. The angular width in the GCS model is equivalent to $\kappa_T$ defined above. In other words, CME angular widths along both toroidal and poloidal directions remain constant as it evolves, and in the meantime, the CME expands radially at the same rate as along the two lateral directions, maintaining a circular cross section, or constant $\kappa_R$. However, the constant aspect ratio along the radial direction will not be true in the interplanetary space as discussed in the next subsection.

A more robust examination of self-similarity can be carried out by comparing expansion speed and bulk speed of CMEs, and a constant ratio with time indicates the self-similarity. Through a statistical study of 475 CMEs from 2007 -2014 that are geometrically well structured and whose geometric centroid and boundary can be well determined from single-viewpoint images (Angelos Vourlidas et al. 2017; Laura A. Balmaceda et al. 2018), Balmaceda et al. (2020) found that (1) the relationship between lateral expansion and radial expansion speeds is linear and does not change with height, and (2) the ratio of the bulk propagation speed to the lateral expansion speed is a function of the angular width that follows the description of self-similar evolution. They also found that most CMEs achieve a self-similar evolution above 4 Rs, which is especially applicable to impulsively accelerated events.

However, in the inner corona (e.g., < 4 Rs), CMEs would not evolve in a self-similar manner and experience the so-called over-expansion, i.e., the aspect ratio increases with height and consequently the angular width also increases with height (S. Patsourakos, Vourlidas, and Kliem 2010; L. A. Balmaceda et al. 2020; Cremades, Iglesias, and Merenda 2020). Studying a sizeable number of CMEs that could be tracked from their inception in the EUV low corona to the outer corona from multiple viewpoints, Cremades, Iglesias, and Merenda (2020) found that CME angular widths, along both toroidal and poloidal directions, increase considerably with height below ~ 3 Rs, and the growth rate along the toroidal direction is higher than that along the poloidal direction. They also found that the ratio of the two expansion speeds is nearly constant after ~4 Rs, implying that CMEs there reach a state of self-similar expansion.

## 2.2.3 CME geometric change in the corona: deflection and rotation

On the top of morphological expansion of CMEs discussed above, the geometry of CMEs evolve in a manner that deviates from the simplest behavior of straight radial motion near the Sun: (1) deflection or non-radial motion, (2) rotation resulting in the change of the tilt angle of the CME. Both deviations pose a challenge for predicting hits or misses for Earth-directed CMEs and eventually whether a given CME would be geoeffective (Kay et al. 2017a).

CME deflection has long been noticed (MacQueen, Hundhausen, and Conover 1986; N. Gopalswamy et al. 2003; Cremades and Bothmer 2004). The deflection found in these observations was restricted along the latitude only, since they were made from single viewpoint observations from spacecraft along the Sun-Earth line. The deflection has a tendency that changes the direction of motion of CMEs from high latitude toward the low latitude equator during the solar minimum. This tendency implies that the deflection during solar minimum is related to the large-scale magnetic field from polar coronal holes (N. Gopalswamy et al. 2003; Cremades and Bothmer 2004). However, during the solar maximum, the directions of deflection can be complex, i.e., toward both higher and lower latitudes from the original position angle of CMEs.

Multiple viewpoint observations from STEREO provide much improved diagnostics of CME deflections, including the time evolution of deflections along both latitudinal and longitudinal directions. (Gopalswamy et al. 2003) found that a slow CME during solar minimum was deflected toward a lower latitude region by ~30°, and demonstrated that such a deflection is caused by a non-uniform distribution of the background magnetic field, and the CME tended to propagate to the region with lower magnetic-energy density. A follow-up study on a larger sample of events further confirmed that the background magnetic field quantitatively described by the magnetic energy density control the deflection of CMEs along both longitude and latitude (Gui et al. 2011) . Kilpua et al. (2009) showed that a CME originating in a high latitude crown prominence was guided by polar coronal hole fields to the equator and produced a clear ICME in the near-ecliptic solar wind at in-situ. Such a scenario of large latitude deflection (e.g., >30°) of a high-latitude CME moving toward the equator and intercepting the Earth was also reported in Byrne et al. (2010). Besides being influenced by coronal holes, Liewer et al. (2015) attributed the rapid initial asymmetric expansion or deflection of some CMEs in the inner corona (< 1.5 Rs in EUVI FOV) to the magnetic pressure of active regions fields in the immediate vicinity of the eruption.

CME deflections in longitude were also recognized and studied, but to a lesser extent than in latitude. One of the earlier clues came from the fact that there was an east-west asymmetry of solar source regions of geoeffective CMEs, i.e., more geoeffective CMEs originated from the western hemisphere than from the eastern hemisphere (J. Zhang et al. 2003; Yuming Wang et al. 2004), thus favoring an interpretation of longitudinal deflection (Yuming Wang et al. 2004). Another line of evidence is related to the finding of "driverless" shocks found at 1 AU whose solar sources were near the solar disk center, indicating that the CME ejecta were deflected away from the Sun-Earth line (N. Gopalswamy, Mäkelä, et al. 2009b).

Direct measurements of longitudinal deflection only became possible with the advent of STEREO (Isavnin, Vourlidas, and Kilpua 2013; 2014a; Christian Möstl et al. 2015; Mays, Thompson, et al. 2015). Based on a sample of 14 events, Isavnin, Vourlidas, and Kilpua (2014) showed that most longitudinal and latitudinal deflections happened within 30 Rs, and a large part of the latitudinal deflection occurred within a few Rs. Möstl et al. (2015) studied a particularly interesting case of the January 7, 2014 CME, which originated near the disk center but was deflected toward the west by ~ 37° in longitude. Thus, this major CME (projected speed of ~2400 km/s and associated with an X1.2 flare) almost entirely missed the Earth and causing a false alarm of prediction by various space weather prediction centers. They also found that such a large longitudinal direction was attained very close to the Sun (<2.1 Rs), likely caused in this particular case by the channeling of nearby active region magnetic fields rather than coronal holes. Such a surprising geomagnetic non-event from a major disk center eruption highlights the importance of knowing the true directionality of CMEs for space weather prediction (Mays, Thompson, et al. 2015).

Another very relevant geometric evolution of CMEs is the rotation, or the change of the tilt angle of the entrained magnetic flux rope. The tilt angle is zero if the toroidal axis of CMEs lies on the equatorial plane of the Sun, and 90 degree if perpendicular to the equatorial plane. The tilt angle is critically important in deciding how much of the southward magnetic field will encounter the Earth for a given impacting CME, thus determining its expected intensity of geoeffectiveness (e.g., Bothmer and Schwenn 1998). Prior to the STEREO era, the evidence of rotation resided on the observations of erupting filaments in EUV coronal images (Ji et al. 2003; G. P. Zhou et al. 2006; L. M. Green et al. 2007). Using the orientation of the elongation of halo CMEs from single viewpoint LASCO observations as a proxy and assuming that the orientation of the post-eruption arcade in the source region is the CME orientation at the beginning of the eruption, Yurchyshyn, Abramenko, and Tripathi (2009) found that most CMEs appeared to rotate by 10°, but up to 30°-50° in some events.

Multiple viewpoint STEREO observations provide direct measurements of CME rotations in the coronagraphic field of view (A. Vourlidas et al. 2011a; Isavnin, Vourlidas, and Kilpua 2013; 2014a; Y. A. Liu et al. 2018; C. Chen et al. 2019). Using the GCS model to define and track the 3D geometry of a slow CME on 2010 June 16, Vourlidas et al. (2011) found that the CME had an initial tilt of about 30° at 2-3 Rs, very similar to the orientation of the neutral line on the surface source, but later rotated by about 60° when the CME traveled from 2 to 15 Rs. Liu et al. (2018) studied a CME on 2015 December 16 and found that the tilt from the GCS model rotated by almost 95° compared with the orientation on the source region; the same CME was also deflected by 45° in longitude and 35° in latitude. Such an extremely large rotation of the main structural axis was also found in an erupting filament based on STEREO observations in Song et al. (2018), who reported a counter-clockwise rotation of about 135° of the filament in ~26 minutes and then reversed to the clockwise rotation of 45° in about 15 minutes. Based on a statistical study of geometry of CMEs, Isavnin, Vourlidas, and Kilpua (2014) noted that the rotation largely occurred below 5 Rs, but continued in the outer corona and the interplanetary space.

## 2.2.4 Geometry of ICMEs in the interplanetary space

As discussed above, in the inner corona (~ < 4 Rs), a CME usually undergoes a super-expansion or non-self-similar increase of sizes in comparison with its distance from the Sun, and also experiences most geometric changes as defined by radial deflection and rotation of tilt angles. In the outer corona (e.g., from ~ 4 Rs to 20 Rs), on the other hand, a CME usually undergoes a self-similar expansion in all three principle directions and some relatively smaller changes in propagation direction and tilt angle. What about morphological and geometric evolution of CMEs in the interplanetary space, i.e., from ~ 20 Rs to 1 AU? Much progress has been made in the last one and half decades, thanks to the Heliospheric Imager (HI) (Eyles et al. 2009) onboard STEREO A/B. Nevertheless, the knowledge that has been gathered is largely limited, as discussed below.

Studies of using HI images from a single spacecraft usually assume that CMEs have a constant propagation direction and speed in HI FOVs. Having expanded into a huge volume thus becoming extremely faint, CMEs in HI images have a much lower signal-to-noise ratio than in coronagraphic images. Instead of forward-fitting CME appearances using a 3D geometric model such as the GCS model, HI studies often make use of time-elongation maps, or so-called J-maps (Sheeley et al. 1999), which are stack plots of slices taken along a given position angle (often along the ecliptic plane) from consecutive images of a single STEREO spacecraft (N. Lugaz, Vourlidas, and Roussev 2009; Davies et al. 2012). The slice provides a direct measure of elongation angles from the inner to the outer edges of HI FOV (Figure 2-3). Such time-elongation maps show enhanced tracks of the leading edges of a CME, but at the expense of its 3D geometry such as aspect ratios and tilt angles. The time-elongation curve of the tracked feature in the map is then used to determine the propagation longitude and speed of the feature along the selected latitude/slice. This assumption of constant propagation longitude and speed, imposed by this time-elongation map method from a single spacecraft, is not unrealistic, since it is known that most changes have occurred near the Sun in coronagraphic FOVs.

Geometrical modeling for CME on 2012 July 12-14

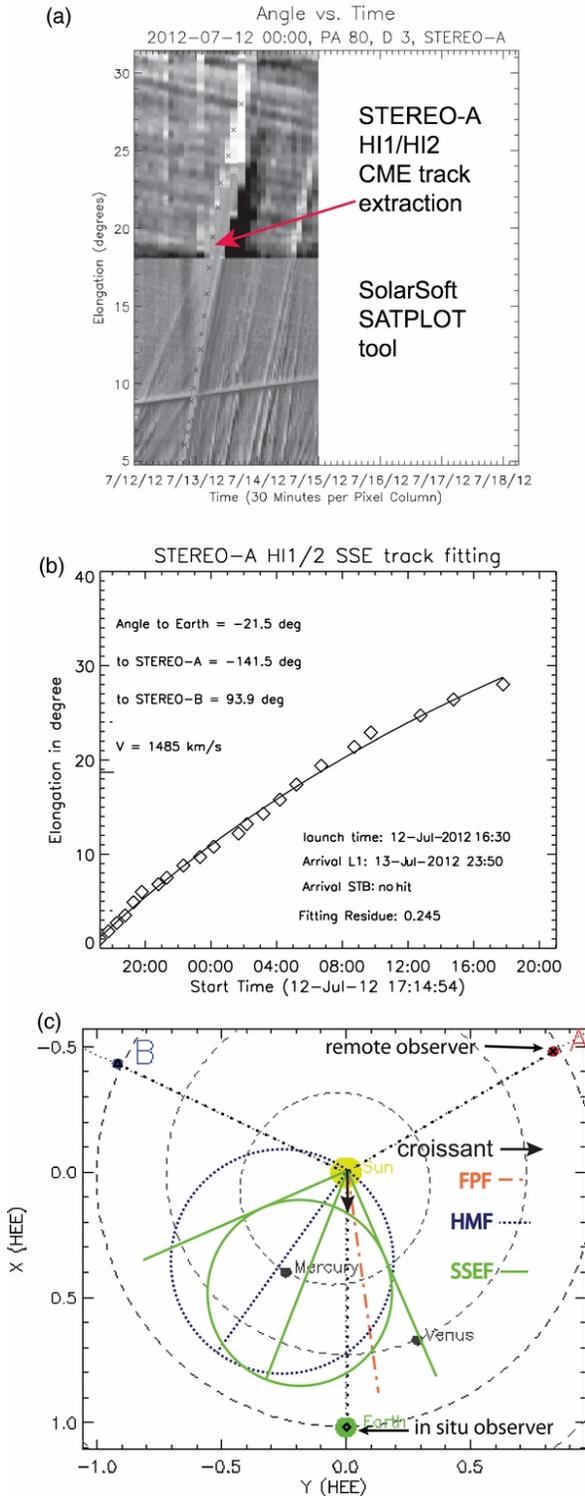

Figure 2-3. *An example of measuring CME (July 12-14, 2012 CME) leading fronts using slice-stacking plot or J-map and geometric models. (a) The density track of the CME viable in a J-map from STEREO-A. (b) Fits of the extracted CME track with the SSEF model. (c) The resulting geometry of the event, with propagation directions derived from FP (dot-dashed red line, zero degree full width), SSEF (solid green line, 90° full width), and HM (dotted blue line, 180° full width) geometric assumption of the CME. Adopted from Möstl et al. (2014).*

The orbital configuration of STEREO A/B is such that the degree to which the same CMEs are imaged by the HI cameras on both spacecraft critically depends on the mission phase (Harrison et al. 2018). The percentage of such so-called coincident events by HI-1 ranges between 40% and 90%. Note that the percentage of coincident events by COR2 is about 80% in total (Angelos Vourlidas et al. 2017). For coincident events of HI observations, one can apply a geometric triangulation technique on the time-elongation data from two spacecraft to extract instantaneous propagation longitude and distance at each time of the observation, thus allowing the time variation of propagation longitude and velocity of CMEs (Ying Liu et al. 2010; N. Lugaz 2010; Davies et al. 2013). Using stereoscopic time-elongation methods and theoretical arguments, several studied suggested possible non-radial motions of CMEs over a large distance in the heliosphere (N. Lugaz, Hernandez-Charpak, et al. 2010; Yuming Wang et al. 2014; Isavnin, Vourlidas, and Kilpua 2014a).

Nevertheless, caution is needed as the time-elongation methods do not provide a converging result on the propagation longitude when different geometric assumption of CMEs are made (Ying D. Liu et al. 2013; Davies et al. 2013) and different spacecraft are used (D. Barnes et al. 2020). Note that a set of commonly used time-elongation methods have been developed, which differ in the assumption of CME geometry on the plane containing the selected slice and the observing spacecraft: as a point or compact source (the fixed-$\varphi$, or FP method) (Sheeley et al. 1999; A. P. Rouillard et al. 2008), as a circle with the feature at the tangent front and the bottom attached to Sun-center (harmonic mean, or HM method) (N. Lugaz 2010), or as a generalized circle of certain half angle $\lambda$ (generalized self-expansion, or SEE method) (Davies et al. 2012); the FP and HM geometries form the limiting cases with $\lambda$ equals 0° and 90° respectively, while $\lambda$ can be chosen between 0° and 90° in the SEE method (Figure 2-3). Davies et al. (2013) found that the derived CME longitude is a function depending on the choice of $\lambda$, and the disparity in longitudes can be significant between the two limiting cases. In a statistical study of 273 coincident events, Barnes et al. (2020) noted that the longitude derived from single-spacecraft are in fairly poor agreement with each other, and moreover, neither agree well with the results from stereoscopic analysis. Such systematic disparity may indicate the incorrectness of the underlying assumption, i.e., the assumed circular front of CMEs may deviate significantly from the actual morphology, which will be discussed below.

## 2.2.5 Morphology of ICMEs in the interplanetary space

In contrast to the largely self-similar expansion pattern of CMEs in the corona, CMEs may undergo significant deformation in the interplanetary space, thanks to the enhanced effect of structured solar wind flows on CMEs (Odstrčil and Pizzo 1999a; Riley and Crooker 2004). It is understood that, in the regime of high plasma beta where plasma pressure dominates magnetic pressure, the magnetic structure of CMEs will be strongly modulated by the pattern of plasma flow. In the interplanetary space, the solar wind plasma flows along the radial direction but in a spherically diverging geometry. Consequently, such a flow pattern introduces the following kinematic effects on the structure of CMEs: (1) self-similar expansion along lateral directions, or directions on the spherical surface (2) no expansion at all along the radial direction that is perpendicular to the spherical surface, leading to the thinning or pancaking of the overall CME morphology (Riley and Crooker 2004). In the following, we discuss the self-similarity of CME evolution along lateral and radial directions respectively.

Observations in HI FOVs show that CMEs maintain a nearly constant angular width, indicating a self-similar expansion or constant aspect ratios along lateral directions with respect to the distance of CMEs (Wood et al. 2009; Wood et al. 2017). Note that the two principal lateral directions for a flux rope CME are along the toroidal and poloidal directions, respectively. To reproduce the observed two-dimensional loop of CME leading fronts in a flexible way, Wood et al. (2009) adopted a geometric shape described by a quasi-Gaussian equation in polar coordinates with a variable power index $\alpha$ regulating the shape of the loop; the shape is a perfect Gaussian for $\alpha=2$, while higher values of $\alpha$ result in loops with flatter tops. Using a statistical survey study of 48 events, Wood et al. (2017) noted that self-similar expansion is a decent, albeit not perfect, approximation for CMEs expanding into the interplanetary space. Such self-similar expansion along lateral directions should have continued from coronagraphic FOVs into HI FOVs.

Nevertheless, the self-similar evolution breaks down for the dimension along the radial direction. The cross section of the CME flux rope can be initially well described by a circle, as in the highly successful GCS models. Into the interplanetary space, the circular shape may evolve into a highly flattened and distorted shape, which has been described as a convex-outward pancake shape (Riley and Crooker 2004), elliptical shape (Savani et al. 2011), or even a concave shape (Savani et al. 2010). As an example of one extreme case, (Savani et al. 2010) clearly showed the observation that a circular-shaped CME in the coronagraphic FOV evolved into a concaved structure in the HI FOV, and suggested that kinematic effect of a bimodal speed solar wind caused such distortion. Therefore, the shape of the leading front of a CME can deviate significantly from a circular shape, and caution needs to be taken when a circular-shape assumption is assumed in modelling ICMEs.

Note that, besides the studies based on coronal and heliospheric imaging observations mentioned above, the geometry and morphology of CMEs can also be inferred from in-situ observations. There is a vast amount of work of fitting in-situ data to infer the structure of shocks and magnetic flux ropes, and such studies are partially reviewed in Section 3.5. In the next sub-section, we provide a review on the studies of kinematic properties of CMEs and ICMEs, which are mostly based on the time tracking of the leading fronts of the CME, instead of the 3D extension of the structure.

## 2.3 Kinematics of CMEs and ICMEs

Our knowledge about the whole kinematic evolution of CMEs from the Sun to the Earth has improved significantly in the last decade, largely thanks to the wide-angle observations of STEREO. Rising from locations above magnetic polarity inversion lines near the surface of the Sun, CMEs accelerate and reach speeds in the outer corona with a wide range of values from tens of km/s up to ~4000 km/s. The subsequent evolution of CMEs depends on their initial speeds in the outer corona relative to the speed of ambient solar wind: faster CMEs decelerate, while slower CMEs accelerate. As CMEs propagate further into the interplanetary space, their speeds tend to equalize with that of the solar wind due to the effect of aerodynamic drag (or more precisely, magnetohydrodynamic drag). For a large fraction of CMEs, the balance in speed and pressure is not established at the distance of 1 AU. The speed of ICMEs at 1 AU ranges from ~ 300 km/s to ~ 1000 km/s, meaning that it can be much faster than the ambient solar wind at 1 AU. In the following, we provide a review on the Sun-to-Earth kinematic evolution, including the phases of evolution, peak velocity, terminal velocity, cessation distance and others. The topic on the prediction of CME Time of Arrival at 1 AU will be given in Section 8.2.

Based on tens of thousands of CMEs observed, it is found that CME speed (i.e., average projected speed measured in LASCO FOV) has a very broad distribution ranging from ~10s km/s to ~ 3000 km/s (S. Yashiro et al. 2004; E. Robbrecht, Berghmans, and Van der Linden 2009; Oscar Olmedo and Zhang 2010; David F. Webb and Howard 2012; Lamy et al. 2019). The average speed of all CMEs in the various observed periods is about 300 km/s during the solar minimum and about 500 km/s during the solar maximum. Further, halo CMEs, which are the ones likely hitting the Earth, have an average speed of about 950 km/s, or about twice of that of all CMEs. Slow CMEs are quite common, as about half of CMEs in the LASCO FOV are slower or near the speed of the ambient solar wind. On the other hand, fast CMEs are equally common. Nevertheless, extremely fast CMEs, i.e., > 1500 km/s, are rather rare, occupying ~0.5% of all CMEs (Yuming Wang and Zhang 2007). The highest CME speed on the record is ~4400 km/s (N. Gopalswamy, Yashiro, et al. 2018).

Recently, Barnes et al. (2019) made a statistical study of CME kinematics in the STEREO HI-1 FOV and compared with that in the LASCO FOV. They found that the velocity distributions are similar in both areas: a sharp peak at the low end of the distribution and a long tail of high-speed CMEs. The yearly mean speeds in HI-1 FOV are consistently higher than that in LASCO; however, the two types of speeds are very similar after projecting HI speeds onto the plane of the sky. In the HI FOV, the range of CME speeds is from ~100 km/s to ~2000 km/s. It is noticed that there are very few CMEs with speeds less than 200 km/s in the HI FOV, which is of a distinct contrast with that of LASCO CMEs. This difference is certainly not surprising, as slow CMEs in the corona are picked up by the drag of ambient solar wind.

A large number of studies on individual events have provided detailed kinematic evolution of CMEs from corona and far into the inner heliosphere (B. E. Wood and Howard 2009a; Poomvises, Zhang, and Olmedo 2010; Ying Liu et al. 2010; R. C. Colaninno, Vourlidas, and Wu 2013; Hess and Zhang 2014b; Ying D. Liu et al. 2016; Yuming Wang, Zhang, et al. 2016; Brian E. Wood et al. 2017a). The observed speed profiles of three typical CMEs, which are of slow, intermediate and fast initial speeds respectively, are shown in Figure 2-4 (adopted from Liu et al. 2016). Clearly, faster CMEs decelerate and slower CMEs accelerate, as also shown in earlier studies (Sheeley et al. 1999; N. Gopalswamy et al. 2000). One of interesting results from observational studies is that there appears the existence of a cessation distance, at which a CME reaches its terminal velocity; after this distance, the CME moves at a nearly constant speed, or too small to be measured by existing imaging instruments. Note that we are cautious on the usage of the term of "terminal speed", as CME speeds will continue to change, albeit in a relatively small rate (e.g, < 1 m/s$^2$). Poomvises, Zhang, and Olmedo (2010) showed that this cessation distance was at about 50 Rs for several events including very fast ones. Using a kinematic model that divides the CME evolution into 2-4 phases of constant acceleration and constant velocity (B. E. Wood and Howard 2009a), a recent statistical study by Wood et al. (2017) showed that the cessation distance ranged from ~10 Rs to ~100 Rs , and the terminal velocity ranged from ~300 km/s to ~1200 km/s. Similar result was found in an earlier study based on Type II radio observations (Reiner, Kaiser, and Bougeret 2007).

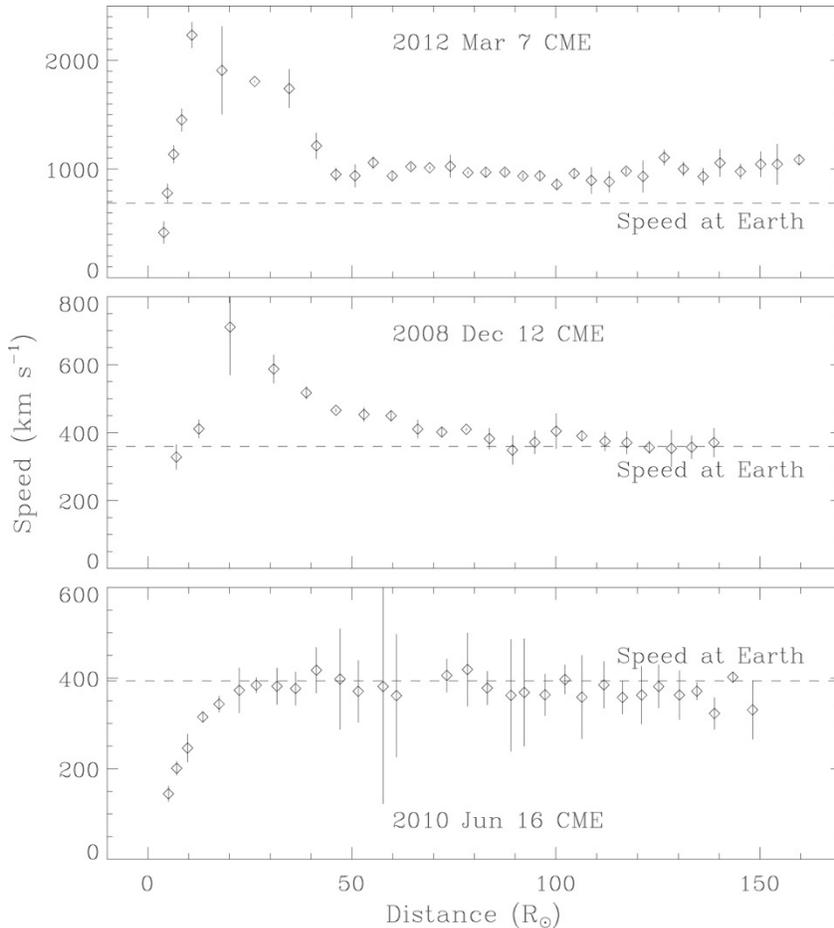

*Figure 2-4: Sun to Earth velocity profiles of a typical fast CME (upper), a typical intermediate-speed one (middle), and a typical slow one (lower). The horizontal dashed line indicates the observed speed at the Earth. Adopted from Liu et al (2016)*

CMEs reach their peak velocity at varying heights from the Sun. In general, fast CMEs reach their peak velocity at a low height, thanks to strong and impulsive acceleration, while slow CMEs reach their peak velocity at a relatively high height (J. Zhang and Dere 2006; Bein et al. 2011; 2012; Brian E. Wood et al. 2017a). Based on a statistical study of 95 events, Bein et al. (2011) found that the heights of peak velocity distribute from a very low height of 1.17 Rs (from disk center) to ~10.5 Rs (close to the border height of STEREO COR2 used in this study). A continued study by Bein et al. (2012) found that CMEs associated with flares, in comparison with CMEs associated with filaments, have on average significantly higher peak acceleration and lower height of peak velocities. Wood et al. (2017) found that the average peak-velocity-height was ~3.2 Rs for fast CMEs that were associated with flares, ~13.9 Rs for intermediate velocity CME associated with erupting filaments, and ~29.4 Rs for slow CMEs that were not associated with any apparent surface source regions.

The full Sun-to-Earth evolution of CMEs can be largely divided into four phases, each of which depends how the velocity varies and what forces drive the velocity change. Near the surface and in the corona, the full kinematic evolution of a CME can be characterized by three distinct phases: (1) a slow rise phase, or initiation phase, (2) a fast acceleration phase, or main phase, (3) a propagation phase with no or small variation of velocity (Zhang et al. 2001); this third phase is called residual acceleration phase in Zhang and Dere (2006). During the first two phases, a CME should be mainly driven by the Lorentz force. However, following the main acceleration, the Lorentz force may become significantly weaker, and the aerodynamic drag force sets in and become important. During this third phase, a CME likely experiences a combined effect of both Lorentz force and aerodynamic drag force, leading to the observed residual acceleration which can be either positive or negative (J. Zhang and Dere 2006). Further moving out, the Lorentz force eventually diminishes and the aerodynamic drag force will dominate; this phase can be considered as the 4th phase of the full evolution, or the drag phase. When only the aerodynamic drag force is considered, the kinematic evolution of a CME can be modeled in a relatively straightforward way (Cargill 2004; B. Vršnak et al. 2013). As the aerodynamic drag force is proportional to the square of the difference between the CME velocity and the ambient solar wind velocity, the CME velocity will asymptotically approach the velocity of the ambient solar wind. In other words, a faster CME decelerates and a slower CME accelerates, and the acceleration rate is not a constant but asymptotically approaches zero. For a slow CME, the full evolution may be reduced to only two phases, a gradual acceleration out to about 20-30 Rs, followed by a nearly constant speed near the solar wind level (Ying D. Liu et al. 2016). A detailed review on theories of CMEs propagation is given in Section 3.

## 2.4 Coronal Sources of Solar eruptions

The initiation and early evolution of CMEs cannot be observed using traditional coronagraphic observations, due to the blockage of the eruption region by the occulting disk. Therefore, various associated phenomena in Hα, extreme-ultraviolet (EUV), X-rays and microwaves on solar disk are linked to specific properties of the eruption and used to infer the origin of CMEs (N. Gopalswamy et al. 1999; H. S. Hudson and Cliver 2001; Harra 2009; David F. Webb and Howard 2012). Over the course of the eruption, the associated activities can be a combination of filament eruptions, solar flares, large-scale coronal EIT waves, post-eruptive arcades, and coronal dimmings. For example, the CME onset is often accompanied with the eruption of filaments/prominences that later form the inner bright core of CMEs observed in coronagraphic data (N. Gopalswamy et al. 2003; Parenti 2014a). The relationship between eruptive prominences and CMEs was investigated in several statistical studies (Munro et al. 1979; D. F. Webb and Hundhausen 1987; Hori and Culhane 2002; N. Gopalswamy et al. 2003), where an association rate of up to 90% was found.

Low coronal observations also revealed the close relationship between solar flares and CMEs (Schmieder, Aulanier, and Vršnak 2015; Vršnak 2016). Strong and powerful flares tend to be associated with fast and massive CMEs (Moon et al. 2002; Burkepile et al. 2004; B. Vršnak, Sudar, and Ruždjak 2005b; Bein et al. 2012), which results in a 90% correspondence for flares above X-class (Yashiro et al. 2006). However, there exist flares without CMEs (i.e. confined flares (e.g., Pallavicini, Serio, and Vaiana 1977; Wang and Zhang 2007; Sun et al. 2015) and vice versa, CMEs without flares (e.g. stealth CMEs) (Eva Robbrecht, Patsourakos, and Vourlidas 2009; Ma et al. 2010; Timothy A. Howard and Harrison 2013; E. D'Huys et al. 2014). A recent study by Nitta and Mulligan (2017) showed that stealth CMEs can result in significant geoeffective disturbances at 1 AU, highlighting their importance in space weather research. If CMEs and flares occur together, they are interpreted to be different parts of the same magnetically driven event (Harrison 1995; Priest and Forbes 2002; Webb and Howard 2012; Green et al. 2018).

Over the past years, it was shown that CMEs and flares are closely related in time; i.e. the SXR peak and the main acceleration phase of the CME are nearly synchronized (Zhang et al. 2001; Neupert et al. 2001; Shanmugaraju et al. 2003; Maričić et al. 2004; Vršnak et al. 2004; Zhang et al. 2004; Zhang and Dere 2006; Cheng et al. 2020). The main acceleration phase of the CME is correlated with the time evolution of the flare-related hard X-ray burst (M. Temmer et al. 2008; Gou et al. 2020) and a close relationship between their onset times was found in statistical studies (Maričić et al. 2007; Bein et al. 2012). Further evidence for a close flare/CME relationship is provided by the strong correlation between characteristic CME parameters, such as the velocity, the acceleration and its kinetic energy with the SXR peak flux, indicating the flare strength, or the integrated flux of the associated flare (Vršnak, Sudar, and Ruždjak 2005b; Maričić et al. 2007; Yashiro and Gopalswamy 2009).

Since the flare energy release rate is closely related to the magnetic reconnection rate (Miklenic, Veronig, and Vršnak 2009), a feedback relationship between the CME and its associated flare is established (J. Zhang et al. 2001b; Bojan Vršnak 2008; M. Temmer et al. 2010). Increasing reconnection rates enhance CME acceleration, and vice versa, enhanced acceleration provides more efficient reconnection. Studies showed the correlation between CME velocities and the total reconnection flux supporting this interpretation (Qiu and Yurchyshyn 2005; Miklenic, Veronig, and Vršnak 2009; Tschernitz et al. 2018; N. Gopalswamy, Akiyama, et al. 2018; Pal et al. 2018). The most recent study by Zhu et al. (2020) even directly proves this interpretation observationally by reporting on a strong correlation between the reconnection rates, estimated by flare ribbons and CME accelerations (c>0.7). Interestingly, they also report on a positive correlation between the maximum speed of CMEs and the total reconnection flux but only for fast CMEs (v>600 km/s). For slow CMEs with weak reconnection other physical processes may play a more important role during acceleration than magnetic reconnection.

The initial lateral expansion of the CME also drives fast-mode magneto-sonic waves observed as large-scale perturbations of enhanced EUV emission, so-called EIT waves (Thompson et al. 1999; Spiros Patsourakos and Vourlidas 2009; Long et al. 2016). Their speeds typically range from 200-400 km/s (Klassen et al. 2000; Thompson and Myers 2009; Muhr et al. 2014), but also EIT waves with speeds up to 1000 km/s have been reported (Nitta et al. 2013; Seaton and Darnel 2018). Statistical studies revealed that fast and wide CMEs are in general accompanied with well-observed EIT waves often associated with shocks and therefore also related with type-II radio bursts (Biesecker et al. 2002; E. W. Cliver et al. 2005; Nitta et al. 2013; 2014; Muhr et al. 2014; Warmuth 2015). Combining type II radio burst observations with EUV waves observed by SOHO and STEREO, Gopalswamy et al. (2013) found that the EUV waves are shocks forming very close to the Sun - as low as 0.2 Rs above the solar surface. For the physical mechanisms leading to the shock wave formation and coronal and chromospheric response, see, e.g., Vrsnak et al. (2016) and references therein.

After the CME has erupted, bright post-eruptive arcades or post-flare loops appear in soft X-ray and EUV (Kahler 1977; McAllister and Hundhausen 1996; Tripathi, Bothmer, and Cremades 2004) as a consequence of magnetic reconnection processes (Kopp and Pneuman 1976). Tripathi, Bothmer, and Cremades (2004) statistically analyzed post-eruptive arcades using data from SOHO/EIT. They found that the majority of post-eruptive arcades (92%) were associated with CMEs identified in SOHO/LASCO.

Due to the expansion of the CME volume and evacuation of plasma during the eruption, regions of decreased emission in soft X-rays and EUV are formed, so-called coronal dimmings (Hudson et al. 1996; Thompson et al. 2000; Harra and Sterling 2001; Vanninathan et al. 2018) . As they represent the lower footprint of CMEs in the low corona, their properties are closely related to the initial properties of the observed CME later on. For instance, several studies tried to relate the mass loss within coronal dimming regions to the CME mass measured from coronagraphic observations (Harrison and Lyons 2000; Zhukov and Auchère 2004; Aschwanden et al. 2009; López et al. 2019).

Recently performed statistical studies confirm the close connection between coronal dimmings and CMEs and found that the dimming area, its total magnetic flux and its brightness are strongly correlated with the CME mass (Dissauer et al. 2018; 2019; Sindhuja and Gopalswamy 2020). Dimming parameters, describing its dynamics, such as the area growth rate, brightness change rate and magnetic flux change rate, are tightly related to the CME speed. This is in agreement with results of J. P. Mason et al. (2016) who studied coronal dimmings extracted from full-disk irradiance light curves of SDO/EVE (EUV Variability Experiment ).

A number of studies also successfully compared magnetic flux rope properties, such as the magnetic flux, the chirality, and its helicity sign determined from post-eruptive arcades, flare ribbons, and coronal dimmings measured close to the Sun with magnetic cloud properties at 1 AU (Qiu et al. 2007; Yurchyshyn 2008; Q. Hu et al. 2014; K. Marubashi et al. 2015; N. Gopalswamy et al. 2017; E. Palmerio et al. 2017; 2018; James et al. 2017; Aparna and Martens 2020). The total amount of magnetic flux ejected during an eruption is estimated by the total reconnection flux in the wake of the CME or sometimes also by the magnetic flux involved in coronal dimming regions, which form the footprint of CMEs in the low corona (Mandrini et al. 2005; Attrill et al. 2006; Qiu et al. 2007; Q. Hu et al. 2014). Especially the total reconnection flux strongly correlates with the magnetic flux of magnetic clouds (Qiu et al. 2007; Q. Hu et al. 2014).

The helicity sign and the total amount of helicity of magnetic clouds at 1 AU seem to be strongly controlled by the location and properties of the solar source region (Cho et al. 2013; Q. Hu et al. 2014; K. Marubashi et al. 2015). CMEs erupting in the southern (northern) hemisphere tend to have a positive (negative) helicity sign (hemispheric helicity rule, e.g. Pevtsov, Balasubramaniam, and Rogers 2003). Recently, Aparna and Martens (2020) investigated the directionality (chirality) of 86 CMEs-ICME pairs by comparing the orientation of their flux rope axes close to the Sun with the direction of the interplanetary magnetic field near Earth at L1. An agreement between the northward/southward orientation of $B_z$ between ICMEs and their CME source regions was found in 85% of the cases, which is comparable to earlier results by Palmerio et al. (2018) and Yurchyshyn (2008), which found agreement for 55% and 77% of their cases.

In recent years, several studies also focused on Sun-to-Earth analysis of CMEs by linking the low coronal behaviour and properties of the eruption with its observed in-situ signature (Christian Möstl et al. 2015; Patsourakos et al. 2016; D'Huys et al. 2017; Manuela Temmer et al. 2017b). A number of studies also compared magnetic flux rope properties, such as the magnetic flux, the chirality, and its helicity sign determined from post-eruptive arcades, flare ribbons, and coronal dimmings measured close to the Sun with magnetic cloud properties at 1 AU (Qiu et al. 2007; Gopalswamy et al. 2017; Palmerio et al. 2017; James et al. 2017). Scolini et al (2019) used proxies of magnetic flux estimates determined from post-flare arcades (Gopalswamy et al. 2017), flare ribbons (Kazachenko et al. 2017; Tschernitz et al. 2018) as well as coronal dimmings (Dissaier et al. 2018), as initial input for the global heliospheric EUHFORIA model, to study the geoeffectiveness of the famous 2017 September events. Good agreement with the observed Dst profile was found for simulations using the optimized input and including CME-CME interactions.

## 2.5 Solar cycle variations of CMEs and ICMEs

Solar cycle 24 is known to be weaker than previous several solar cycles, which is the focus of many studies during the VarSITI program. A weak solar cycle 24 is understood to be due to the weak polar magnetic field in the preceding solar minimum according to the Babcock Leighton Mechanism of solar cycle (see e.g., Petrovay 2010). A weak cycle implies mild space weather that helps satellites in Earth orbit live longer. A weak cycle also means less total solar irradiance reaching Earth (e.g., Krivova and Solanki 2008). Here we focus on the effect of weak solar cycle on solar wind magnetic structures originating from the Sun and their space weather consequences.

Both solar source and impact of CMEs showed significant variations in cycle 24. The overall rate of CMEs increased in solar cycle 24 relative to cycle 23, although the rate of fast and wide CMEs decreased. Accordingly, the phenomena that are linked to fast and wide (FW) CMEs appeared subdued in cycle 24. The rate of occurrence of CMEs is known to be correlated with the sunspot number (SSN) for a long time. However, the slope of the regression line is significantly different in cycle 24. The relation between CME width and speed is also different in cycle 24: for a given speed, cycle-24 CMEs are significantly wider. CMEs are the main source of severe space weather. Weakened solar activity is reflected in the weak heliospheric state in terms of magnetic field strength, temperature, density, speed, and consequently the total pressure. The backreaction of the weakened heliosphere had led to the changed properties of CMEs and hence affected the space weather consequences. Marked reductions are observed in the number of intense (Dst ≤ -100 nT) geomagnetic storms and high-energy (≥ 500 MeV) solar energetic particle (SEP) events. The number of halo CMEs in cycle 24 did not decrease significantly. In fact, the number of halo CMEs normalized to the sunspot number is larger in cycle 24. One would have expected enhanced geomagnetic activity in cycle 24 because of the higher abundance of halo CMEs, but it did not occur. In this section, we summarize some of the key observational results that describe the compound effect of the weak solar activity and heliospheric backreaction on CME properties.

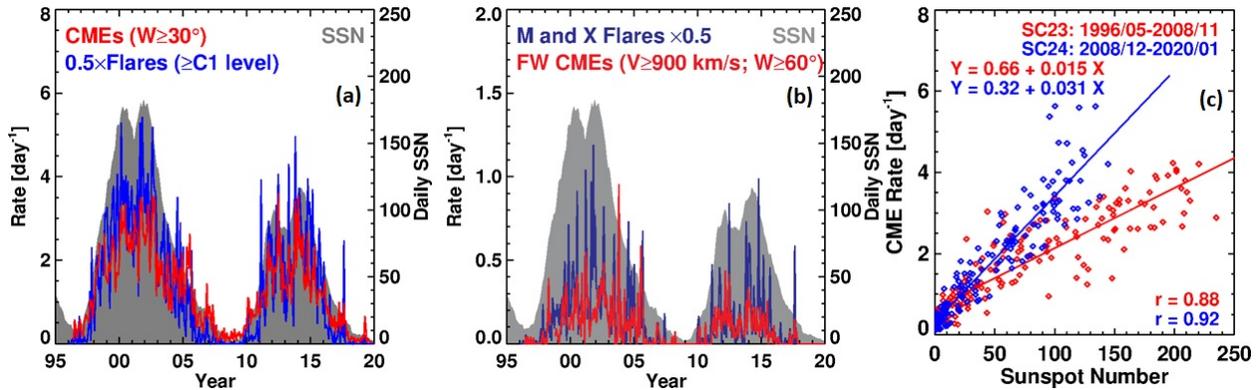

*Figure 2-5: Solar cycle variation of eruptive phenomena (flares and CMEs): (a) daily CME rate for the general population (width ≥30º) in red and the number of soft X-ray flares with size ≥C1.0 in blue, (b) fast and wide CMEs (speed ≥900 km/s; width ≥60º) in red and major (M- and X-class) soft X-ray flares in blue, and (c) scatter plots between the sunspot number (SSN, V2.0) and the daily CME rate (general population) for cycle 23 (red) and 24 (blue). In (a) and (b), SSN is shown in gray background. The CME and flare rates are averaged over the Carrington rotation (CR) periods. The flare rates are multiplied by 0.5 to fit the scale. Updated from Gopalswamy et al. (2020); Gopalswamy, Akiyama, and Yashiro (2020).*

## 2.5.1 Solar activity and eruption properties

Figure 2-5 shows the solar-cycle variation of CMEs and flares compared to SSN cycles 23 and 24 updated from (Gopalswamy et al. 2020; Gopalswamy, Akiyama, and Yashiro 2020). We can readily infer the following: (i) the daily rate of the general population of CMEs (width ≥30º) and that of the soft X-ray flares (size ≥C1.0) did not decline in cycle 24, (ii) the FW CME rate declines significantly in cycle 24 (as opposed to the general population), (iii) the CME daily rates have different relationship with SSN in the two cycles, (iv) the variation in the number of FW CMEs is similar to that of major soft X-ray flares (M- and X-class flares), and (v) the CME rate increases more rapidly as the activity increased, indicated by the steeper slope in the cycle-23 CME rate – SSN scatter plot. The reduction in FW CMEs is significant because they are the ones that are relevant for space weather consequences (geomagnetic storms and SEP events).

Petrie (2015) studied SOHO/LASCO CMEs with angular widths >30° listed in the manual (CDAW) and automatic (CACTus and SEEDS) catalogs. He found that the CME rate relative to the sunspot number began an upward divergence with respect to the SSN in 2004 after the polar field reversal, while the interplanetary magnetic field decreased by ~30% around the same time (see Fig. 2.3 a). These results are consistent with the enhanced halo CME detections due to the increased CME expansion in a heliosphere with diminished total pressure (Gopalswamy, Xie, et al. 2015). Petrie (2015) also showed that the increased CME rate in cycle 24 is not due to the LASCO cadence change that occurred in August 2010 (Wang and Colaninno 2014; Hess and Colaninno 2017). The cadence change was also found to be not important for halo CME rates (Gopalswamy, Xie, et al. 2015). Michalek, Gopalswamy and Yashiro (2019) showed that the higher rate of the narrow CMEs can be attributed to the global magnetic structure in cycle 24 coupled with the reduced total pressure in the heliosphere, in agreement with Petrie (2015).

## 2.5.2 Phenomena associated with energetic CMEs

Figure 2-6 shows the solar cycle variation of the numbers of halo CMEs, intense geomagnetic storms, large SEP events, and decameter-hectometric (DH) type II bursts. All the numbers are summed over Carrington rotation periods. The properties of halo CMEs have a good overlap with FW CMEs, although some halos are wide, but not fast. Halo CMEs, when front-sided, are important because they can affect space weather. Intense geomagnetic storms are mostly due to energetic CMEs heading toward Earth containing southward magnetic field component either in the ejecta part or in the shock sheath. On the other hand, the shock at the leading edge of CMEs accelerate electrons and ions. The accelerated electrons produce type II radio bursts that are a good indicator of shock-driving CMEs near the Sun and in the interplanetary medium. Accelerated particles traveling along interplanetary field lines are detected as SEP events. We see that the number of events in each case generally follows the solar cycle with more events occurring during solar maxima. A notable exception is the number of intense geomagnetic storms in cycle 24 that remained flat.

### 2.5.2.1 Halo CMEs in solar cycle 24

As reported in (Gopalswamy, Xie, et al. 2015; Gopalswamy et al. 2020; Gopalswamy, Akiyama, and Yashiro 2020), the halo CME rate in cycle 24 did not decline commensurate with the SSN. There are roughly the same number of halos in the two cycles. When normalized to SSN, cycle 24 has ~30% more halos (per SSN). As with the general population of CMEs, the halo CMEs are slower in cycle 24. Furthermore, CMEs in cycle 24 become halos at shorter heliocentric distances than in cycle 23. These authors attributed the peculiar behavior of halo CMEs to the diminished heliospheric total pressure in cycle 24 that made relatively slower CMEs and those originating at larger central meridian distances become halos. Solar wind parameters measured at Sun-Earth L1 confirm that most of the parameters have lower values in cycle 24. In particular, the decline is the heliospheric field strength resulted in the diminished total pressure and the Alfven speed (Gopalswamy et al. 2020).

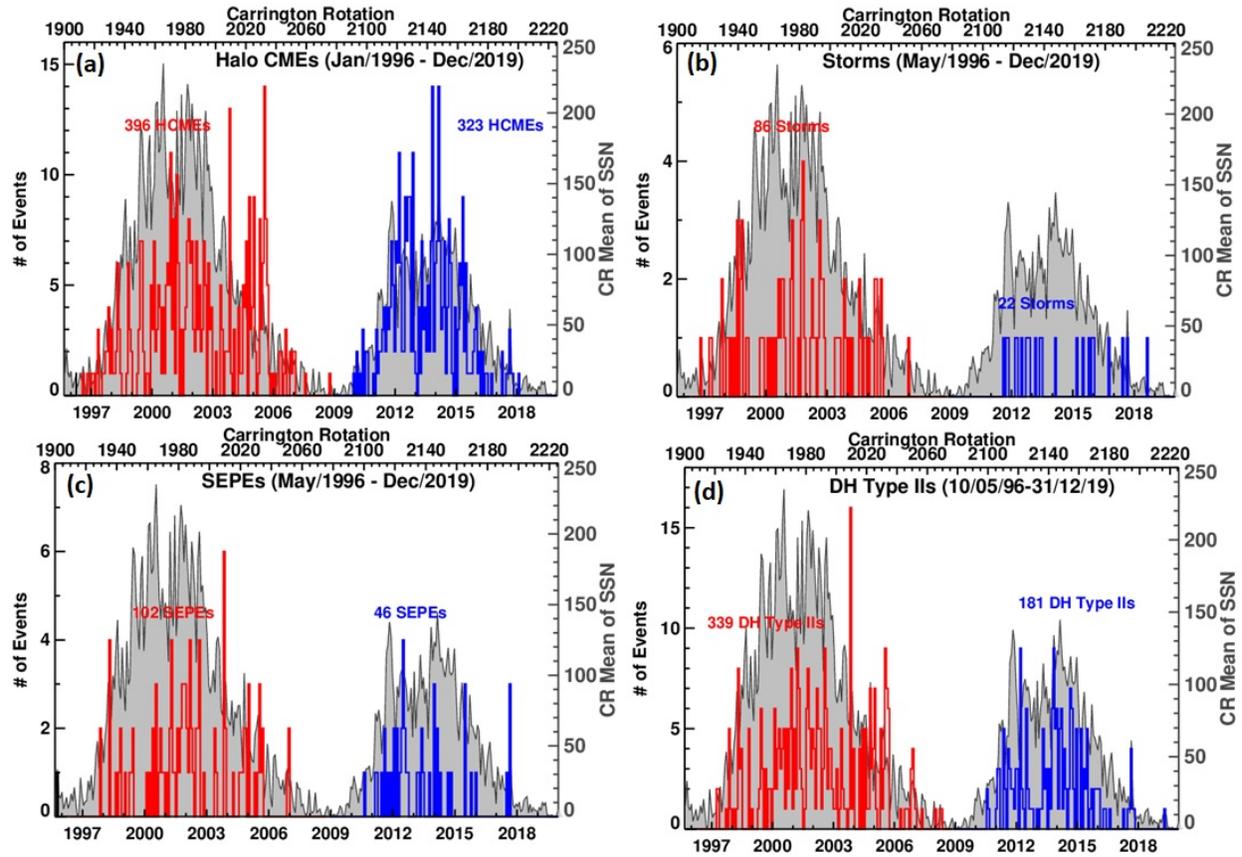

*Figure 2-6. Solar cycle variation of the key CME related phenomena during cycles 23 and 24 (from 1996 to the end of 2019): (a) SOHO/LASCO halo CME number from https://cdaw.gsfc.nasa.gov/CME_list/halo/halo.html. (b) the number of large geomagnetic storms (Dst ≤ -100 nT), (c) the number of large solar energetic particle events (SEPEs) detected in the >10 MeV GOES channel with intensity exceeding 10 pfu, and (d) the number of decameter-hectometric (DH) type II radio bursts from Wind/WAVES. For comparison with the solar cycle, the sunspot number averaged over Carrington rotation (CR) periods is also shown.  Updated from Gopalswamy et al. (2015).*

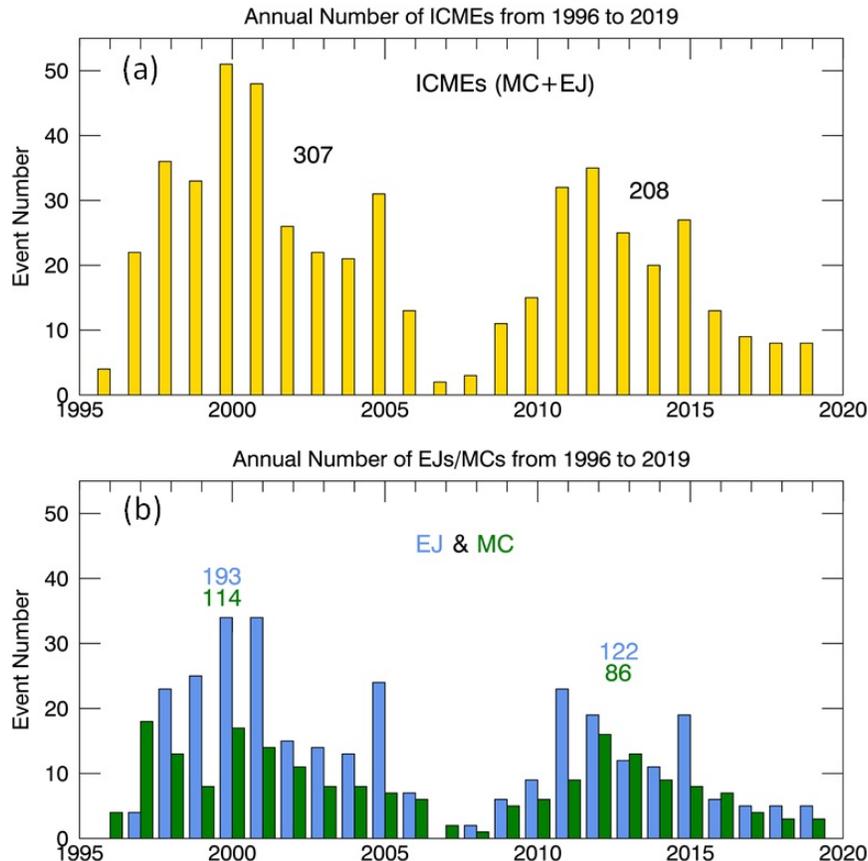

*Figure 2-7. (a) Solar cycle variation of the annual number of ICMEs (magnetic clouds – MCs and non-cloud ejecta - EJs) observed near Earth as compiled in http://www.srl.caltech.edu/ACE/ASC/DATA/level3/icmetable2.htm. The total number of ICMEs in cycles 23 (307) and 24 (208) are noted on the plot. (b) Annual number of EJs and MCs shown separately. The number of EJs and MCs are noted on the plot.*

### 2.5.2.2 Interplanetary CMEs in solar cycle 24

The reduction in the number of FW CMEs in cycle 24 is expected to be reflected in the number of interplanetary CMEs (ICMEs) observed at 1 AU because the latter are known to be associated with energetic CMEs. The annual number of ICMEs in Figure 2-7 shows clear solar-cycle variation, similar to other phenomena. The total number of ICMEs declined from 307 in cycle 23 to 208 in cycle 24, representing a reduction of 32%. The decline is smaller than that in SSN. One possible explanation is that ICMEs also originate from non-spot regions where quiescent filaments erupt and the associated CMEs become ICMEs. When normalized to the cycle-averaged SSN in each cycle (81 in cycle 23 dropping to 49 in cycle 24), we see a 12% increase in the number of ICMEs. If we separate the ICMEs into magnetic clouds (MCs) and non-cloud ejecta (EJ), we see a similar trend, but the decline in EJ (37%) is steeper than in MCs (25%). When normalized to SSN, we see an increase of 4% and 25% for EJs and MCs, respectively. This behavior was noted previously for MCs detected during the rise and maximum phases of cycles 23 and 24 (Gopalswamy, Yashiro, et al. 2015). The fraction of MCs is also slightly higher in cycle 24: 37% (114 out of 397 ICMEs) vs. 41% (86 out of 208 ICMEs) in cycle 23.

## 2.5.3 Stream Interaction Regions

While CMEs cause most of the intense geomagnetic storms, stream interaction regions (SIRs) and especially co-rotating interaction regions (CIR) can cause relatively weaker but more frequent geomagnetic storms. The solar sources of SIRs and CIRs also have been shown to be different in cycles 23 and 24. Nakagawa, Nozawa, and Shinbori (2019) studied the temporal and spatial variations of the low-latitude coronal hole (CH) area related to high-speed solar (HSS) wind during solar cycles 23 (1996–2008) and 24 (2009–2016). They found that (i) the CHs in solar cycle 24 appeared over a wider latitude range than in solar cycle 23, and (ii) the maximum values of the CH area and the solar wind speed in solar cycle 24 were smaller than those in solar cycle 23. Jian et al. (2019) compared the annual occurrence rates of SIRs and CIRs using Wind/ACE and STEREO in situ observations from 1995 to 2016 as displayed in Figure 2-8. They found a higher occurrence rate of SIRs in cycle 24, which they attribute to the presence of persistent equatorial coronal holes as well as weaker CMEs in that cycle. The fraction of CIRs is higher in the declining to minimum phase of each cycle. Grandin, Aikio, and Kozlovsky (2019) developed a catalog of 588 HSS and SIR events that occurred during the interval 1995 to 2017. Their list is largely in agreement with the list of SIRs identified manually by Jian et al. (2019).

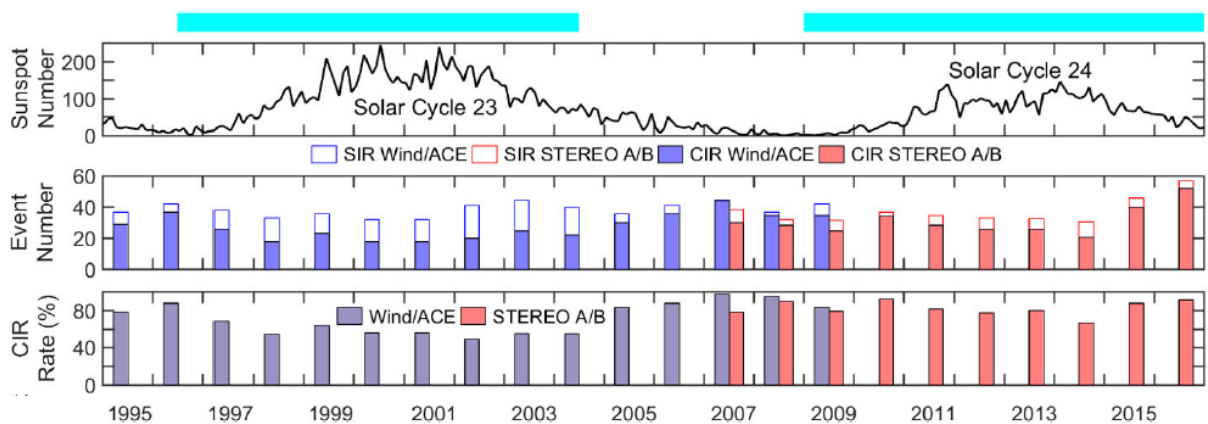

*Figure 2-8. Monthly sunspot number (top), number of SIR/CIR (middle), and CIR rate (bottom) in solar cycles 23 and 24. Corresponding epochs in cycles 23 and 24 are indicated by the cyan bars at the tope (adapted from Jian et al. 2019).*

## 2.5.4 Geoeffectiveness

A severe reduction in the geoeffectiveness of CMEs in cycle 24 as measured by the number of intense (Dst ≤ -100 nT) geomagnetic storms has been reported  (Gopalswamy, Akiyama, et al. 2014; Gopalswamy, Yashiro, et al. 2015) in the rise to the maximum phases. In the updated version covering two full cycles shown in Fig. 2-4 b we see that there are 86 intense storms in cycle 23 compared to just 22 in cycle 24. These include 11 CIR storms in cycle 23 and 2 in cycle 24 (an 82% reduction). The first intense geomagnetic storm occurred during the maximum phase of cycle 24 on 2013 June 1 due to a low-latitude coronal hole that was at the central meridian on 2013 May 30 (Gopalswamy, Tsurutani, and Yan 2015). The second storm occurred on 2015 October 6 due to a coronal hole that was present at the central meridian on 2015 October 5 (Watari 2018). The number of intense storms due to CMEs declined by ~73% from 75 to 20. Both these reductions can be attributed to the dilution of the CME and CIR magnetic content.

Considering MCs that occurred during the first 6 years in each of cycles 23 and 24, it was found that the MC-associated storms were weaker in cycle 24: the average value of Dst increased from −66 to −55 nT. Scolini et al. (2018) investigated the geoeffectiveness of halo CMEs in cycle 23 and 24. They found that during the first 85 months of Cycle 23 the geoeffectiveness rate of the disk-center full-halo CMEs was 58% compared to 35% in cycle 24. The average minimum value of the Dst index was −146 nT in cycle 23 compared to -97 nT in cycle 24. These results are consistent with the reduced geoeffectiveness of cycle-24 MCs (Gopalswamy, Yashiro, et al. 2015). Hess and Zhang (2017) studied 70 Earth-affecting ICMEs in Solar Cycle 24 and found the lack of events resulting in extreme geomagnetic storms.

Selvakumaran et al. (2016) considered the moderate and intense geomagnetic storms that occurred during the first 77 months of solar cycles 23 and 24. While they confirmed an 80% reduction in the occurrence of intense storms, they found that the number of moderate storms decreased by only ~40%. When moderate storms from CIRs are considered separately, these authors found that the reduction in the number of geomagnetic storms is more drastic (see also Grandin, Aikio, and Kozlovsky 2019). In cycle 23, there were 43 CIR storms during the first 77 months of cycle 23 compared to 15 in cycle 24, amounting to a reduction of 63%. This is about the same when the CIR storms are separated into moderate and intense storms. Chi et al. (2018) studied the geoeffectiveness of SIRs from 1995 to 2016 and found that about 52% of the SIRs caused geomagnetic storms with Dst ≤−30 nT, but only 3% of them caused intense geomagnetic storms (Dst ≤−100 nT). They also reported that the possibility of SIR-ICME interaction structures causing geomagnetic storms is significantly higher than that of isolated SIRs or isolated ICMEs.

## 2.5.5. Solar Cycle Variation of Large SEP events

As of December 2019, there were 46 large SEP events (>10 MeV proton intensity ≥10 pfu) in cycle 24 compared to 102 in cycle 23 (see Fig. 2-4 c). This is a 55% reduction, more than the reduction in the SSN. Considering the period up to the middle of the maximum phase, the reduction reported previously was smaller (~ 26%, Gopalswamy, Tsurutani, and Yan 2015), most likely due to the lower Alfven speed in the rise phase of cycle 24. If we consider the highest energy particle events, viz., the ground level enhancement (GLE) events, the reduction is very drastic: there were 16 GLE events in cycle 23 compared to just 2 in cycle 24, amounting to a reduction of 88%. All large SEP events are associated with DH Type II bursts because the same CME-driven shock accelerates electrons (producing type II bursts) and ions (observed in space as SEP events). CMEs associated with SEP events and DH type II bursts are typically fast and wide (N. Gopalswamy, Mäkelä, and Yashiro 2019). Not all DH type II bursts are associated with SEP events because of connectivity issues, high particle background, and <10 pfu events. The number of DH type II bursts decreased from 339 in cycle 23 to 181 in cycle 24 (see Fig. 2.4.2d), indicating a decline by 47%, very similar to the reduction in FW CMEs (50%) because most of the type II bursts are due to FW CMEs.

*Table 2-1. Earth-affecting events in solar cycles 23 and 24 in comparison with the SSN*

| Property | Cycle 23 | Cycle 24 | Change | Change/SSN |
|---|---|---|---|---|
| SSN | 81 | 49 | - 39% | 0% |
| Major Flares (M&X) | 1584 | 809 | -49% | -16% |
| X-class flares | 128 | 49 | -62% | -37% |
| M-class flares | 1456 | 760 | -48% | -14% |
| All CMEs (Width ≥ 30º)[a] | 8429[c] | 8470 | +1% | +66% |
| Fast & Wide CMEs[a] | 501[c] | 253 | -50% | -17% |
| Halo CMEs[a] | 409[c] | 323 | - 21% | +30% |
| ICMEs[b] | 307 | 208 | - 32% | +12% |
| Magnetic clouds (MC) | 114 | 86 | - 25% | +25% |
| Ejecta (EJ) | 193 | 122 | - 37% | +4% |
| DH Type II bursts | 339 | 181 | -47% | -12% |
| Large SEP events | 102 | 46 | -55% | -26% |
| GLE events | 16 | 2 | -88% | -79% |
| Magnetic Storms (Dst < -100 nT) | 86 | 22 | - 74% | -58% |
| CIR storms | 11 | 2 | - 82% | -70% |
| CME storms | 75 | 20 | -73% | -56% |

[a]from the search engine in the CDAW catalog (http://www.lmsal.com/solarsoft/www_getcme_list.html)
[b]from  http://www.srl.caltech.edu/ACE/ASC/DATA/level3/icmetable2.htm
[c]includes 223 Width ≥ 30º CMEs, 16 FW CMEs, and 13 halo CMEs estimated to have occurred during the 4-month SOHO/LASCO data gap

## 2.5.6 Summary of Solar Cycle Variation

Table 2-1 summarizes the Earth-affecting events in solar cycles 23 and 24 in comparison with the SSN. In addition to SSN, we have shown major soft X-ray flares, regular CMEs, fast and wide CMEs, halo CMEs, and ICMEs followed by the heliospheric consequences (DH type II bursts, large SEP events, and major geomagnetic storms). The decline in solar activity in cycle 24 is represented by the 39% drop in the cycle-averaged SSN.  Most events declined more than the SSN did in cycle 24, except for ICMEs (declined by 32%), halo CMEs (declined by 21%) and the general population (W ≥ 30º) of CMEs (increased by 1%). When the numbers are normalized to SSN, these three types of events show clear increase in cycle 24. Fast and wide CMEs declined in cycle 24 by ~47%, slightly more than the SSN did. When normalized to SSN, the number of FW CMEs per SSN declined only by 17% in cycle 24, somewhat similar to the decline in major soft X-ray flares and DH type II bursts. The general population of CMEs with width >30º clearly showed no decrease in number; when normalized to SSN, the average number of CMEs per SSN increased by 66%.  The two phenomena that showed the deepest decline are the major geomagnetic storms (74%) and GLE events (88%). Both these events are related to the heliospheric magnetic field strength, which significantly declined in cycle 24. The reduced heliospheric magnetic field results in the weaker heliospheric pressure leading to the anomalous expansion of CMEs and the attendant magnetic dilution in CMEs. In the case of SEP events, the particle acceleration efficiency depends on the strength of the heliospheric magnetic field, hence a reduction in the latter results in less efficient acceleration and hence particles do not attain high energies (Gopalswamy, Akiyama, et al. 2014). The reduction in the heliospheric magnetic field

is also likely to be responsible for weaker CIR storms: the decreased MHD compression of the weaker field does not increase it to high levels.

## 2.6 Summary

To conclude this section, we provide a list of catalogs of Earth-affecting transient events that have been compiled and maintained online by many workers in the past. The catalogs include lists of observed solar flares, CMEs, ICMEs in the inner heliosphere, ICMEs at in-situ, interplanetary shocks, ICMEs with solar sources, SIRs and SEPs. These event catalogs are useful resources for facilitating research for the wide community.

*Table 2-2. Catalogs of Earth-affecting solar transients, including flares, CMEs, ICMEs-IH (Inner Heliosphere), ICMEs-IS (In-situ), Shocks, SIRs and SEPs.*

| Type | Acronym | Description and link | Reference(s) |
|------|---------|----------------------|--------------|
| Flares | -- | Solarsoft latest events on solar flares https://www.lmsal.com/solarsoft/latest_events/ | -- |
| | | | |
| CMEs | CDAW | SOHO CME catalog https://cdaw.gsfc.nasa.gov/CME_list/ | (S. Yashiro et al. 2004) |
| CMEs | SEEDS | SOHO and STEREO CME catalogs based on automated method. http://spaceweather.gmu.edu/seeds/ | (O. Olmedo et al. 2008) |
| CMEs | CACTUS | SOHO and STEREO CME catalogs based on automated method http://sidc.oma.be/cactus/ | (E. Robbrecht and Berghmans 2004) |
| CMEs | ARTEMIS | SOHO CME catalog based on automated method http://cesam.lam.fr/lascomission/ARTEMIS/index.html | (Boursier et al. 2005) |
| CMEs | CORIMP | SOHO CME catalog based on automated method http://alshamess.ifa.hawaii.edu/CORIMP/ | (Byrne et al. 2012) |
| CMEs | -- | STEREO COR1 catalog, including CMEs and other events https://cor1.gsfc.nasa.gov/catalog/ | -- |
| CMEs | MVCC | STEREO Dual-viewpoint CME catalog http://solar.jhuapl.edu/Data-Products/COR-CME-Catalog.php | (Angelos Vourlidas et al. 2017) |
| CMEs | KINCAT | STEREO COR2 CMEs (2007-2013) with GCS model results http://www.affects-fp7.eu/cme-database/index.php | (Bosman et al. 2012) |
| | | | |
| ICMEs-IH | HELCATS | STEREO HI event catalogs including HICAT, HIJoinCAT, HIGeoCAT http://www.helcats-fp7.eu/ | (Harrison et al. 2018) |
| | | | |
| ICMEs-IS | -- | ACE ICMEs since 1996 complied by Richardson & Cane http://www.srl.caltech.edu/ACE/ASC/DATA/level3/icmetable2.htm | (I. G. Richardson and Cane 2010) |
| ICMEs-IS | -- | WIND ICME catalog (1995-2015) https://wind.nasa.gov/ICME_catalog/ICME_catalog_viewer.php | (T. Nieves-Chinchilla, Vourlidas, et al. 2018) |
| ICMEs-IS | -- | WIND Magnetic Cloud list (1995-2006) https://wind.nasa.gov/mfi/mag_cloud_pub1.html | (Lepping and Wu 2007) |
| ICMEs-IS | -- | WIND ICME catalog (1995-2015) http://space.ustc.edu.cn/dreams/wind_icmes/ | (Chi et al. 2016) |
| ICMEs-IS | -- | ICMEs and other large scale structures in solar wind ftp://www.iki.rssi.ru/pub/omni/ | (Yu. I. Yermolaev et al. 2009) |
| | | | |
| Shocks | -- | CfA Interplanetary Shock Database https://www.cfa.harvard.edu/shocks/ | -- |
| Shocks | -- | Heliospheric shock database at the University of Helsinki http://ipshocks.fi/ | (E. K. J. Kilpua et al. 2015) |
| | | | |
| ICMEs-CMEs | -- | ICMEs and their solar sources in solar cycle 24 from GMU http://solar.gmu.edu/heliophysics/index.php/GMU_CME/ICME_List | (Hess and Zhang 2017) |
| | | | |
| Coronal Holes | -- | Coronal holes during SDO era list https://cdsarc.unistra.fr/viz-bin/cat/J/other/SoPh/294.144 | (Stephan G. Heinemann, |

| | | | Temmer, Heinemann, et al. 2019) |
|---|---|---|---|
| SIRs | -- | STEREO SIR list http://www-ssc.igpp.ucla.edu/forms/stereo/stereo_level_3.html | (Jian et al. 2019) |
| | | | |
| SEPs | -- | Solar Proton Events from SWPC https://umbra.nascom.nasa.gov/SEP/ | -- |
| SEPs | -- | > 25 MeV Proton Events Observed by the High Energy Telescopes on the STEREO A and B Spacecraft and/or at Earth | (I. Richardson et al. 2014) |
| SEPs | -- | Catalogue of 55-80 MeV solar proton events extending through solar cycles 23 and 24 | (Paassilta et al. 2017) |
| SEPs | -- | STEREO/SEPT Solar Energetic Electron Event List http://www2.physik.uni-kiel.de/stereo/downloads/sept_electron_events.pdf | (Nina Dresing et al. 2020) |

## 3. Progress in Theories of CMEs and ICMEs

## 3.1 Introduction

It is generally accepted that CMEs are driven by the free magnetic energy stored in the non-potential magnetic fields. There is a general consensus that the erupted structure is twisted, where the most common magnetic structure employed in modeling is a flux rope, i.e., a cylindrical plasma structure with magnetic field draped around the central axis (Lepping, Jones, and Burlaga 1990). The eruption of the twisted magnetic structure is interrelated with the magnetic reconnection of the surrounding coronal magnetic field, releasing both thermal and non-thermal energy and producing a number of effects (Priest and Forbes 2002). The magnetic dips of the flux rope can support cool plasma, in which case also an eruptive filament can be observed. The whole process is known as "the standard flare-CME model" and is sketched in the left panel of Figure 3-1. In "the standard magnetic cloud model", the erupting flux rope propagates away from the Sun, expanding at the same time, but stays attached to the Sun, i.e. remains a closed structure, as shown in the right panel of Figure 3-1.

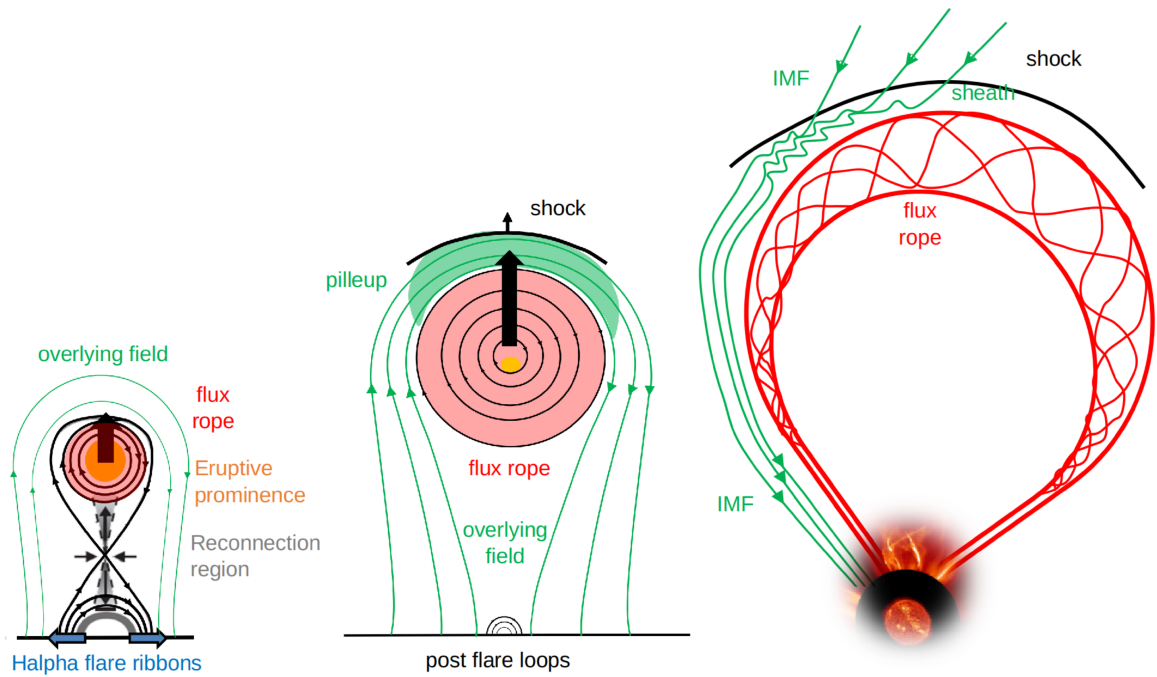

*Figure 3-1. Cartoon describing three different stages of CME evolution based on "the standard CME-flare model" and "the standard magnetic cloud model": the onset (left), post-eruption phase (middle) and interplanetary propagation (right)*

In order to understand the full picture of the CMEs, i.e. the magnetic structure of CMEs and processes involved in its initiation, evolution and propagation, we need theories to explain the observed properties of CMEs. Both numerical and analytical models need to be employed to reach deeper understanding of the origin and evolution of CMEs and provide a theoretical basis to the CME/ICME observations. The numerical modelling will be addressed in Section 4, while this section is devoted to analytical models and divided into subsections related to CME initiation, propagation, forward modelling, fitting to in situ measurements and finally interactions with other magnetic structures.

## 3.2 Mechanisms, processes and forces governing the CME take-off

Comprehensive theoretical work on solar eruptive events started in seventies and early eighties of the past century (Green et al. 2018). Many models were focused on the flux-rope configurations and their instabilities (see references in Chen 1989; Vršnak 1990), identifying Lorentz force as the main driving element of the eruption, and emphasizing the importance of the amount of the magnetic field twist. For example, using the analytical approach, Chen (1989) and Vršnak (1990), independently considered in their semi-toroidal flux-rope models a number of effects that were before taken only partially or were neglected. The flux-rope models progressively developed, becoming more sophisticated. A particular important step was the analytical model presented by Titov & Demoulin (1999), which took into account relevant effects and provides an idealized, but quite realistic configuration of a semi-toroidal flux rope embedded in a coronal arcade. This model was later on included in a number of analytical studies, or numerical simulations of the loss of equilibrium in general, and in particular, in studies of kink and torus instabilities (Török and Kliem 2003; 2005; Török, Kliem, and Titov 2004; Török, Berger, and Kliem 2010; Kliem and Török 2006; Kliem et al. 2014).

More recently in the past decade, which includes also the VarSITI/ISEST era, there was a significant progress in physical understanding of solar eruptions. A number of papers were published considering the pre-eruptive storage of free energy and helicity, processes that cause the evolution of the system towards an unstable state, criteria defining stable/unstable configurations, initiation of the eruption, the dynamics of the eruption itself, as well as the analysis of the effects of the eruption in the ambient corona (see recent reviews by Aulanier 2013; Schmieder, Aulanier, and Vršnak 2015; Green et al. 2018; Patsourakos et al. 2020). Although it was clearly demonstrated that the active region magnetic complexity and its dynamics, free-energy and helicity content are essential parameters, a number of open questions appeared. All of them can be summarized in the fact that we still do not understand why active regions that are quite similar according to the mentioned basic characteristics sometimes produce an eruption and sometimes not. Obviously, some key parameters are still missing, and consequently, this makes the eruption forecasting still highly unreliable.

Apart from general theoretical, numerical or observational considerations regarding the mentioned pre-eruption configurations, a number of papers were published being focused more on specific processes that lead to the eruption than to general properties of active regions. For example, Török et al (2013) analyzed in detail consequences of rotational motions at the footpoints of a flux rope in one well-observed event. The analysis provided a very detailed physical interpretation of characteristics of the pre-eruptive evolution of the magnetic configuration, the initiation of the eruption, the dynamics of the eruption, as well as the evolution of ambient magnetic system. On the other hand, Vršnak (2019) considered the rotating flux-rope leg process in more general terms, and showed that except in quite extreme cases, such as considered by Török et al (2013), the twisting motion does not allow for poloidal flux injection sufficient to explain the observed speeds of the slow rise of the flux rope in the gradual pre-eruptive stage. On the other hand, it was shown that the emerging flux process is viable to cause rising speed close to that observed in pre-eruptive phase. Let us note that both processes eventually lead to a loss of equilibrium of the system if persistent long enough (Green et al. 2018).

In the last decade, a special attention was paid to the role of the reconnection below the erupting flux rope in the dynamics of the eruption, primarily based on a number of observational results directly relating the eruption-related flare energy release and the eruption acceleration. This type of reconnection has three very important consequences. First, it reduces the tension of the overlying field, which tries to prevent the eruption and, in some cases, can cause a failed eruption. Second, it supplies the flux rope with a fresh poloidal flux that has a twofold effect. One is a direct enhancement of the upward directed component of the hoop-force. The other is weakening of the self-inductive effect of the expanding structure that should lead to fast electric-current decrease, and thus a weakening of the Lorentz force. In this respect it is worth mentioning the paper by Vršnak (2016), where it was clearly demonstrated that the peak acceleration of the flux rope is dependent on the reconnection rate and the peak velocity is proportional to the total reconnected flux. Moreover, it was shown that the flux-rope acceleration time-profile, as well as the velocity time-profile, are closely synchronized with the time evolution of the reconnection rate. These theoretical results are able to fully explain the effects observed in CME-flare relationship, as discussed in Section 2.4.

## 3.3 CME propagation from corona through interplanetary space

The kinematical evolution of CMEs can be divided into three phases: (1) slow rising phase during the CME initiation; (2) impulsive or main acceleration phase; (3) interplanetary propagation phase (Zhang and Dere 2006; Temmer 2016). There are three competing forces governing CME kinematical evolution, the gravitational force, Lorentz force and the MHD drag. The Lorentz force mainly provided by the magnetic field in the corona is introduced in Section 3.2, while the MHD drag can be well represented by the aerodynamic drag equation and is believed to be dominant in the interplanetary propagation phase (Cargill 2004; Vršnak and Žic 2007). The latter is supported by observations showing deceleration of fast CMEs and acceleration of slow CMEs (Sheeley et al. 1999; N. Gopalswamy et al. 2000; N. Sachdeva et al. 2015). This concept was previously introduced into a simple Empirical Shock Arrival (ESA) model (Gopalswamy et al. 2001), Expansion Speed Model (ESM) (Schwenn et al. 2005), and more recently in an Effective Acceleration Model (EAM) (Paouris and Mavromichalaki 2017), an analytical observation-driven model by (Ying D. Liu et al. 2017) and an empirically driven piston shock model by (Corona-Romero et al. 2017). Based on the concept of CME propagation being governed by MHD drag, an analytical drag-based model (DBM) was introduced by (B. Vršnak et al. 2013). DBM was found to very successfully describe heliospheric propagation of CMEs (B. Vršnak et al. 2014; Hess and Zhang 2014b), therefore it was expanded to different geometries in the recent years: most notably using a 2D-Cone geometry (Žic, Vršnak, and Temmer 2015) and 2D ellipse front (Möstl et al. 2015). Note that the interplanetary acceleration obtained by Gopalswamy et al. (2001) is proportional to CME speed and hence related to the Stokes drag, as opposed to the aerodynamic drag proportional to the square of the CME speed.

Observations by STEREO-HI instrument lent support to empirical models. The elongation conversion models were previously typically combined with fitting algorithms to determine CME arrival, assuming constant speed, as discussed in Section 2.2.4. However, the constant speed assumption was recently substituted with DBM and combined with a newly developed Ellipse Conversion method for the HI observations into the ElEvoHI model (Rollett et al., 2016).

CMEs typically propagate radially, although deviations were found, namely rotations (Lynch et al., 2009; Vourlidas et al., 2011) and deflections (Wang et al., 2004; Lugaz et al. 2010b; Isavnin et al., 2014). Deflections by coronal holes have been previously modelled empirically, where a deflection from the original direction by a certain angle was quantified by so-called coronal hole influence parameter (CHIP), assuming that CME continues to propagate radially (Gopalswamy et al., 2009; Mohamed et al., 2012). Interplanetary deflections were regarded previously as well, in a kinematic model assuming interplanetary spiral magnetic field and the background solar wind.  As a result, CMEs faster than background solar wind are deflected to the east, whereas CMEs slower than background solar wind are deflected to the west (Wang et al. 2004; 2006). Recently, both deflections and rotations were included into a 3D CME propagation model, where the CME flux rope is represented with a rigid, un-deformable torus and deflections and rotations are calculated using the magnetic tension and magnetic pressure gradients calculated from magnetometer input (Kay et al., 2013; Kay and Opher, 2015).

Analytical models are easy to run and are not time consuming and therefore ideal for ensemble modelling (for details on ensemble modelling see section 4). Therefore, it is not surprising that several recently developed analytical models have their ensemble version (see Table 3-1). Finally, recently a novel approach was adapted in semi-empirical CME propagation modelling, machine learning (Sudar, Vršnak, and Dumbović 2016; J. Liu et al. 2018), which in future might prove to be a powerful predictive tool.

## 3.4 Forward modeling of CMEs

Recently, there has been quite development in the forward modelling of the magnetic structures using physics-based empirical models. The forward models typically assume a specific morphology of a magnetic structure, i.e. flux rope, and evolve it, assuming specific propagation and expansion. For instance, Wood et al. (2017) use the croissant-like morphology described by Wood and Howard (2009), apply self-similar expansion and propagate it to Earth using either the Harmonic mean or Fixed Phi method as a kinematic model. With this method, it is possible to obtain a global shape of the structure, as well as its local size and orientation.

A step further is to include a specific flux rope magnetic field topology. Möstl et al. (2018) in their 3-Dimensional COronal Rope Ejection (3DCORE) model used a tapered torus geometry and a Gold-Hoyle magnetic field topology (Gold and Hoyle, 1960) and evolve it using DBM and self-similar expansion. They constrain the magnetic field values using measurements at Mercury. In their Flux Rope from Eruption Data (FRED) method, Gopalswamy et al. (2018) evolve self-similarly a croissant shaped flux rope obtained by the Graduated Cylindrical Shell (GCS) model and assume a Lundquist-type magnetic field topology (Lundquist, 1951) constrained using reconnection flux computed from the area under post-eruption arcades (N. Gopalswamy et al. 2017; Pal et al. 2018; Sarkar, Gopalswamy, and Srivastava 2020). It should be noted that a similar approach was studied previously by Savani et al. (2015) and  Temmer et al. (2017), where several options were regarded to constrain the magnetic field (reconnection flux computed from H-alpha ribbons, dimming flux and non-linear force free modelling). Kay et al. (2017) also use a Lundquist-type magnetic structure along with the shape and size represented with a rigid, un-deformable torus, which is the output of their ForeCAT model. Their ForeCAT In situ Data Observer (FIDO) model primarily assumes self-similar expansion and constant time-shift, however, with addition of some free parameters they are able to also simulate distortion, i.e. pancaking. They also extensively study sensitivity of the model using ensembles (Kay and Gopalswamy, 2018), although in its current form the magnetic structure is free-fitted to the in situ measurements. Recently, FIDO model is improved by incorporating the forward model of the CME-driven shock and sheath, as FIDO-SIT (Sheath Induced by Transient) (Kay, Nieves-Chinchilla, and Jian 2020). Flux Rope in 3D (FRi3D) model by Isavnin (2016) uses a croissant-like shape, but allow it to deform while expanding due to front flattening, pancaking and skewing. The assumed magnetic structure is of Gold-Hoyle type and the whole FR is free-fitted to the in-situ measurements. Finally, probably the most extensive forward modelling study was performed by Patsourakos et al. (2016), who used GCS croissant for the CME shape, propagated it using DBM, and then applied several different combinations for estimating the magnetic structure topology (including non-force free magnetic structures by Hidalgo et al., 2000 and Cid et al., 2002), as well as the initial helicity and expansion.

Although most of the forward modelling procedures are focused in reproducing the magnetic structure at a certain heliospheric distance, other applications might be noteworthy as well. Similar procedure was recently adopted in Forbush decrease model (ForbMod) by (Dumbović, Heber, et al. 2018), who expanded the GCS croissant using empirical power-law relations for size and magnetic field, assuming constant-speed propagation and homogeneous magnetic field in order to derive galactic cosmic ray counts. The evolution of the internal properties (e.g., the plasma temperature, density, velocity and heating) of CMEs is also one of the important research aspects, but it is limited to a certain position or a certain time by using remote sensing and in situ observations (e.g. Wang et al., 2005; Liu et al., 2006; Bemporad and Mancuso, 2010; Susino et al., 2013; Susino and Bemporad, 2016). In a novel approach, GCS is used to calculate the CME volume and density to derive the evolution of the magnetic ejecta and sheath density from Sun to Earth (M. Temmer et al. 2020). To continuously figure out the internal state of an individual CME during its heliospheric propagation, a self-similar flux rope internal state (FRIS) model was proposed by Wang, et al. (2009), providing the variations of the polytropic index of the CME plasma, the average Lorentz force and the thermal pressure force inside a CME with heliocentric distance. Recently, Mishra and Wang (2018) improved FRIS model, by constraining it with the observed propagation and expansion behavior of a CME and deriving a few additional parameters (absorbed heat, entropy, heating rate, and entropy changing rate). This model was then implemented to a slow (Mishra and Wang, 2018) and fast (Mishra et al., 2020) CME respectively, which showed that during the propagation (1) the expansion was driven by the thermal force inside the CME but prevented by the Lorentz force, (2) the slow CME released heat before it reached an adiabatic state and then absorbed heat, and (3) the fast CME was in the heat-releasing state throughout its journey.

## 3.5 In-situ fitting of ICMEs

ICMEs can be identified by a number of typical properties that differ from those of the ambient solar wind (Gosling, 1990; Wimmer-Schweingruber et al., 2006; Zurbuchen and Richardson, 2006; Rouillard, 2011; Kilpua et al., 2017). Among ICMEs there is a subset, called magnetic clouds (MCs), which were first identified by Burlaga et al. (1981), and then studied widely in the past decades. MCs exhibit a smooth rotation of the magnetic field direction through a large angle, an enhanced magnetic field strength and a low proton temperature (Burlaga et al., 1981). MCs play an important role in understanding the evolution of CMEs in the heliosphere and are one of the main drives of space weather ecents (Wilson 1987; N. Gopalswamy et al. 2008; N. Gopalswamy, Yashiro, et al. 2015; B. T. Tsurutani et al. 1988; K. Emilia J. Huttunen, Koskinen, and Schwenn 2002; C.-C. Wu and Lepping 2002; H. V. Cane and Richardson 2003; J. Zhang et al. 2007; Hidalgo 2011a; Y. Li, Luhmann, and Lynch 2018).

MCs are believed to have magnetic flux rope structure with two ends rooting on the Sun (Burlaga et al., 1981; Larson et al., 1997; Janvier et al., 2013). So far, the observations about MCs could only rely on the *in situ* data along the MC pass path. To reconstruct the global configuration of MCs in 2D or 3D space, a variety of techniques has been developed. Based on the idea of force-free ($\nabla \times \mathbf{B} = \alpha \mathbf{B}$) magnetic configuration of MCs (Goldstein, 1983), the symmetric-cylindrical models, with linear (Burlaga, 1988; Lepping et al., 1990; Lepping, 2003) and non-linear (Farrugia et al., 1999) force-free fields, were proposed. Other models were also developed by adopting different assumptions, e.g., (1) the expanding model (Farrugia et al., 1993; Vandas et al., 2006), (2) the models with distorted cross section (Romashets and Vandas, 2003; Vandas et al., 2006; Démoulin and Dasso, 2009), and (3) the torus model (Romashets and Vandas, 2003; Marubashi and Lepping, 2007). Based on the observations of pressure gradients inside MCs, some non force-free models were proposed and then improved (Hidalgo et al., 2000; Mulligan and Russell, 2001; Hidalgo et al., 2002a; 2002b; Cid et al., 2002; Hidalgo, 2003).

Nowadays, with our deeper understanding about MCs, several new models have emerged, which are still based on the well-developed description of (non-)force-free magnetic configuration, but can provide extra information about MCs.

**Force-free magnetic field.** The assumptions of the cross section and the symmetry along the MC axis are two keys in the models. Keeping the cylindrical symmetry and the circular cross section, Wang et al., (2015; 2016) developed the velocity-modified force-free flux rope models, which can provide the MC kinetic information by incorporating the linear propagating motion away from the Sun, the expanding, and the poloidal motion of the plasma inside MCs with respect to the MC axis. Recently, Lepping et al. (2018) improved their previous model (Lepping et al., 1990) by modifying the magnetic field magnitude based on a B-modification scheme presented by Lepping et al., (2017). This model improves the fitted B-profile, but is applicable for use with data originating only at/near 1 AU. The models of linear force-free field with different boundary pitch-angle treatments and of non-linear force-free field with varying prescribed twist were derived by Nishimura et al. (2019) and Vandas and Romashets (2019), respectively. Discarding the assumption of locally straight MC axis, the non-cylindrical models were developed. For example, Owens et al. (2012) developed a curved flux rope (CFR) model, assuming a circular cross section but bending the axial field in a similar manner to a Parker spiral magnetic field. This allows the radius of curvature of the axis and the cross-sectional extent of the MC to vary along the length of the axis. To model irregularities in MCs, Romashets and Vandas (2013) added a local irregularity in the form of a compact toroid into a cylindrical linear force-free magnetic structure. Furthermore, MCs in toroidal geometry can also be constructed by the constant-alpha force-free magnetic field with elliptical cross sections (Vandas and Romashets, 2017) or uniform-twist force-free field with circular cross sections (Vandas and Romashets, 2017b).

**Non-force-free magnetic field.** Inherited from Hidalgo (2003; 2011), a much more complicated model was improved in series (Hidalgo and Nieves-Chinchilla 2012b; Hidalgo 2013; 2014b; 2016b). Hidalgo and Nieves-Chinchilla (2012) described MC topology with torus geometry and a non uniform cross section. They established an intrinsic coordinate system for that topology, and then analytically solved the Maxwell equations in terms of it. The model was tested by applying it to the observations of multiple spacecraft by Hidalgo (2013). The model was further improved with inclusion of the plasma pressure (Hidalgo, 2014) and the proton current density (Hidalgo, 2016). Extending the concept of Hidalgo et al. (2002a), Nieves-Chinchilla et al. (2016) presented a circular-cylindrical flux-rope model by introducing a general form for the radial dependence of the current density, which can give the information on the force distribution inside MCs. The circular cross section was later improved to an ellipse (new elliptic-cylindrical flux rope model, Nieves-Chinchilla et al., 2018), which, for the first time, allows us to completely describe MCs by nonorthogonal geometry.

Different from the above introduced techniques, the Grad-Shafranov (GS) reconstruction technique (Hu and Sonnerup, 2001; 2002) doesn't presume a specific magnetic structure or cross section of MCs. It assumes that an asymmetric-cylindrical MC is in an approximately magnetostatic equilibrium with an invariant direction, and uses the GS equation to recover the magnetic field as well as the plasma pressure. With this technique, the boundaries of the MC need not first be identified in the data. Recently, trying to approach a more real flux rope topology, the GS reconstruction of for MCs in toroidal geometry with rotational symmetry was developed (Hu, 2017; Hu et al., 2017).

The development of all the introduced techniques provides invaluable tools for extracting information about the properties of MCs and leads to our understanding of the underlying physics of MCs. However, it was shown that assessing their accuracy based on in situ data is challenging. Different comparisons of the MC properties have been performed between: (1) MHD simulations and fitting techniques (Riley et al., 2004; Vandas et al., 2010; Al-Haddad et al., 2011; 2019) and (2) different fitting models (Al-Haddad et al., 2013; Démoulin et al., 2013; Hu et al., 2013; Vandas et al., 2015; Vandas and Romashets, 2015; Lepping et al., 2018; Nishimura et al., 2019). While these studies largely support the applicability of examined techniques and methods to gain insight into the MC structure, they also reveal their limitations as well as reliability issues, which should be tackled with in the future.

## 3.6 CME-CME interactions

Erupting from the Sun, coronal mass ejections (CMEs) will interact with different structures. The interaction between a CME and the magnetic fields in the corona and interplanetary space, and the solar wind govern the propagation and evolution of the CME itself (see Section 3.3 and 3.4). Magnetic reconnection between CMEs and ambient solar wind can lead to the peeling-off of the magnetic field lines and the erosion of the axial magnetic flux from the CMEs (Dasso et al., 2006; Gosling, 2012; Ruffenach et al., 2012; 2015; Manchester et al., 2014; Lavraud et al., 2014; Wang et al., 2018). Furthermore, CMEs were found to be interacting with other CMEs by many observations (e.g. Gopalswamy et al., 2001; Lugaz et al., 2012; Shen et al., 2012; Harrison et al., 2012; Liu et al., 2012; Liu et al., 2014; Martínez Oliveros et al., 2012; Möstl et al., 2012; Temmer et al., 2012; 2014; Webb et al., 2013; Shanmugaraju et al., 2014; Colaninno and Vourlidas, 2015; Mishra et al., 2014; 2015; 2016; 2017), involving complicated physical processes, resulting in the changes of CME properties, forming complex structures, and playing an important role in leading to large solar energetic particle events (see Section 6) and intense geomagnetic storms. CME-CME interaction can result in the changes in CME speed, propagation direction (Xiong et al., 2009; Lugaz et al., 2012; Shen et al., 2012; Mishra et al., 2016), radial extent as well as internal magnetic strength (Schmidt and Cargill, 2004; Xiong et al. ,2006; Lugaz et al., 2005; 2012; 2013). The speed change, or more physical, the natures of the collision of one CME by another, were widely studied, namely, inelastic (Lugaz et al., 2012; Maričić et al., 2014; Mishra et al., 2015) vs. elastic (Mishra et al., 2014; 2015) vs. super-elastic collision (Shen et al. 2012; Shen et al. 2013; Colaninno and Vourlidas, 2015). Note that the term of collision used here refers to that the main bodies of two CMEs are touching.

In determining the nature of collisions, most of the earlier studies have used a simplistic approach that CMEs are propagating exactly in the same direction, i.e., 1D head-on collision (with momentum conservation constraint, Mishra and Srivastava, 2014). However, a relatively more precise analysis should be performed in 3D space. For the first time, Shen et al. (2012) studied an oblique collision of CMEs in 3D using imaging observations (Figure 3-2a) and took several uncertainties into consideration, but did not constrain the conservation of momentum instead of indirectly evaluating it by analyzing the effect of the solar wind on the acceleration of the first CME. Recently, Mishra et al. (2016) addressed previous limitation and proposed a method for the 3D oblique collision (Figure 3-2b), in which the post-collision directions and speeds are theoretically measured with different pre-set restitution coefficients, together satisfying the momentum conservation law, and the best-matched parameters with the observations are extracted. Based on this, Mishra et al. (2016; 2017) calculated the uncertainties in determining the nature of collisions quantitively, with different CME observed parameters considered, which then emphasized the possibility of a large uncertainty (see one example in Figure 3-2c). Furthermore, Shen et al. (2017) presented four definitions of different types of collisions, i.e., a classical Newtonian definition, an energy definition, Poissons definition, and Stronges definition, helping with a deeper understanding of the determination of collisions in theory. They focused on the first two used in observational and numerical studies, and found out that these two definitions are not equivalent to each other when the colliding objects are expanding or contracting with a changing rate.

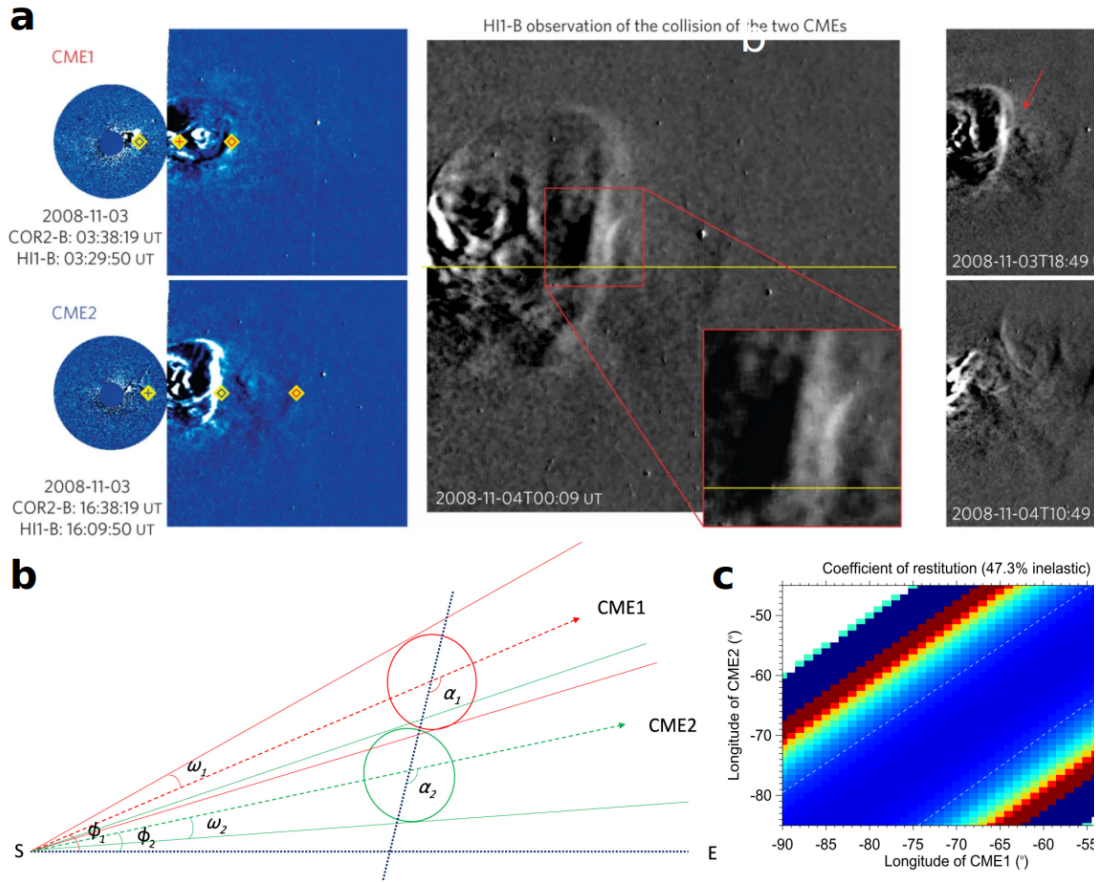

*Figure 3-2. (a) The super-elastic collision of two interacting CMEs observed in STEREO/HI (Shen et al 2012). (b) Oblique collision of two CMEs assumed as spherical bubbles (Mishra et al. 2016).(c) The variation of the restitution coefficient with the uncertainties of pre-collision longitudes of CMEs considered (Mishra et al. 2016).*

Including collision, CME-CME interaction could be involved in four forms: 1) the two-CME driven shock waves interact without the ejecta interacting, 2) one shock wave interacts with a preceding magnetic ejecta, 3) the direct interaction between two ejecta, and 4) the reconnection between successive magnetic ejecta. The last three may result in a variety of complex structures at 1 AU: from 1) partial ongoing interaction of the preceding ejecta with a shock wave inside (Lugaz et al., 2015a; 2015b; Wang, 2003; Liu et al., 2018; Wang et al., 2018; Xu et al., 2019), to full interaction of 2) the multiple magnetic cloud (MC) events (Wang et al., 2002; Wang, 2003; Shen et al., 2011; Lugaz et al., 2013), 3) a complex ejecta or compound stream (Burlaga et al., 2003), and 4) long-duration events (Dasso et al., 2009; Lugaz et al., 2013), leading to a hard understanding of the undisturbed conditions by in situ observations. Wang et al. (2018) proposed a recovery model based on Rankine-Hugoniot jump conditions to recover the shocked structure back to the uncompressed state in the first type of the interacted structure, which was later used to estimate the related interacting effects in causing geoeffectiveness (C. Shen et al. 2018; M. Xu et al. 2019b). The resulting structure from CME-CME interaction was discussed in detail in the review of Lugaz et al. (2017).

## 3.7 Summary of recent analytical and semi-empirical models

Table 3-1 shows the analytical and semi-empirical models of CMEs/ICMEs introduced above, which were proposed roughly during the VarSITI period (ca. 2014-2018). Models are listed according to their type: P=propagation, EC=elongation conversion, FM=forward modelling, FIT=in situ fitting, GS=Grad-Shafranov reconstruction.

*Table 3-1: list of recent analytical and semi-empirical models of CMEs/ICMEs*

| Type | Acronym | Short description | Reference |
|------|---------|------------------|-----------|
| P | EAM | Empirical; shock propagation | Paouris and Mavromichalaki (2017) |
| P | -- | Empirical; shock propagation | Liu, Zhao, and Zhu (2017) |
| P | -- | Empirical; shock propagation | (Corona-Romero et al. (2017) |
| P | ElEvo | DBM+2D ellipse; shock propagation | Möstl et al. (2015) |
| P | DBM/DBEM | DBM + 2D Cone; CME propagation | (Žic, Vršnak, and Temmer 2015); (Dumbović, Heber, et al. 2018) (ensemble) |
| P | PDBM | 1D DBM; probabilistic; CME propagation | (Napoletano et al. 2018) (ensemble) |
| P | ElEvoHI | DBM+ElEvo+HI fitting; shock propagation | Rollett et al. (2016); (Amerstorfer et al. 2018) (ensemble) |
| P | ForeCAT | CME propagation; deflection; rotation | (Kay, Opher, and Evans 2013b); (Kay and Opher 2015a) |
| P | -- | Machine learning; CME propagation | (Sudar, Vršnak, and Dumbović 2016) |
| P | CAT-PUMA | Machine learning, shock propagation | Liu et al. (2018) |
| P | -- | Oblique collision in 3D + constrain conservation of momentum | Mishra, Wang, and Srivastava (2016); Mishra et al. (2017) |
| EC | -- | triangulation | Liu et al. (2017) |
| FM | -- | Croissant + self-similar expansion + HM/Fixed Phi => size, orientation | Wood et al. (2017) |
| FM | 3DCORE | Torus+self-similar expansion + DBM + Gold-Hoyle => magnetic structure | Möstl et al. (2018) |
| FM | FRED | GCS + self-similar expansion + Lundquist => magnetic structure | Gopalswamy et al. (2017) |
| FM | FIDO | Torus + self-similar expansion + Lundquist => magnetic structure | Kay et al. (2017); Kay and Gopalswamy (2018) (ensemble) |
| FM | FIDO-SIT | FIDO + forward model of shock and sheath | Kay, Nieves-Chinchilla, and Jian (2020) |
| FM | FRi3D | Deformable croissant + Gold-Hoyle | Isavnin (2016) |

| | | => magnetic structure | |
|---|---|---|---|
| FM | -- | GCS+power-law expansion + DBM +(non) force free FR => magnetic structure | (Patsourakos, et al. 2016) |
| FM | ForbMod | GCS + power-law expansion => Forbush decrease | (Dumbović, Heber, et al. 2018) |
| FM | FRIS | CME internal properties | Mishra and Wang (2018) Mishra et al. (2020) |
| FIT | -- | Force free + circular-cylindrical + velocity modified; | Wang et al. (2015) Wang et al. (2016) |
| FIT | -- | Force free + circular-cylindrical + **B** modified | Lepping et al. (2018) |
| FIT | -- | Force free + circular-cylindrical + boundary pitch-angle treatments | Nishimura, Marubashi, and Tokumaru (2019) |
| FIT | -- | Force free; circular-cylindrical + varying prescribed twist | Vandas and Romashets (2019) |
| FIT | -- | Non force free; torus + non-uniform cross-section; plasma pressure + proton current density | Hidalgo and Nieves-Chinchilla (2012) Hidalgo (2014) Hidalgo (2016) |
| FIT | -- | Non force free; force distribution + circular-cylindrical;  elliptic-cylindrical | Nieves-Chinchilla et al. (2016) Nieves-Chinchilla et al. (2018) |
| GS | -- | GS reconstruction + toroidal geometry | Hu (2017) |

# 4. Progress in Numerical Modeling of CMEs/ICMEs

Magnetohydrodynamic (MHD) simulations have proved to be one of the most important tools to study the evolution of a coronal mass ejection (CME) in both corona and interplanetary space, and the modeled results can be used to analyze the initiation, propagation characteristics observed by ground-based and space-based instruments. Lugaz and Roussev (2011) gave a detailed review and discussion on the efforts to use numerical simulations of ICMEs to investigate the magnetic topology, density structure, energetics and kinematics of ICMEs in the interplanetary space. In the book of Feng (2020), the author has provided a recent in-depth review of the field focusing primarily on the current status of MHD modeling for space weather with a thorough collection at the time of writing the book. Here, the review is devoted to recent progress of time-dependent MHD space weather modeling with the focus on such topics: the ambient solar wind, CME initiation and CME propagation, CME-solar-wind interaction, and CME-CME interaction, especially on the work performed in the second half of the 2010s.

## 4.1 Modeling the background solar wind

Numerical simulations have shown the importance of an accurate modelling of the background medium in which the disturbances propagate (Odstrčil and Pizzo 1999a; 1999b; Chané et al. 2005; Fang Shen et al. 2007). In the past decade, due to the vast improvement in computational resources, the usage of 3D MHD models for reconstructing the solar corona and interplanetary solar wind has become almost routine. Moreover, based on much improved observations, it is possible to produce more realistic simulations, e.g., by the inclusion of the observational data through the boundary conditions or through data assimilation. Hayashi (2012) presented a treatment of observation-based time-dependent boundary conditions for the inner boundary sphere in the 3D MHD simulations of the global co-rotating solar wind structures. In order to adjust the model to the time-varying magnetic field on the bottom boundary, developed the model of the confined differential potential field (CDPF) to prescribe the bottom boundary condition, In addition, a modified version of this model was adopted as the module of the time-dependent 3D MHD simulation at the Joint Science Operation Center (JSOC) of SDO (K. Hayashi et al. 2015). The module could routinely generate 3D data of the time-relaxed minimum-energy state of the solar corona using the full-disk magnetogram data from HMI/SDO.

In parallel, data assimilation has been included in the WSA model through ADAPT (Hickmann et al. 2015), which allows photospheric simulations to agree better with available observations from magnetograms. This model has been coupled to the 3D MHD LFM-Helio (Merkin, Kondrashov, et al. 2016) to perform time-dependent simulations of the background solar wind. These simulations are able to reproduce more accurate details of small-scale of the heliospheric current sheet and corotating interaction regions.

Recently, by using magnetogram synoptic map images from GONG and theoretical/empirical models such as the PFSS model and WSA model, Shen et al. (2018) applied a new boundary treatment to the 3D MHD simulation of solar wind and established the 3D IN (INterplanetary)-TVD-MHD model. The boundary conditions depend on five tunable parameters when simulating the solar wind for different phases of the solar cycle, and the simulated solar wind parameters are in good agreement with the observations most of the time. However, we know that models tend to fail when solar activity increases. The comparison of their modeled results with the in-situ data throughout 2007 is shown in Figure 4-1, which demonstrates that the simulation retrieves most of the high-speed streams (HSSs), and the duration time and the magnitude of the HSSs are largely consistent with those of the observations. Later, <u>Yang and Shen (2019)</u> presented a new method to construct the global distribution of solar wind parameters at the source surface using multiple observations and the ANN (Artificial Neural Network) technique, which could be used to provide a more realistic boundary condition for 3D MHD solar wind modeling.

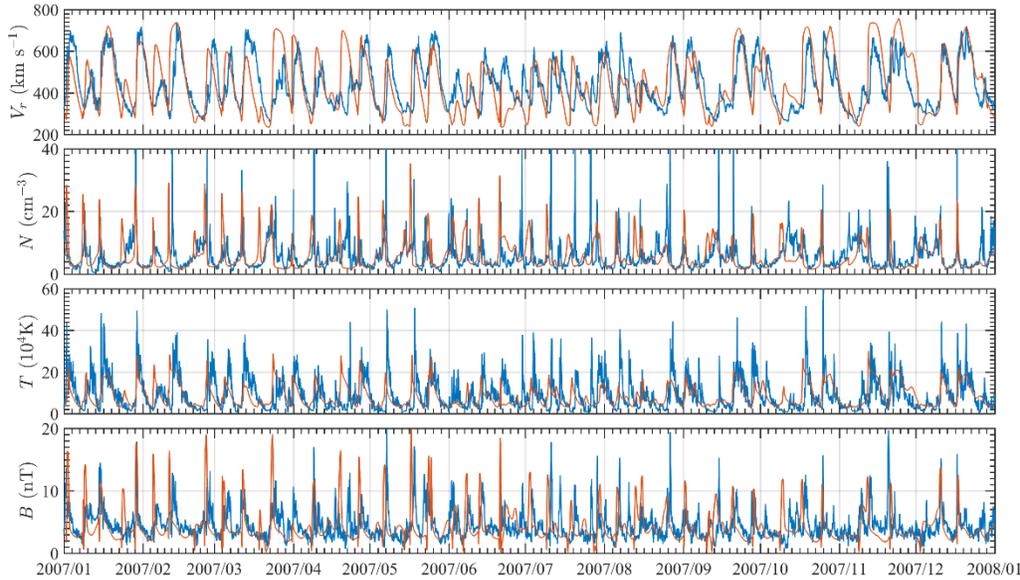

Figure 4-1. Modeled (red lines) and observed (blue lines) time profiles of solar wind parameters at 1.0 au through all of 2007. From the top to the bottom, the panels show the speed $V_r$, number density N, temperature T, and total magnetic field strength B (From Shen et al. 2018). Besides the improvement in the treatment of the inner boundary condition, there have been significant improvement on other aspects of background solar wind simulations in recent years. Some of these are detailed below.

By using CORHEL (CORona-HELiosphere), <u>Linker et al. (2016)</u> further developed a time-dependent study of the solar wind empirically driven by magnetic maps at a daily cadence using ADAPT. Their simulation showed both classic features of stream structure in the interplanetary medium often seen in steady-state models and evolutionary features unable to be captured in a steady-state approach. Their model results also compared reasonably well with 1 AU OMNI observations. As a rather mature space weather model, CORHEL is a couple suite of models for simulating the solar corona and solar wind in 3D space, which provides three solar coronal models at present, including the Magnetohydrodynamic Algorithm outside a Sphere (MAS) polytropic model, MAS thermodynamic model and the potential field source surface and Wang-Sheeley-Arge (PFSS-WSA) model. The heliospheric models involved in CORHEL are ENLIL (Odstrcil, Riley, and Zhao 2004), the MAS-Heliosphere (MAS-H) models (Lionello et al. 2013) and the LFM-Helio (Merkin, Lionello, et al. 2016).

Based on the SWMF (Space Weather Modeling Framework), van der Holst et al. (2014)

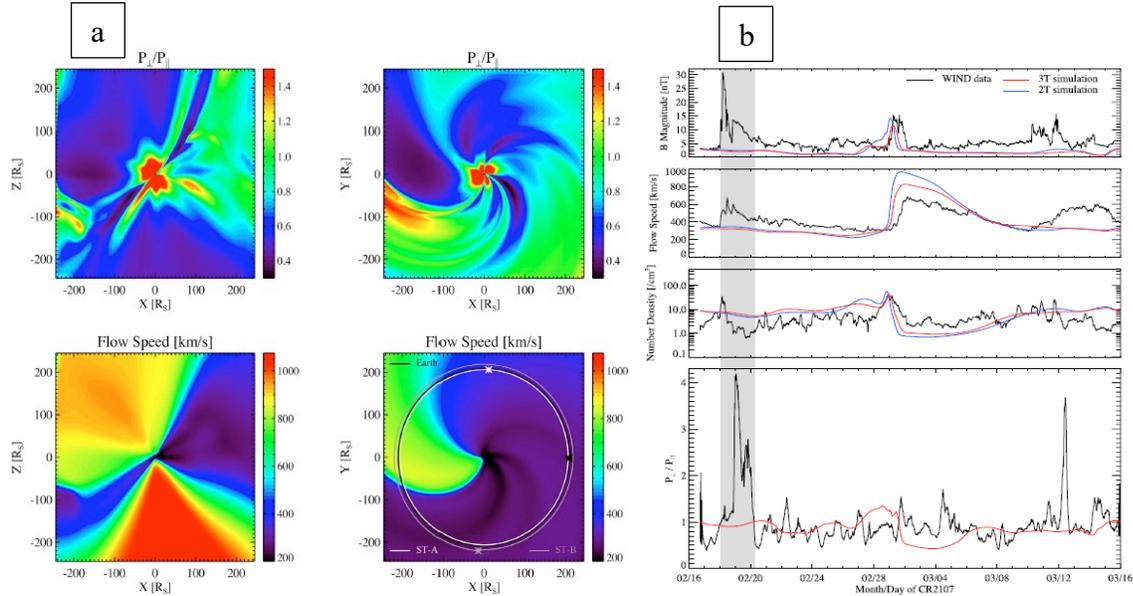

*Figure 4-2. (a) The simulated proton pressure anisotropy ratio p⊥/p_ and the simulated solar wind speed in the Y = 0 and Z = 0 planes for CR2107 by the AWSoM model, and the plot on the right bottom panel also shows the trajectories of the Earth, STEREO-A and B satellites projected to the Z = 0 plane; (b) The simulated solar wind properties along the WIND orbit and the WIND data during CR2107 (From Meng et al. 2015).*

developed the Alfvén wave solar model named AWSoM, which is a 3D MHD model that considers the anisotropy of ion temperature in the solar corona and the inner heliosphere. In Meng et al. (2015), the AWSoM model has been applied to simulate the steady solar wind from the solar corona to 1 AU of CRs 2107 and 2123. Figure 4-2(a) shows the simulated proton pressure anisotropy ratio and the simulated solar wind speed in the Y = 0 and Z = 0 planes for CR2107. The simulation results also show a reasonable agreement with in-situ observations at 1 AU, as shown in Figure 4-2(b). Together with a coronal charge state evolution model, the Michigan Ionization Code (MIC, see Landi et al. 2012), Oran et al. (2015) employed the AWSoM model to calculate the elemental charge state evolution along the modeled open magnetic field lines. The charge state evolution model was initiated with the electron density, temperature, and speed simulated by the AWSoM wind model, and provided the first charge state calculation covering all latitudes in a realistic magnetic field. Evans et al. (2012) self-consistently coupled the Alfvén wave energy transport with the MHD equations. In this solar wind model, they introduced an additional dissipation mechanism: surface Alfvén wave (SAW) damping, which was weak in the polar regions, and strong in subpolar latitudes and the boundaries of open and closed magnetic fields. Their simulated results showed that SAW damping could reproduce regions of enhanced temperature at the boundaries of open and closed magnetic fields seen in both tomographic reconstructions in the low corona and Ulysses data in the heliosphere. Sokolov et al. (2013) presented a combined global model of the solar corona, the low corona, the transition region, and the top of the chromosphere. Their model used MHD Alfvén wave turbulence as the only momentum and energy source to heat the coronal plasma and drive the solar wind with different turbulence dissipation efficiencies in coronal holes and closed field regions. Recent developments include further validations of the AWSoM model (Gombosi et al. 2018; Nishtha Sachdeva et al. 2019). It is known that the SWMF couples the models of Lower Corona (LC), Solar Corona (SC), Inner Heliosphere (IH), and other integrated components (Gábor Tóth et al. 2012). The models of SC, IH, and several other components are modeled by the BATS-R-US code. In the SC model, van der Holst et al. (2010) solved the two-temperature MHD equations with Alfvén wave heating and heat conduction on either Cartesian or spherical grid in a frame corotating with the Sun.

Feng et al. (2010) employed SIP-CESE MHD model within a six-component overset grid for solar wind simulation. They numerically investigated the large-scale structures of interplanetary solar wind and the evolution of the heliospheric magnetic field. Feng et al. (2012) carried out the numerical studies for the solar wind background of different solar-activity phases by using the SIP-AMR-CESE MHD model, and their modeled results could reproduce many features near the Sun and in interplanetary space, e.g., the changing trends of the solar-wind parameters for the selected CRs, and the IMF polarities and their changes. Furthermore, Feng, Ma, and Xiang (2015) investigated the solar wind evolution between the solar surface to the Earth's orbit from 1 July to 11 August 2008 with the SIP-AMR-CESE MHD model driven by the consecutive synoptic maps from GONG. Similarly, Li and Feng (2018) simulated the evolution of solar wind from the solar surface to the Earth's orbit during year 2008, and evaluated simulated results quantitatively by comparison with in-situ measurements.

Merkin et al. (2011) adapted the Lyon-Fedder-Mobarry (LFM) model for the inner heliosphere, which was referred as LFM-helio model. They simulated the solar wind and heliospheric magnetic field between 0.1 and 2 AU to study the disruption of a heliospheric current sheet fold during CR 1892 in the decline phase of solar cycle 22. (Merkin, Kondrashov, et al. 2016) also presented a simulation study exploring heliospheric consequences of time-dependent changes at the Sun during a 2-month period in the beginning of year 2008, in which they used the Air Force Data Assimilate Photospheric Flux Transport (ADAPT) model to obtain daily updated photospheric magnetograms and drive the WSA model of the corona. The results of the WSA model was used as a time-dependent boundary condition for the LFM-helio model. They compared the simulation results with ACE, STEREO-A and B near 1 AU, and MESSENGER spacecraft orbiting between 0.35 and 0.6 AU and their simulations showed that time-dependent simulations could reproduce the gross-scale structure of the heliosphere.

The CRONOS model was employed to solve the equations of ideal MHD in a one-fluid model and to obtain realistic modeling solar wind conditions from 0.1 AU to 2 AU (Wiengarten et al. 2013; 2014). Additionally, Wiengarten et al. (2015) incorporated turbulence transport into the Reynolds-averaged MHD equations in the framework of the CRONOS, which was used to investigate the effects on the turbulence evolution for transient events from 0.1 AU to 1 AU by injecting a CME from the inner boundary.

Shiota et al. (2014)) developed the SUSANOO model (also see Shiota and Kataoka 2016), and they used it to simulate the ambient solar wind structure from 25 to 425 Rs covering 3 years (2007-2009). Their numerical results were in reasonable agreement with *in situ* measurements at Venus and Mars by Venus Express and Mars Express, respectively (Shiota et al. 2014).

The "European heliospheric forecasting information asset" (EUHFORIA) is another space weather forecasting model focus on the inner heliosphere, which is capable to provide MHD modeling of the ambient solar wind and the CME eruptions from 0.1 AU to 2 AU (Moschou et al. 2017; Pomoell and Poedts 2018). EUHFORIA consists of an empirical coronal model and an MHD heliosphere model. The coronal model provides plasma and magnetic field parameters at $R_b$ =0.1AU, which was then used as boundary condition to drive the heliospheric model. The performance of the solar wind model was recently performed in the study of Hinterreiter et al. (2019).

Usmanov et al. (2018) presented a fully 3D MHD model of the solar corona and solar wind by coupling Reynolds-averaged solar wind equations with transport equations for turbulence energy, cross helicity, and correlation scale, from the coronal base to 5 AU. Their simulation results showed that the model could reproduce most of the solar wind parameters compared with Ulysses data during its first and third fast latitude transits.

Piantschitsch et al. (2017) developed a 2.5D MHD code to simulate the corona wave propagation and its interaction with a low-density region, such as the coronal hole (CH). By using this new code, they also made a comprehensive analysis on the dependence on different alfvén speed inside the CH and initial amplitude of the incoming wave (Piantschitsch et al. 2018a; 2018b). Their results depicted that the density value inside the CH influenced the phase speed and the amplitude values of density and magnetic field for all different secondary waves; and there existed correlation between the initial amplitude of the incoming wave and the amplitudes of the secondary waves as well as the peak values of the stationary features.

## 4.2 Modeling CME Initiation and Propagation

Jacobs and Poedts (2011) gave a detailed review about the state-of-art models for CME simulation before 2011. Here, we focus on the progress on the CME initiation, propagation, CME-CME interaction and CME- solar wind interaction, mainly after 2011, especially from years 2014 to 2019. The review by Manchester et al. (2017) includes a section about simulations and additional information about the main physical processes occurring during CME evolution in the inner heliosphere.

### 4.2.1 Modeling CME Initialization

As mentioned by Jacobs and Poedts (2011), there is no CME model sufficiently well developed to fully explain all of the observed features of solar eruptions and related phenomena (dimming regions, ribbons, post-eruption arcades, EUV waves, solar energetic particles, etc.). In addition, the basic pre-eruption configuration and the topological changes in the magnetic field that result in the conversion of a large fraction of the magnetic energy into kinetic energy are still not well understood. Most of the existing CME models are candidates for mimicking the morphology near the Sun, with the purpose of reproducing the plasma parameters comparable with 1 AU observations, with the stated goal for many of them to move towards real-time space weather forecasting simulations. Presently, significant progress has been made towards improving the performance of the existing CME initialization models.

### 4.2.1.1 Cone model

The cone model (Fisher and Munro 1984; X. P. Zhao, Plunkett, and Liu 2002; Xie, Ofman, and Lawrence 2004) is one of the popular CME initiation models because of its simplicity and relatively good match with CME arrival time observations. In most implementations of the model, the CME does not possess internal magnetic field, but the input size, speed and location is determined from coronal observations, typically from coronagraphs. In addition, due to its geometry and lack of internal magnetic field, the initiation does not include parameters related to CME orientation. By using cone model as CME initialization, Odstrcil, Pizzo, and Arge (2005) applied the 3-D MHD simulation to the 12 May 1997 interplanetary event to analyze possible interactions of the ICME propagating in various steady state and evolving configurations of the background solar wind.

Taktakishvili et al. (2011) reported the simulated results of selected well-observed halo CME events using combined model of the WSA/Enlil and the cone models. Their simulation results demonstrated that the combination of numerical models with the observations from coronagraph as input could give reasonably good results for the CMEs' arrival times for the selected set of "geoeffective" CME events. Bain et al. (2016) also combined the WSA-ENLIL and cone models to discuss shock connectivity in the August 2010 and July 2012 events. Dewey et al. (2015) integrated the cone model into the WSA-Enlil model to study the CME-related solar wind perturbations on the Mercury system. Their simulation results demonstrated that the modeled results could be compared with the observations by the spacecraft of MErcury Surface, Space ENvironment, GEochemistry and Ranging (MESSENGER) during the period from March 2011 to December 2012.

Pomoell and Poedts (2018) integrated the cone model into EUHRORIA model to simulate the CME events in the inner heliosphere during 17-29 July 2015. Also by combining the cone model and the EUHRORIA model, Scolini et al. (2018) tested the effect of different CME shapes on the simulation results, and their simulation results showed that all the parameters specifying the CME shape in the model significantly affect simulation results at 1 AU as well as the predicted CME geoeffectiveness.

### 4.2.1.2 Flux rope models

Flux rope models have been shown to self-consistently reproduce many observed properties of CMEs, including the three-part density structure (W. Manchester et al. 2017). Contrary to the cone models, they include internal magnetic field and may therefore reproduce not only the arrival time but also the magnetic field components when a CME impacts Earth. Due to their more complex nature, additional parameters are required to initialize such models, including the internal magnetic field strength or flux and the orientation of the flux rope. These models were first implemented in 3D MHD simulations by Roussev et al. (2003).

Lionello et al. (2013) improved the MAS-ENLIL model by inserting an out-of-equilibrium flux rope in the coronal model within 7 Rs as CME initiation model, and they simulated the propagation of an interplanetary CME (ICME) from 18 Rs to 1.1 AU. Their simulation results showed that the improved model could follow the propagation of the CME accurately. By using the MAS/MAS-H model combined the modified Titov–Démoulin (TDm) model (Titov et al.

2014), Török et al. (2018) inserted a magnetically stable flux rope to generate a CME close to the observed properties of the 2000 July 14 "Bastille Day" eruption. The properties of the CME as it propagates were studied based on MHD simulations of solar eruptions from near the Sun to the

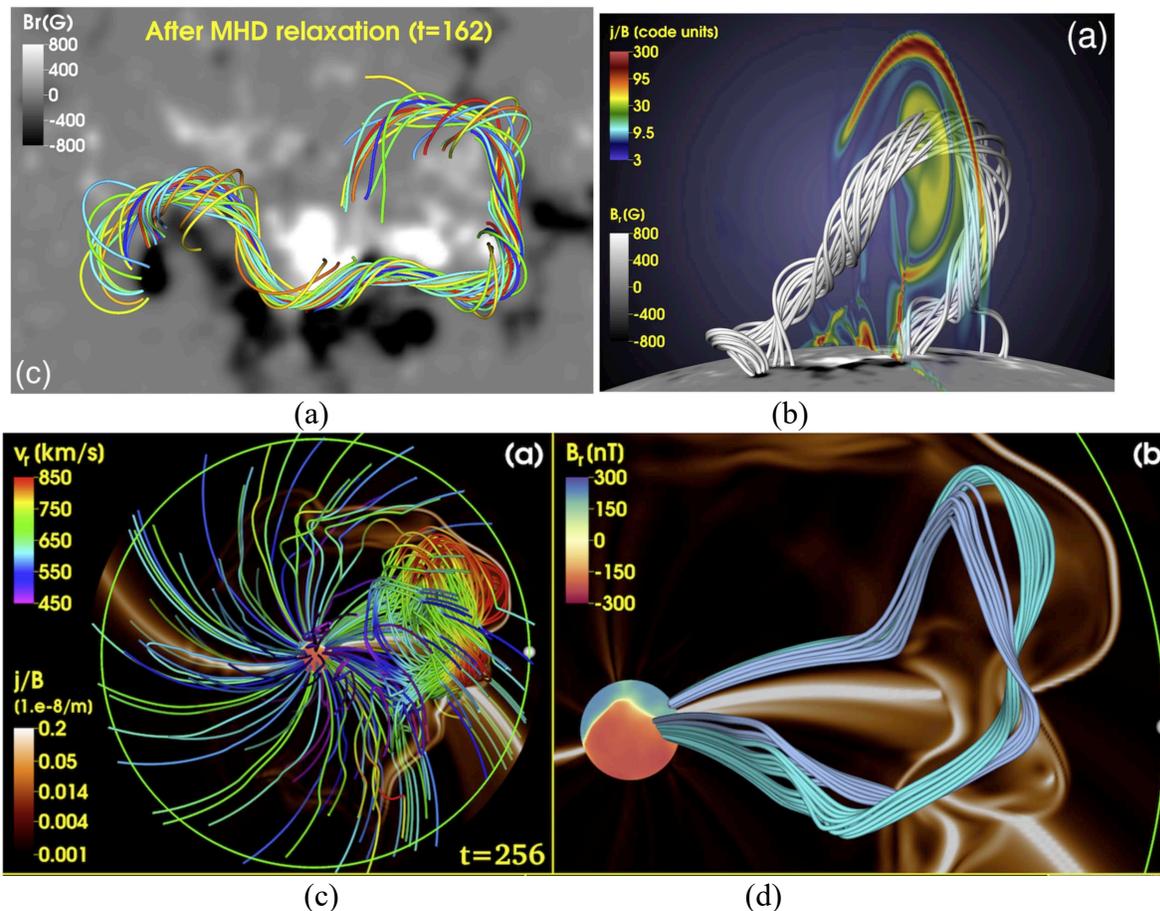

*Figure 4-3. (a) Initial flux-rope field lines with zero-β relaxation; (b) Field lines of the flux-rope core at t=164.10; (c) Interplanetary magnetic field and ICME flux rope at t=256; (d) Close-up view on (c), showing two flux bundles at the core of the flux rope (From Török et al. 2018).*

Earth. Figure 4-3(a) showed the initial flux-rope field lines, 3(b) depicted the field lines of the flux-rope core at t=164.10, shortly after eruption onset, 3(c) and 3(d) demonstrated the interplanetary magnetic field and ICME flux rope at t=256, shortly before it reached 1 AU.

Using the Titov-Démoulin (TD) flux-rope model to initiate the CME, Jin et al. (2013) simulated a fast CME erupted from active region NOAA AR 11164 during CR2107. Simulations of this CME event were conducted with 1T (one-temperature) and 2T (two-temperature: coupled election and proton) MHD models. The authors compared the propagation of this fast CME and the thermodynamics of CME-driven shocks in both the 1T and 2T CME simulations, and their results demonstrated the importance of the electron heat conduction in conjunction with proton shock heating in order to produce the physically correct CME structures and CME-driven shocks.

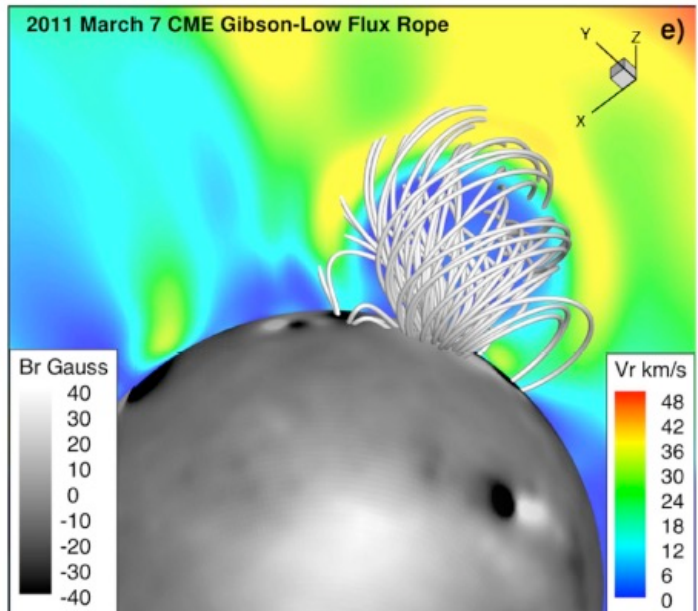

(a)

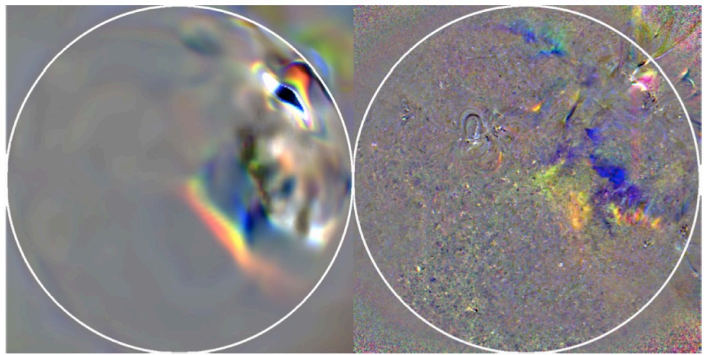

(b)

*Figure 4-4. (a) Initial GL flux rope configuration for 7 March 2011 CME; (b) EUV waves in the simulation (left) and in the SDO/AIA observation (right) (From Jin, Manchester, Holst et al. 2017).*

Based on the solar wind background constructed by AWSoM SC model (van der Holst et al. 2014), Jin et al. (2016) presented a numerical simulation on the CME which occurred at 00:04 UT on 15 February 2011, which was initiated by using the analytical Gibson-Low (GL) flux rope model (S. E. Gibson and Low 1998) with different parameters. Their simulation results showed that a CME's impact on the surrounding solar wind structures would be influenced by the magnetic strength of these structures, their distance to the source region, and the interaction between the CME with the large-scale magnetic field. Jin et al. (2017) developed a new data-driven tool called Eruptive Event Generator Gibson-Low (EEGGL) to automatically determine the GL flux rope parameters using synoptic magnetogram data from GONG and CME speed derived from the observations of SOHO/LASCO. By combining the EEGGL model and the AWSoM solar wind model, Jin et al. (2017) conducted a comprehensive study of CME propagation on the 7 March 2011 by performing a simulation from the chromosphere to 1 AU. Their simulated results could reproduce many of the observed features both near the Sun and in the heliosphere. Figure 4-4 (a) depicted the initial GL flux rope configuration for 7 March 2011 CME event with central plane showing the radial velocity, and Figure 4-4(b) compared EUV waves in the simulation and in the SDO/AIA observation.

The Versatile Advection Code (VAC) is a general tool for solving MHD and hydrodynamical problems with astrophysical applications. A variety of numerical schemes are available for users to solve hyperbolic differential equations, including, e.g., TVD-Roe, TVDLF, and flux correction of transport (FCT) method. By using VAC model, Jacobs and Poedts (2012) solved the MHD equations with the inner boundary of the domain locating at the low solar corona. They investigated the effect of new flux emergence on a magnetic system that possessed a 3D topology favorable for the breakout scenario, which was suitable for the 'breakout' CME scenario to work. Keppens et al. (2012) implemented a block-based AMR on the parallel VAC model using the Message Passing Interface library (MPI-AMRVAC), which has been used to provide interplanetary space weather forecasting models with relative accurate time dependent boundary conditions of erupting magnetic flux ropes in the upper solar corona. Pagano, Bemporad, and Mackay (2015) performed a 3D MHD simulation of a flux rope ejection where a CME was produced by using the MPI-AMRVAC. Their results showed that the polarization ratio technique could reproduce the position of the center of mass along the line of sight with relative high accuracy and studied the propagation of the CME on the real 3D direction.

Singh et al. (2019; 2020) modified the Gibson-Low flux rope by constraining the poloidal and toroidal fluxes of the initial flux rope using eruption data such as total reconnected flux (N. Gopalswamy et al. 2017) and flux in the core dimming regions (D. F. Webb et al. 2000; Dissauer et al. 2018; Kay and Gopalswamy 2018b). The modified spheromak has the option to control the helicity sign of flux ropes, which can be derived from line-of-sight magnetograms. Singh et al. (2020) simulated the 2012 July 12 CME and showed that they can reproduced the properties of the CME in the coronagraph FOV. They created a solar wind background from 1.03 Rs to 30 Rs using solar synoptic magnetograms. Then the flux rope model is inserted into the domain, allowing it to erupt as a CME due to pressure imbalance. They used the Multi Scale Fluid Kinetic Simulation Suite (MS-FLUKSS, Yalim, Pogorelov, and Liu 2017), which is a highly parallelized code suitable for MHD treatment of plasma and fluid.

Wu et al. (2016) presented a 3D MHD simulation based on an observed eruptive twisted flux rope deduced from solar vector magnetograms. They combined a data-driven flux rope model for the CME initiation and a global coronal-heliosphere evolution model to track the propagation of the CME. They selected the CME event on 6 September 2011 to test this model, and their simulation results suggested that the flux rope evolution model produced the physical properties of a CME, and the morphology resembled the observations made by STEREO/COR1.
By using the 3D IN-TVD-MHD model (Fang Shen et al. 2018), Liu, Shen, and Yang (2019) established a CME flux rope model based on the graduated cylindrical shell (GCS) model and applied it into the numerical simulation on the propagation and deflection of the fast CMEs in the interplanetary space from 0.1 AU to 1 AU.

### 4.2.1.3 Spherical plasmoid / Magnetized plasma blob

Besides the cone model and the flux rope models, the spherical plasmoid model and the magnetized plasma blob are also popular CME initialization models used in the recent years (F. Shen, Feng, Wu, et al. 2011; F. Shen, Feng, Wang, et al. 2011b; Y. F. Zhou et al. 2012; C. Shen et al. 2012a; Y. F. Zhou and Feng 2013; Fang Shen et al. 2014; Y. Zhou, Feng, and Zhao 2014; Kataoka et al. 2009; Shiota and Kataoka 2016; Chané et al. 2006). Similar to the flux rope model, they incorporate internal magnetic field and require the associated parameters. Initial 3D MHD simulations were performed by Groth et al. (2000) and Manchester et al. (2004).

By using the 3D SIP-CESE MHD model with the spherical plasmoid mimicking CME initiation model, the time-dependent propagation of the Sun-Earth connection CME events, such as 4 November 1997, 12 May 1997, and 2010 April 3 CME events were investigated (Y. F. Zhou et al. 2012; Y. F. Zhou and Feng 2013; Y. Zhou, Feng, and Zhao 2014). And their simulated results provided a relatively satisfactory comparison with the Wind spacecraft observations.

By using the 3D COIN-TVD MHD model with the magnetized plasma blob as CME initialization model, Shen et al. (2011) and Shen et al. (2014) simulated the time-dependent propagation of single CME events, and the interaction of two CMEs events, such as 4 April 2000 and 12 July 2012 CME events, and 28 March 2001 CME-CME interaction event. Their simulation could reproduce relatively well the real 3-D nature of the CME in morphology and their evolution from the Sun to the Earth.

A spheromak-type magnetic flux rope was also taken as the magnetic field structure of the initial CME model by Kataoka et al. (2009) and Shiota and Kataoka (2016), to simulate the propagation of the CME by using the SUSANOO model.

Recently, a spheromak model was included in EUFHORIA(Verbeke, Pomoell, and Poedts 2019a) and shown to work in term of comparison with in situ measurements for specific case studies for Sun-to-Earth propagation of CMEs (C. Scolini, Rodriguez, et al. 2019a; Erika Palmerio et al. 2019).

### 4.2.1.4 Reconstructions of Coronal Magnetic Fields

In order to extrapolate the coronal magnetic field from photospheric vector magnetograms based on the nonlinear force-free method, Jiang et al. (2011) and  Jiang and Feng (2012) Jiang and Feng (2012) exploited the CESE-MHD model to solve the zero-beta MHD equations with a fictitious frictional force and make reconstructions of coronal magnetic field, which was called as CESE-MHD nonlinear force-free field (CESE-MHD-NLFFF) model (also see (Jiang, Feng, et al. 2013; Jiang et al. 2014; Jiang and Feng 2014). By using the CESE-MHD-NLFFF model combing the vector magnetograms observations, a series of simulations were carried out to investigate, among others, the 3D magnetic field of NOAA AR 11117 on 25 October 2010, formation and eruption of the active region sigmoid in AR 11283, a large-scale pre-flare current sheet in NOAA AR 11967, and the evolving magnetic topology for an X9.3 eruptive flare from geoeffective AR 12673 that occurred on September 6, 2017 (Jiang et al. 2012; Jiang, Wu, et al. 2013; Jiang et al. 2016; 2017; 2018). Their simulations could qualitatively reproduce the basic structures of the 3D magnetic field; the current sheet in the corona as well as providing insight into the magnetic mechanism of solar flares; the spatial location, the temporal separation of the observed flare ribbons, as well as the dynamic boundary of the flux rope's feet by mapping footpoints of the newly reconnected field lines. Figure 4-5 presented the comparison of the modeled magnetic field with the observed features of the solar corona prior to the flare field lines (Jiang et al. 2018).

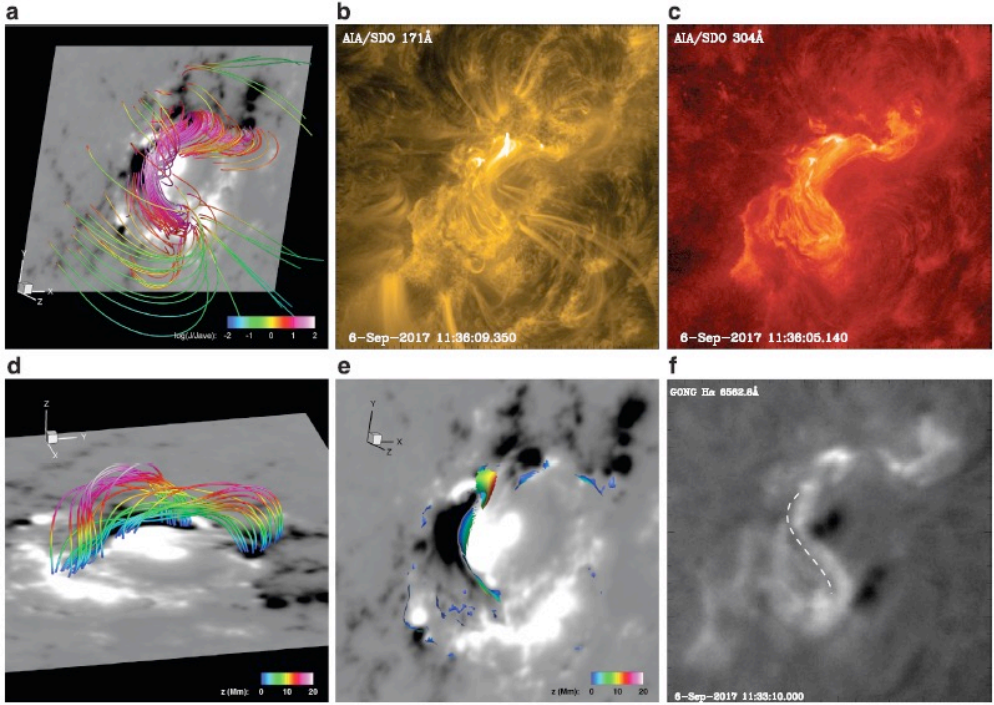

Figure 4-5. Comparison of the reconstructed magnetic field with the observed features of the solar corona prior to the flare. (a) SDO view of sampled magnetic field lines of the CESE-MHD-NLFFF reconstruction. (b) and (c) SDO/AIA 171 Å and 304 Å images of the pre-flare corona. (d) The low-lying magnetic field lines in the core region. The field lines are color-coded by the value of height z. (e) Locations of dips in the magnetic field lines; the color indicates the value of height z. (f) GONG Hα image of the active region. The dashed curve denotes the location of a long filament (From Jiang et al. 2018).

By using MPI-AMRVAC MHD code to the reduced MHD equations with only the density, velocity, and the magnetic field, and without the gradient of gas pressure and gravity, and the energy equation, Guo et al. (2019) developed a data-driven MHD model with the zero-β approximation. The initial condition is provided by a nonlinear force-free field derived from the magnetofrictional method based on vector magnetic field observation from SDO. Their MHD simulation was carried out for AR 11123 observed on 11 November 2010 and could reproduce the eruption process of the magnetic flux rope.

Other work that have more realistic magnetic field evolution models into CME model include Price et al. (2019), Pomoell, Lumme, and Kilpua (2019), Hayashi et al. (2018) and Hayashi et al. (2019) among others.

### 4.2.1.5 Other CME models

The HAFv.2+3D MHD model has been used to study a variety of solar eruptive events, such as the interplanetary evolution of the observed geoeffective CME during 1-4 August 2010 (C.-C. Wu et al. 2011), and the effects of coronal hole on CME/shock morphology in the inner heliosphere with 7 March 2011 solar events (B. E. Wood et al. 2012). Liou et al. (2014) employed the model to investigate the propagation of the extremely fast backside CME event on 23 July 2012 and the modeled results were in agreement with the in-situ measurement from STEREO-A. Specially, Wu et al. (2017) investigated the CME encountered by the Wind spacecraft on 9 September 2011 in detail and verified the association of the short-duration ($\sim 35$ minutes) extremely dense pulse (with a peak of $\sim 94$ cm$^{-3}$) with the heliospheric plasma sheet compressed by the interplanetary shock.

By injecting a CME from the bottom boundary, Wiengarten et al. (2015) incorporated turbulence transport into the CRONOS model, and investigated the effects on the turbulence evolution for transient events from 0.1 AU to 1 AU. Their study found that the CME-associated shock increased the turbulence levels and inhibited the cross helicity. They also indicated that researches on the large-scale structures associated with CMEs did not need to consider the turbulence transport effects due to the absence of strong back-reaction of the turbulence on the large scale structures.

## 4.2.2. Modeling the interaction between CMEs and solar wind structure (CIR, HCS)

Previous numerical studies have shown that both the corotating interaction region (CIR) and heliospheric current sheet (HCS) structures of the background solar wind could play a substantial role in the propagation of CMEs and their geoeffectiveness (Odstrčil, Dryer, and Smith 1996; Odstrcil, Riley, and Zhao 2004). Therefore, the MHD simulation on the interaction between CMEs and the solar wind structure (e.g., CIR, HCS) is one of the important aspects in the CME simulations, and has achieved a lot of progress in recent years.

By using the 3D SIP-CESE MHD model, Zhou and Feng (2017) simulated the propagation characteristics of CMEs launched at different positions in a realistic structured ambient solar wind. By using the HAFv.2+3DMHD model, a time series of synoptic photospheric magnetic maps, and the recording of CMEs from STEREO/COR2, Wu et al. (2016) simulated the Sun-to-Earth propagation of multiple CMEs and their associated shocks in September 2011. Their simulation found that the evolution of the CME-driven shock and its interaction with the HCS and the non-uniform solar wind could explain time-intensity profile of the high-energy (> 10 MeV) solar energetic particles (SEPs), and the sector boundary acted as an obstacle to the propagation of SEPs. Further, Wu et al. (2016) employed the model to study 12 CMEs and their associated shocks in September 2011. The results demonstrated that the background solar wind speed was an important controlling parameter in the propagation of interplanetary shocks and CMEs.

Using 2.5D version of VAC MHD model, Zuccarello et al. (2012) and (Bemporad et al. 2012) numerically studied the role of streamers in the deflection of CMEs or multiple CMEs. Their results showed that the CME deflected toward the current sheet of the larger northern helmet streamer due to an imbalance in the magnetic pressure and tension forces and finally gets into the streamer. As pointed out by (Zuccarello et al. 2012), during solar minima, even CMEs originating from high latitude could be easily deflected toward the HCS, eventually resulting in geoeffective events, and that this latitudinal migration depended on both the strength of the large-scale coronal magnetic field and the magnetic flux of the erupting filament.

Zhuang et al. (2019) simulated the deflection of CMEs with different speeds in the interplanetary space using a 2.5D MHD simulation. Their simulation confirmed the existence of the CME deflection in the interplanetary space, which was related to the difference between the CME speed and the solar wind speed. They found that a CME, which traveled slower or faster than the solar wind medium, would be deflected to the west or east; and the greater the difference was, the larger the deflection angle would be. Liu, Shen, and Yang (2019) simulated the propagation and deflection of the fast CMEs interacting with CIR in the interplanetary space by using the 3D IN-TVD-MHD model and the CME flux rope model based on the GCS reconstruction. Their simulation results showed that when the fast CME hit the CIR on its west side, it would deflect eastward, and the deflection angle would increase compared with the situation without CIR.

## 4.3.  Modeling CME-CME interactions

Observational and numerical studies have shown that the kinematic characteristics of two or more CMEs may change significantly after the CMEs interaction. The CME-CME interaction is always associated with complex phenomena, including magnetic reconnection, momentum exchange, energy transfer, the propagation of a fast magnetosonic shock through a magnetic ejecta, and changes in the CME expansion, and so on (Noé Lugaz et al. 2017; Fang Shen et al. 2017b). Numerical modeling, which always yield the observed complexity, have been proved to be a useful tool to understand and determine the dynamical evolutionary processes of the CME-CME interaction.

Webb et al. (2013) tracked the propagation of multiple CMEs of late July to early August 2010 in the inner heliosphere by comparing the results from the ENLIL model, 3D reconstruction techniques based on a kinematic solar wind model, and in situ results from multiple spacecraft. By using WSA-ENLIL+Cone model based on coronagraph image observations, Werner et al. (2019) modeled the multiple CME interaction event on 6-9 September 2017. The predicted arrival time of the first interplanetary shock was drastically improved, while the background solar wind preconditioned by the passage of the first interplanetary shock likely caused the last CME to experience insignificant deceleration and led to the early arrival of the second interplanetary shock.

Using the COIN-TVD MHD model, Shen et al. (2011) and Shen et al. (2012) simulated the interaction of two CMEs in interplanetary space, analyzed variations of different forces during the interaction, and found that the momentum exchange during the collision of two CMEs was very important for the deceleration and acceleration of the CMEs. The 3D COIN-TVD model was also used to study the super-elastic collisions of CMEs in the heliosphere (Fang Shen et al. 2013b). Results showed that the collision led to extra kinetic energy gain by 3%-4% of the initial kinetic energy of the two CMEs, which suggested that the collision of CMEs could be superelastic. Shen et al. (2016) furthered the dependence of CMEs' collision type on the ratio of the CME's kinetic energy to the CME's total energy.

By employing the SUSANOO model, Shiota and Kataoka (2016) reproduced the propagation and interaction process of multiple CMEs associated with the highly complex active region NOAA AR 10486 from 30 R$s$ to 430 R$s$ in October to November 2003. Their simulation results could successfully provide reasonably good results for velocity and the profile of southward magnetic field component of the Halloween Event on 29 October 2003. The simulation also indicated that for the propagation of the following CME was significantly affected by the trails of the preceding CMEs. Lugaz et al. (2013) used the SWMF to study the influence of the relative orientation of the two interacting CMEs on their interaction and the resulting structure. In addition to the well-studied multiple-MC event, they described other potential structures, further compared with actual CME measurements in Lugaz and Farrugia (2014).

The September 2017 series of events, which resulted in one of the largest geomagnetic storms of solar cycle 24, has been investigated by means of numerical simulations, focusing on the CME-CME interaction by a number of groups. Scolini et al. (2019) used EUHFORIA to study how complex interactions between multiple interacting CMEs on their way to Earth may result in an intensification of the geo-effectiveness potential of such multiple-CME events. Werner et al. (2019) used the ENLIL model to investigate how the succession of CMEs in early September 2017 resulted in one of the largest geo-effective periods of solar cycle 24. They specifically focused on the importance of pre-conditioning of previous, non-interacting CMEs on the propagation of the following fast CMEs, confirming past work, both based on simulations (N. Lugaz, Manchester IV, and Gombosi 2005b; GáBor Tóth et al. 2007) and measurements Liu, Zhu, and Zhao (2019).

## 4.4. Conclusions and Future Prospects

In the past five years, the main developments in the investigation of CME propagation by mean of numerical simulation have been as follows.

There has been a significant increase in the number of 3-D MHD codes that have been successfully used to simulate the Sun-to-Earth propagation of CMEs; this has been the case most notably in Europe with EUHFORIA, Japan with SUSANOO and China with IN-TVD MHD. In addition, a number of existing MHD codes have been adapted to investigate the heliospheric propagation of CMEs, including LFM into LFM-Helio and MAS into MAS-Helio. Heliospheric codes (starting typically at 0.1 AU) have been used with spheromak and/or flux rope CMEs, which bridges the gap between computationally intensive Sun-to-Earth simulations and heliospheric simulations with cone models. These types of simulations may be used to investigate the magnetic field configuration inside CMEs as well as their arrival time and are more physically consistent when investigating CME-CME interaction than simulations where the CMEs do not have internal magnetic fields.

In parallel, there has been an effort to make the CME initialization quicker and easier to perform in coronal codes using out-of-equilibrium flux ropes, especially with EEGL in the SWMF and within the MAS code. This paves the way for future, real-time Sun-to-Earth simulations with magnetized CMEs initiated based on magnetograms, EUV images, and early coronagraphic images. It is well known that major changes in the CME properties, including its speed and orientation, may occur below 0.1 AU where heliospheric models are initiated. At this time, it is however unclear whether simulations with magnetized CMEs initiated at 0.1 AU using multi-viewpoints coronagraphic measurements (as described in the point 2) will perform worse than simulations with magnetized CMEs initiated at the solar surface in term of space weather forecasting capabilities. The number of Sun-to-Earth simulations of CMEs initiated at the solar surface with a realistic model is still relatively low, even though there has been effort in presenting the results near 1 AU of more complex initiation mechanisms, as done for example in Török et al. (2018).

There has been significant new physics included in the solar wind models, including more advanced thermodynamics treatment and the inclusion of Alfvén waves. There has not been significant work quantifying how these new additions affect the CME propagation and the resulting structure near 1 AU.

Lastly, there has been progress towards coupling time-dependent magnetic field models with coronal models and heliospheric models. This is already the case for the background coronal and interplanetary magnetic field with ADAPT coupled to a number of MHD models, which has been shown to result in more accurately modeled heliospheric current sheets. Initiating CMEs by means of magnetofrictional or other self-consistent models based on solar observations or flux emergence may lead the way for a better physical understanding of CMEs and is probably the only way space weather forecasting could provide information before the launch of a CME. We expect further improvements towards this coupling in the next few years.

# 5. Campaign Study of Sun-Earth Connection Events

## 5.1 Introduction

The task of ISEST Working Group 4 (Campaign Events) was to integrate theory, simulations and observations to better understand the chain of cause-effect activity from the Sun to Earth for carefully selected events. ISEST provided "textbook", or well-understood, Sun to Earth cases to the community, but WG 4 also examined more controversial events, such as stealth CMEs and problem ICMEs, to enhance our understanding. This includes analyzing the difficulties in linking CMEs to ICMEs, which are usually observed only *in situ*.

WG 4 classified the studied events into three general categories: (1) Possible "textbook" cases in which the complete chain of a well-observed event is relatively well understood from its solar source, through its heliospheric propagation, to its geo-effects. These cases involve forecasts that are successful in a general way. (2) Cases in which there were problems understanding the complete chain, but which we think we now understand. Thus, something was missing in the chain of a well-observed event but, *in retrospect after analysis*, we now understand why. These cases usually involve forecasts that failed because they were not geoeffective, or were otherwise not accurate. (3) Finally, there are problem cases in which the chain is not complete and we still do not understand why. In the next two sub-sections, we briefly summarize the results for each type of the events, which are discussed in detail in Webb and Nitta (2017) (hereafter WN17) and Nitta and Mulligan (2017) (hereafter NM17). WN17 studied six cases during the rise of Solar Cycle 24 that highlight forecasting problems. The six events were chosen to illustrate some key problems in understanding the chain from solar cause to geoeffect. NM17 studied stealth, or problem CMEs that have no clear Low Coronal Signatures (LCS).

Table 5-1 is a summary of the 14 campaign events that were discussed and analyzed by WG 4. These studies have resulted in many presentations and papers in the literature. The first column group gives the Event Number and the range of dates from the solar source to any geo-effect. The second column group summarizes the source activity, the third the geo-response, followed by the storm peak Dst, if any, and the peak Kp and G indices (see below). Finally, at the right is given our estimate of the degree of forecast success. The first six events were chosen as VarSITI-wide Campaign Study Events because they had certain space weather effects of interest to one or more of the other three VarSITI projects. A focus of the WG 4 studies was to understand the Sun-to-Earth cause-effect chain for five of these 6 campaign events. The other 8 of the 14 events were chosen because of particular aspects of interest to ISEST WG 4 that help elucidate the Sun-to-Earth chain.

Table 5-1. ISEST/MiniMax WG 4 Campaign Events

## ISEST / MiniMax WG 4 Event List

| Dates | Source | Geo-response* | Dst | Kp/G Level | Forecast Success |
|---|---|---|---|---|---|
| **VarSITI-wide Campaign Study Events** | | | | | |
| 1) 2012 July 12-14 | X1 flare, wave, fast CME | Shock, MC, Strong storm | -127 | 7/G3 | Under-predicted |
| 2) 2012 Oct. 4-8 | CME; weak surface signs. | Shock, MC, HSS, Moderate stm | -105 | 6+/G2 | Under-predicted |
| 3) 2013 March 15-17 | M1 fl, wave, EF, IV, fast halo | Shock, MC? SEP, Strong storm | -132 | 6+/G2 | |
| 4) 2013 June 1 | Slow CME on 27 May? CH influence? | Cause of Strong stm unclear; CIR? | -119 | 7/G3 | Failed-not pred. |
| 5) 2015 March 15-17 | C9;C2 fl, wave, EF, fast CME | Shock, sheath, MC, Severe storm | -222 | 8+/G4 | Under-predicted |
| 6) 2015 June 22-24 | 2 M-fls, waves, fast halo CMEs | Shock, sheath, MC, SEP, Severe storm | -204 | 8+/G4 | Mostly successful |
| **Other ISEST/MiniMax Study Events** | | | | | |
| 7) 2012 March 7-9 | X5 flare, wave, fast CME | Shock, MC, Strong storm | -131 | 8/G4 | |
| 8) 2012 July 23-24 | 2 flares? Wave, EFs | Extreme ST-A event; "Strong storm" (Carr.-type) | --- | | |
| 9) 2014 January 6 | CME <2000 km/s, over WL | GLE at Earth | No | --- | |
| 10) 2014 January 7-9 | X1 fl, wave, fast asym halo | Shock, SEP. No storm- CH deflection; AR channeling? | No | ≤3 | |
| 11) 2014 Sept. 10-13 | X2 flare, wave, sym halo | Shock, MC, Moderate storm | -88 | 7/G3 | Over-predicted |
| 12) 2015 January 3-7 | Slow CME | Brief ICME, MC, HSS, Mod. stm | -99 | 6+/G2 | |
| 13) 2016 October 8-12 | Slow CME | Shock, MC, HSS, Moderate stm | -104 | 6+/G2 | |
| 14) 2017 Sept 4-10 | Act. series; ~ 3 S-E evts. | Shocks, MCs, Strong storm(s), FD | -124 | 8/G4 | |

CME = coronal mass ejection; AR = active region; EF = erupting filament; CH = coronal hole; MC = magnetic cloud; SEP = solar energetic particle event; CIR = corotating interaction region; GLE = ground-level event; HSS = high speed stream

xx) Events featured in Webb & Nitta (2017)
xx) Problem events featured in Nitta & Mulligan (2017)

Only the first 11 events were included at the time of writing of the WN17 paper. The 5 events studied by WN17 are highlighted in red. Three of the events, as well as others, were described by NM17 and are marked in purple. The October 2012 problem event was discussed in both papers.

This Section is organized as follows. The next subsection provides a summary of the Campaign events WG 4 studied during the rise of Solar Cycle 24 that highlight forecasting problems. Subsection 5.3 summarizes the results from the Nitta and Mulligan study of stealth CMEs. The results are discussed in subsection 5.4.

## 5.2 Understanding Problem Forecasts

WN17's six events were selected to illustrate the range of problems that can occur in understanding the complete chain of activity from its source region(s) at the Sun, its propagation through the heliosphere, to its effects at Earth. Likely source CMEs were identified in all six cases, but related solar surface activity ranged from uncertain or weak to X-class flares. The geoeffects ranged from no effects to severe effects, such as the two Sun-Earth events in 2015 that caused "superstorms". For each event they noted the official NOAA forecast that was issued after the solar source eruption but before its arrival at Earth, and whether the forecast was successful or was problematic in some important manner. Summaries of these forecasts are available under Reports of Solar and Geophysical Activity (RSGA) through the Space Weather Prediction Center (SWPC) site: ftp://ftp.swpc.noaa.gov/pub/warehouse/.

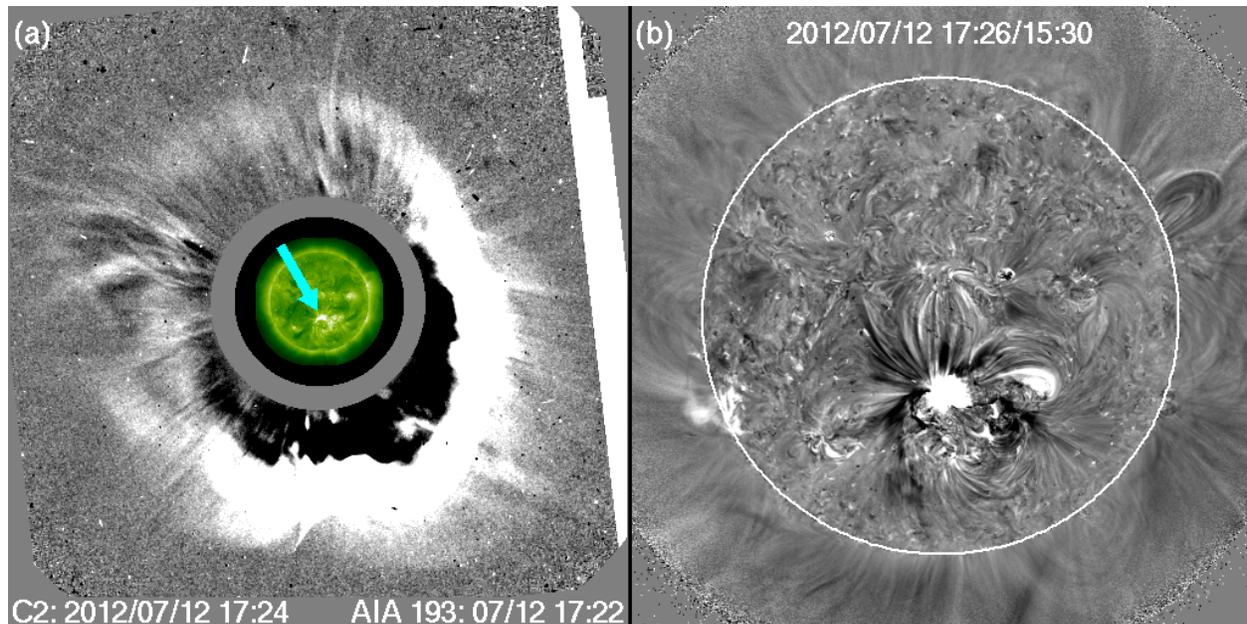

*Figure 5-1. Images showing the solar source region of Event #1 on 12 July 2012. Left: SOHO/LASCO running difference image showing the full halo CME on 12 July, 17:24 UT. Superimposed within the occulting disk area is a near-simultaneous SDO/AIA 193 Å image with an arrow pointing to the near-disk center flare and dimming in AR 11520. This image is from the SOHO/LASCO CDAW CME catalog. Right: Enlarged AIA base difference image of the source region.*

Event #1, 12-14 July 2012, was considered a classic textbook event, in that we observed the complete chain of a well-observed Sun-to-Earth event, from its solar source, through heliospheric propagation, to its geoeffects (Figure 2-1 and Figure 5-1). The propagation kinematics, flux rope eruption, and MHD modeling for this event were well studied by WG 4 members (Gopalswamy et al. 2013; Hess and Zhang 2014; Shen et al. 2014b; Möstl et al. 2014; Cheng et al. 2014; Hu et al. 2016; Marubashi, Cho, and Ishibashi 2017) and others. On 12 July 2012 an eruptive X1.4 flare occurred in AR 11520 (S17 W08) with an X-ray peak ~ 16:45 UT. Later during its rotation, this same active region produced several strong flares and CMEs. One was the 23-24 July CME (#8), aimed at the STEREO-A and one of the fastest, most energetic CME ever observed (*e.g.,* Baker *et al.*, 2013; Liu *et al.*, 2014). On 14 July the CME arrived at L1 with a shock observed by the *Wind* spacecraft at 17:38 UT, followed by the shock sheath and a 2-day long magnetic cloud. This ICME drove a moderate, long-lived geomagnetic storm with peak Dst = −127 nT on 15 July and with a duration of several days. The NOAA forecast was mostly successful for this event, but the storm was only moderate level so slightly under-predicted.

The 21-24 June 2015 case, #6, was also possibly a textbook event, but it was a <u>compound</u> *in situ* event at 1 AU resulting from a series of four shocks arriving over a 3-day span, and one likely ICME on 23-24 June. The third shock and ICME were likely produced by a symmetric halo CME on 21 June. Southward field in multiple shock sheaths and the ICME drove a powerful multi-step geomagnetic storm reaching Kp = 8+, G4 and Dst = -204 nT on 22-23 June. The NOAA forecast of a severe storm was accurate, but that level was reached a day later than predicted. This severe level was reached because there were multiple shocks and sheaths, strong southward MC fields, and high speed solar wind that acted to compress and enhance the wind structures. Publications of this event include the following (Ying D. Liu et al. 2015; Manoharan et al. 2016; N. Lugaz et al. 2016; K. Marubashi, Cho, and Ishibashi 2017; N. Gopalswamy, Mäkelä, et al. 2018)

The 15-18 March 2015 case, #5, was initially a problem event, but we now understand why. A slow (350 km s$^{-1}$) CME occurred to the south-southwest late on 14 March likely associated with a small C2.6 flare from AR 12297 at S21°W20° with a small filament eruption. Then on 15 March, ~01:36 UT, a fast (1120 km s$^{-1}$), asymmetric halo CME associated with a C9.1 flare erupted from the same active region but was brightest over the west limb. *In situ* L1 observations showed a strong shock at *Wind* on 17 March at 04:01 UT, followed by an extended sheath then an ICME and a MC later on 17 March. Behind the cloud were a corotating interaction region (CIR) and its high speed stream (HSS) that likely enhanced the solar wind parameters. This most severe storm of Solar Cycle 24 was very much under-predicted, in terms of both its magnitude and early time of arrival. Thus, this was a problem forecast. The CMEs may or may not have interacted near the Sun. There were two detailed papers by Liu *et al.* (2015) and Wang *et al.* (2016) arguing either side of that dispute. It is likely that during transport to Earth there was interaction with a CIR and deflection toward Earth. Other papers analyzing the MC/flux rope at 1 AU include (Katsuhide Marubashi et al. 2016; K. Marubashi, Cho, and Ishibashi 2017; Wu et al. 2016).

Although not one of WN17's primary study events, Event #10, 7-9 January 2014, was at the time considered a problem event, but is now understood. It was a problem because a large storm was predicted but none occurred! Unlike the March 2015 case in which the CME was deflected toward Earth, in this case the source flare and EUV wave were Sun-centered but the CME was offset to southwest, possibly deflected by a CH and/or channeled by strong AR magnetic fields, and thus missed the Earth (Gopalswamy, Xie, et al. 2014; Möstl et al. 2015; Wang et al. 2015a; Mays, Thompson, et al. 2015).

The 10-12 September 2014 case, #11, was initially a problem event, but we now understand why. On 10 September 2014, an X1.6 flare erupted in AR 12158, associated with a fast (1400 km s$^{-1}$) symmetric halo CME. The event was centered on the disk and had a large dimming region and a rapidly expanding coronal wave. This event seemed like a textbook example, with a major storm predicted, followed by a strong shock and long-duration MC hitting Earth on 12-13 September. However, the storm was minor because the sheath and MC magnetic fields were northward (+$B_Z$), so the storm was over-predicted. WG 4 members tried to use polarity inversion line data to predict the expected flux rope orientation at 1 AU, but no consensus was reached. It was found that the flux rope fit better to a later flare in the same AR (Marubashi, Cho, and Ishibashi 2017; Cho et al. 2017).

The 4-9 October 2012 case, #2, was initially a problem event but we now understand why. The source CME and resulting ICME that drove a small, two-step geostorm (Kp = 6+, G2) were identified, but the storm was slightly under-predicted. There were weak and multiple surface signatures and the CME was initially very slow, leading to uncertainties in the arrival time. Marubashi, Cho and Ishibashi (2017) studied the ICME for this event and fit a portion of it as a flux rope.

Finally, the 27 May-1 June 2013 case #4, was a problem event, which we still do not fully understand. During this entire period, NOAA/SWPC did not forecast any important geoactivity. But on 1 June, there was a brief but strong storm (G3) that reached Kp=7 and Dst = -119nT. A possible source was a slow CME on 27 May, but the associated surface features were unclear. There was also likely influence from a large coronal hole that led to interaction with a CIR or HSS at Earth (Gopalswamy, Tsurutani, and Yan 2015), with a likely embedded ICME and flux rope (Marubashi, Cho, and Ishibashi 2017; Nitta and Mulligan 2017). The ultimate cause of the strong storm remains unclear.

Events #12 and 13 in the WG 4 table are discussed as stealth CMEs in the next section. The last "event", #14, was actually a series of major flare (M and X-class)-CME events from 4-10 September 2017 resulting in shocks, MCs and a Forbush decrease at 1 AU. X flares occurred on 6, 7 and 10 September. One notable feature of this series of events was the timing of the 1 AU arrival of the shock from the second CME (associated with a X9.3 flare) that was nearly simultaneous with when the magnetic field of the ICME from the earlier CME (associated with a M5.5 flare) turned southward. This apparently enhanced the net geoeffectiveness of these individual events, but it is presently very challenging to forecast the timings of the successive phenomena, let alone individually. This event period occurred too late in the ISEST interval to be extensively studied by WG 4, but WG 4 members contributed to several papers mostly analyzing the X flares and CMEs at the Sun.

## 5.3 Study of Stealth CMEs; Those without Clear Low Coronal Signatures

A related study of the Campaign group was of the origin of CMEs that were not accompanied by obvious low coronal signatures (LCSs), but produced appreciable geoeffects at 1 AU. These CMEs characteristically start slowly. In several examples, extreme ultraviolet (EUV) images taken by the Solar Dynamics Observatory (SDO)/Atmospheric Imaging Assembly (AIA) revealed coronal dimmings and post-eruption arcades using difference images with sufficiently long temporal separations, which are commensurate with the slow initial development of the CME. Images from SECCHI EUVI and COR coronagraphs provided limb views of Earth-bound CMEs. Combined with SOHO observations, these helped limit the time interval in which the CME forms and undergoes initial acceleration. For other CMEs, we found similar dimming, although with lower confidence of its link to the CME. We note that even these unclear events can result in unambiguous magnetic cloud/flux rope signatures in *in situ* data at 1 AU. In addition, there was a tendency for the CME source regions to be located near coronal holes. *i.e.,* open field regions. This may have implications for both the initiation of a stealth CME in the corona and its outcome in the heliosphere.

The fact that some events without obvious LCSs produce appreciable geostorms was one of the motivations of the work by NM17, who not only discussed Event #2 in detail but also described Events #12 and #13 and also included Event #4 in their event list. It is notable that four of the 14 events dealt with by WG 4 were in the category of geomagnetic storms without obvious LCSs. Stealth CMEs are of great scientific interest because they may represent a different class of eruptions than normal CMEs, whose LCSs are unambiguous. Howard and Harrison (2013), however, cautioned that stealth CMEs may be due largely to observational effects (such as limited sensitivity and temperature coverage). Indeed it has been shown that initially unseen LCSs may be revealed after image enhancement or processing (Alzate and Morgan 2017; Nitta and Mulligan 2017). Therefore, it may be more appropriate to use the adjective "stealthy" when the LCSs of the CME are not clearly identified. Regardless of whether they are fundamentally different from normal CMEs or simply represent the low-energy end of a continuous spectrum of events triggered in similar ways (Lynch et al. 2016), stealthy events pose challenges to space weather prediction.

In Event #2, NM17 found a post-eruption arcade (PEA) sandwiched by coronal dimming regions in AIA images around the time of the first appearance of the CME in LASCO C2 images, using difference image cubes with long (~several hours) temporal separations (Figure 5-2). It was found that the PEA and dimming regions delineated a polarity inversion line that looked like a filament channel without a filament. Apart from the base difference images that needed compensation of solar rotation as a result of the long temporal separations, another important point was to use COR-1 data that allowed the determination of the CME lift-off time by observing the eruption from the side. Using AIA base difference images and COR-1 data for several stealth events, NM17 identified similar patterns of a PEA with dimming regions on either side. Figure 5-3 shows the ICME with MC and fitted flux rope at 1 AU.

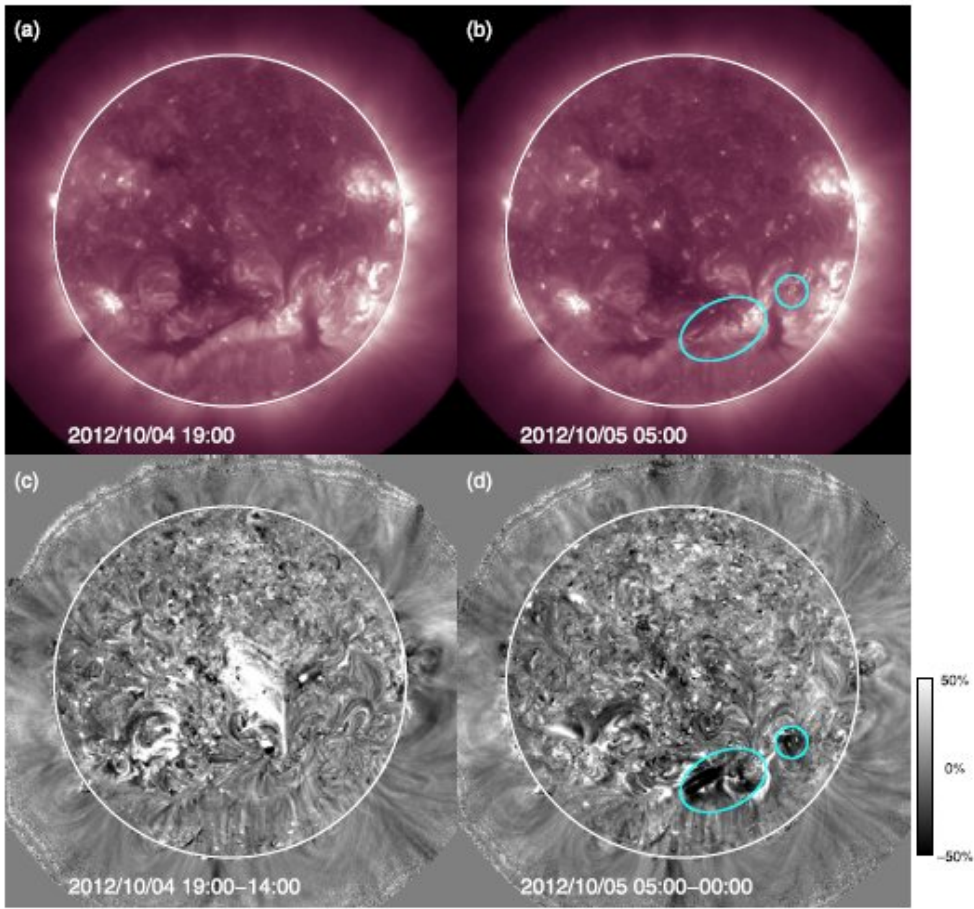

*Figure 5-2. AIA 211 Å intensity and percent difference images in the upper and lower panels, respectively, taken before and during the CME on 5 October 2012. Differences are made relative to an image five hours earlier. The dimming regions found in difference images are encircled in cyan. From Nitta & Mulligan (2017).*

However, Events #12 and #13 were more problematic. In Event #12, a partial halo CME was linked to the ICME responsible for the strong (Dst = 99 nT) geostorm. This CME was both very slow and diffuse. Even though one dimming region was clear not only in base difference images but also in intensity images in AIA's 193Å and 211Å channels as an augmentation of the south polar coronal hole, its mapping to the CME was not straightforward. Assuming that the dimming regions represent the legs of the erupting flux rope responsible for the CME, we would expect two dimming regions in opposite magnetic polarities. However, no second dimming region was found. Moreover, multiple regions became brighter in base difference images, and it is not possible to determine which ones may represent PEAs. This event also lacked STEREO observations, making it difficult to know the time of the CME liftoff.

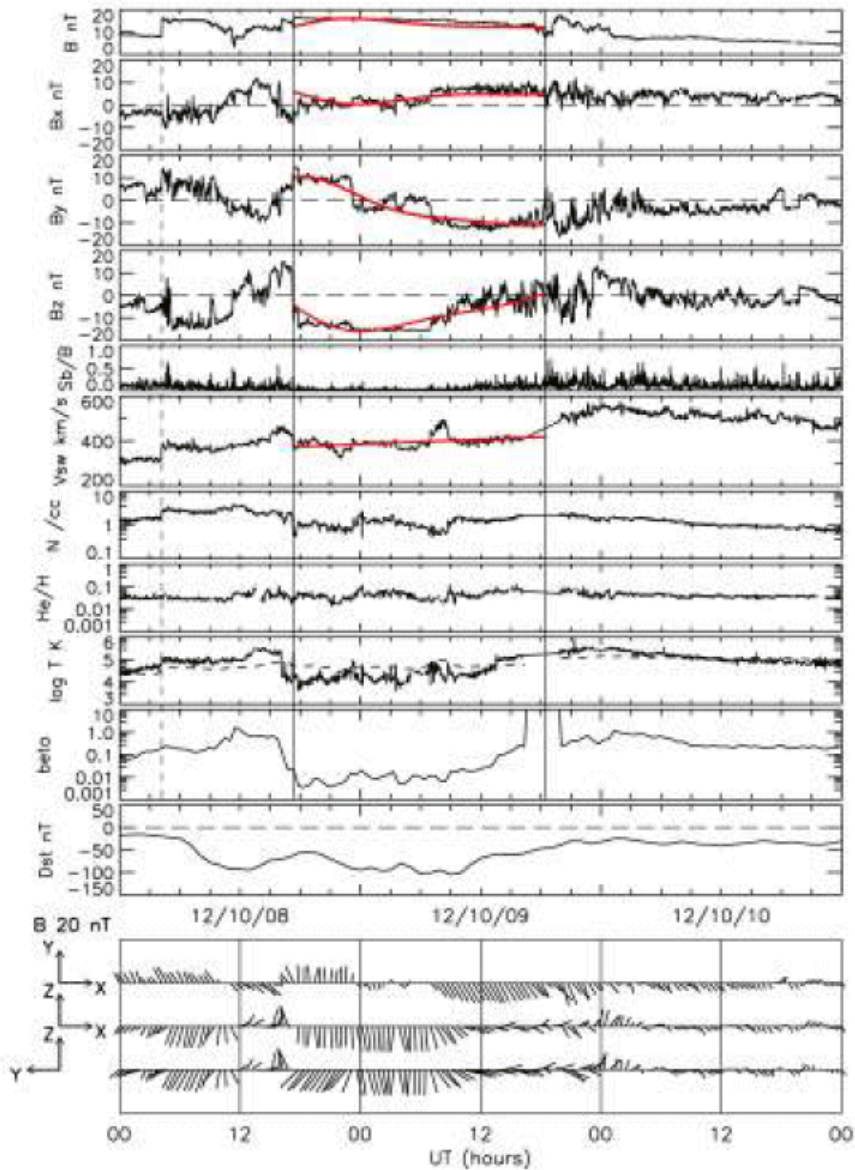

*Figure 5-3. Solar wind data from Wind and the Dst variation for three days, 8-10 October 2012. From top to bottom are plotted the IMF intensity (B), X-, Y-, Z- components in GSE coordinates (Bx, By, Bz), the degree of field fluctuations defined by the standard deviation divided by averaged intensity obtained from higher time resolution data (Sb/B), solar wind speed (Vsw), proton number density (N), number density ratio of He++ to H+ (He/H), proton temperature, plasma β, and the Dst index. The vertical dashed line indicates the shock arrival time (on 8 October at 04:12 UT) and the two vertical lines indicate the flux rope (from 8 October at 17:20 UT to 9 October at 18:30 UT). The red curves show the model values obtained from the fitting with a toroidal flux rope model. Adapted from Marubashi, Cho and Ishibashi (2017); courtesy K. Marubashi (2017, priv. comm.).*

Event #13 involved a full halo CME, which was again slow and diffuse. Without STEREO observations, this could have easily been taken as a backside event because there were no clear changes in the low corona around the time of the CME. NM17 showed two marginal dimming regions in AIA difference images taken 14 hours apart, but did not attach high confidence to them. Event #4 is controversial as to whether the strong (Dst = 119 nT) geomagnetic storm was purely CIR-related or enhanced by a small ICME embedded in the solar wind. The most likely CME in the time range in question was observed to head northward, giving an impression that it was not Earth-directed, but it could have deflected equatorward as indicated in STEREO HI data. Marubashi, Cho and Ishibashi (2017) showed a flux rope fitting, possibly supporting the latter possibility. NM17 located a dimming region next to the coronal hole as was the case for Event #12.

## 5.4 Discussion

The goal of ISEST WG 4 was to integrate observations, theory, and simulations to understand the chain of cause-effect dynamics from the Sun to Earth for a few carefully selected (Campaign) events. This should help us develop and/or improve the prediction capability for the arrival of these transient and their potential impacts at Earth.

WG 4 also examined controversial events, such as stealth or "silent" CMEs and problem ICMEs to enhance our understanding. One focus of WG 4 was on why do forecasts fail and how can we improve our predictions. This included analyzing the complications in linking CMEs to ICMEs, usually observed only *in situ* at 1 AU. Our July 2012 (#1) and June 2015 (#6) cases were considered "textbook", but the forecasts were not fully accurate. The June 2015 case involved a compound event that likely enhanced the level to a severe storm. The next three cases, March 2015, #5, September 2014, #1, and October 2012, #2, were all considered problem events that we now understand. In March 2015 two CMEs possibly interacted near the Sun and were deflected by a CIR. In September 2014 the storm was much over-predicted because the shock sheath and MC fields were almost entirely northward ($+B_Z$). For October 2012 a CME was identified but the surface signatures were multiple and weak leading to uncertainties in the arrival time. Finally, the last case in May-June 2013 (#4) was a problem event that we still do not fully understand. No storm was forecast but a brief, strong storm occurred. The surface activities associated with a slow CME were unclear as was the cause of the southward field ($-B_Z$) at 1 AU.

As in several of our cases, we note that about 20% of important geomagnetic storms have identified ICMEs but no compelling solar signatures. Likewise, Earth-affecting CMEs are sometimes "stealthily" launched without clear LCSs.  In our stealth CME study, we demonstrated the need to compare AIA images with long temporal separations to find weak LCSs, especially coronal dimmings and PEAs, in stealthy eruptions or slow CMEs. In addition, STEREO COR data provided the time range to examine AIA data that matches CME formation and acceleration. We found a tendency for the CME source regions to be located near coronal holes, or open field regions.

Finally, about 10% of intense storms are due the compression of fields and plasma by CIRs and their HSSs (J. Zhang et al. 2007). CIRs played a role in two of our cases. The shock sheath region can also be very important for driving storms as was the case in at least two of our events. A problem is that the sheath fields consist of swept-up coronal and heliospheric material which are hard to predict in advance. Thus, studies of sheath regions are an important, but not poorly understood aspect of space weather forecasting.

# 6. Solar Energetic Particle (SEP) Events

## 6.1 Introduction

Solar Energetic Particles (SEPs) from suprathermal (few keV) up to relativistic (few GeV) energies constitute an important contributor to the characterization of the space environment. They are emitted from the Sun in association with solar flares and Coronal Mass Ejection (CME)-driven shock waves. SEP radiation storms may have durations from a period of hours to days or even weeks and have a large range of energy spectrum profiles. These events pose a threat to modern technology strongly relying on spacecraft and are a serious radiation hazard to humans in space, and additionally of concern for avionics and commercial aviation in extreme circumstances (O. E. Malandraki and Crosby 2018a; 2018b). This section is divided into subsections, devoted to the progress of SEP research from both the observational as well as the theoretical and modeling perspective.

## 6.2 State of SEP Observations and Theory until 2014

By the end of the 1990s, a two-class paradigm (see Figure 6-1) for SEP events was generally accepted (Reames 1999a; 2013; M. Desai and Giacalone 2016a). In this paradigm, the gradual events occurred as a result of diffusive acceleration at CME-driven coronal and interplanetary (IP) shocks, while the impulsive events were attributed to acceleration during magnetic reconnection in solar flares. The gradual or CME-related events typically lasted several days and had larger fluences, while the impulsive or flare-related events lasted a few hours and had smaller fluences. Impulsive events were typically observed when the observer was magnetically connected to the flare site, while ions accelerated at the expanding large-scale CME-driven shocks can populate magnetic field lines over a significantly broad range of longitudes (H. Cane, Reames, and Von Rosenvinge 1988). The distinction between impulsive and gradual SEP events was further justified on the basis of the energetic particle composition and radio observations (H. Cane, McGuire, and Von Rosenvinge 1986). For instance, the flare-related impulsive SEP events were electron-rich and associated with type III radio bursts. These events also had $^3$He/$^4$He ratios enhanced between factors of $\sim 10^3$–$10^4$, Fe/O ratios enhanced by up to a factor of 10 over the corresponding SW values and had Fe with ionization states up to $\sim 20$. In contrast, the gradual events were proton-rich, had average Fe/O ratios of $\sim 0.1$ with Fe ionization states of $\sim 14$, had no measurable enhancements in the $^3$He/$^4$He ratio, and were associated with type II bursts (Reames 1999a; Edward W Cliver 2000).

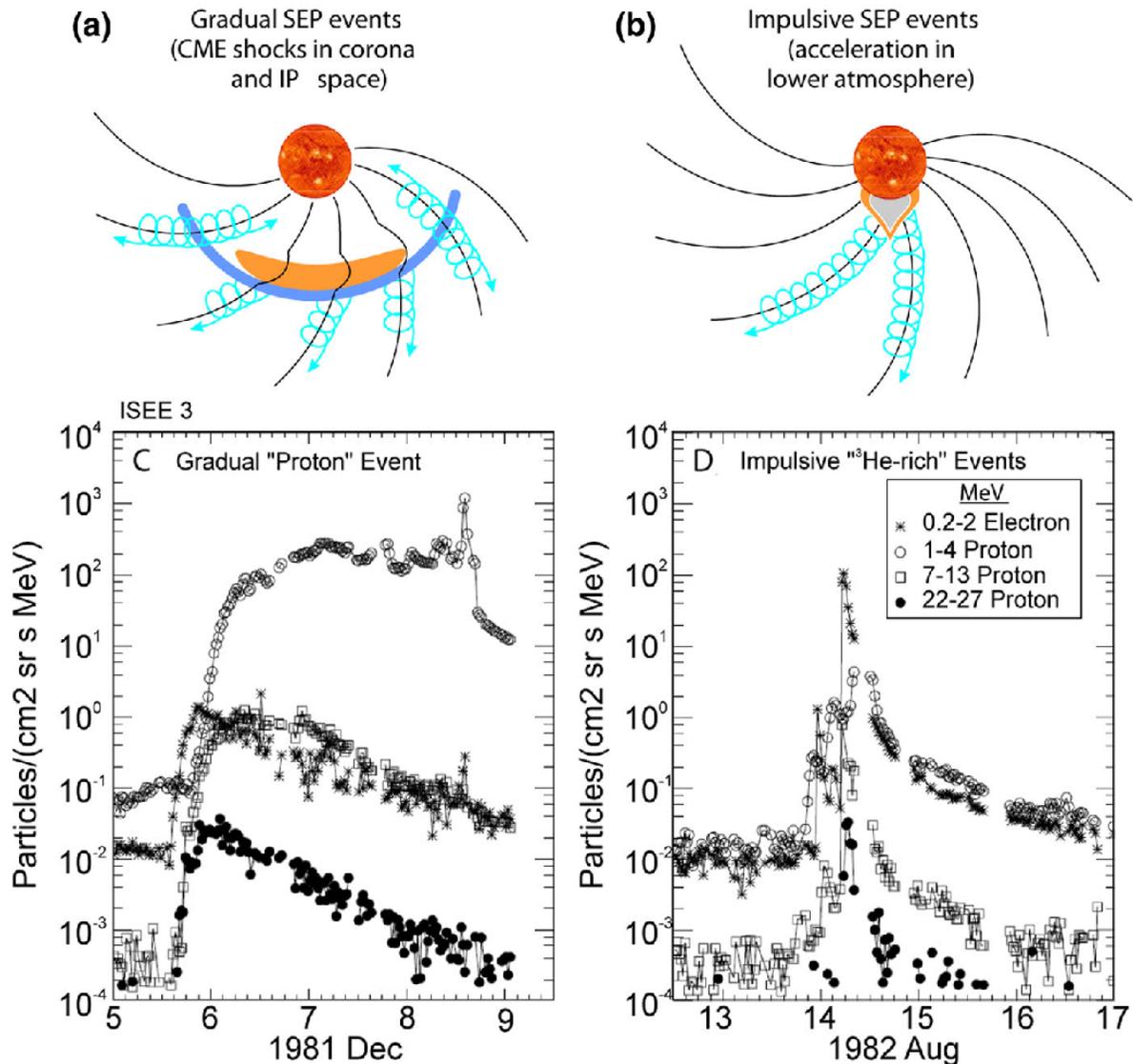

*Figure 6-1. The two-class paradigm for SEP events in which a) gradual SEP events are produced as a result of diffusive acceleration by large-scale CME-driven coronal and interplanetary (IP) shock waves. The accelerated SEPs populate Interplanetary Magnetic Field (IMF) lines over a wide range of longitudes. b) Impulsive SEP events are attributed to acceleration during magnetic reconnection in solar flares and observed when the observer is magnetically connected to the flare site. Intensity-time profiles of electrons and protons in c) gradual and d) impulsive SEP events. (Reproduced from Desai and Giacalone, 2016, after Reames 1999)*

Based on these early measurements, most researchers accepted the notion that CME-shock associated large gradual SEP and ESP events result from the diffusive shock acceleration (DSA) of thermal solar wind (M. A. Lee 1983; L. Tan et al. 1989), although others pointed out that the suprathermal tail of the solar wind may be the source (Gosling et al. 1981; B. Tsurutani and Lin 1985; L. Tan et al. 1989). Indeed, based on the correlation between the particle intensities, abundances, and energy spectra during ESP events and pre-event ion populations, Tsurutani and Lin (1985) and Tan et al. (1989) had suggested that the concomitant solar flares might provide the suprathermal seed particles accelerated at the IP shocks.

Since the mid-1990s, instruments with greater sensitivity and resolution on board Wind (C. Russell et al. 1995) and ACE (Stone et al. 1998) have provided major observational advances in terms of measuring the solar wind ion composition and its variations (Von Steiger et al. 2000) and comparing them with the energy-dependence and event-to-event variability of the ionic charge state, and elemental and isotopic composition in ESP and SEP events over a broad energy range
(Oetliker et al. 1997; J. Mazur et al. 1999; Möbius et al. 1999; Cohen et al. 2005; Mihir I Desai et al. 2006; Klecker, Möbius, and Popecki 2007). These new observations have made it possible to re-examine questions about the origin of the seed populations, and improve understanding of how SEPs are accelerated and transported to 1 AU. The following subsections highlight major advances and insights into the origin, acceleration and propagation of SEPs that have resulted from two decades of research. In particular, we present the state of knowledge of SEP studies by the end of the 2014 and discuss open questions that are yet to be fully resolved.

## 6.2.1 SEP origin

Observations of extremely rare elements and rare tracer ions like impulsive SEP-associated $^3$He and interstellar pickup He+ ions in SEP and ESP events have provided compelling evidence that CME-driven shocks accelerate material preferentially out of a suprathermal "seed" population that comprises contributions from the heated solar wind, coronal material, and remnants of solar transient events (G. Mason, Mazur, and Dwyer 1999; Gloeckler 2003; Mihir I Desai et al. 2006; G. M. Mason et al. 2004; Allegrini et al. 2008; Dayeh et al. 2009). Other studies have shown that the abundances of heavy ions accelerated in SEP and ESP events are not well organized by any physical quantity such as the ion's Q/M ratio or its First Ionization Potential (FIP) when compared with the corresponding SW abundances (R. Mewaldt et al. 2002; Mihir I Desai et al. 2003; 2006; SW Kahler, Tylka, and Reames 2009). Further, Mewaldt et al. (2012) found that the suprathermal Fe densities at 1 AU are generally significantly greater one day before the occurrence of these large SEP events compared to all other days, perhaps indicating that the presence of high-density suprathermal Fe is necessary for SEP events with large Fe fluences to occur. Finally, Desai et al. (2003) found that the IP shock abundances were well correlated with the average abundances measured at the same energy (~1 MeV/nucleon) in the interplanetary medium prior to the arrival of the IP shocks. In particular, elements with higher M/Q ratios are systematically depleted, which is consistent with shock acceleration models wherein ions with higher M/Q ratios are accelerated less efficiently than those with lower M/Q values (M. A. Lee 2005a). Collectively, these results indicate that the material accelerated in large SEP events is quite distinct from that measured in the solar wind. Therefore, the SEP heavy ions are unlikely to originate from the bulk solar wind, but rather from a suprathermal tail that comprises ions from multiple sources, including $^3$He and Fe-enriched material accelerated in flares and suprathermal material accelerated at previous CME shocks (G. Mason, Mazur, and Dwyer 1999; Mihir I Desai et al. 2006; R. A. Mewaldt, Looper, et al. 2012; R. A. Mewaldt, Mason, and Cohen 2012).

## 6.2.2 SEP acceleration

DSA comprises two main mechanisms, namely, shock-drift mechanism at quasi-perpendicular shocks (Decker 1981), and first-order Fermi mechanism at quasi-parallel shocks(M. A. Lee 1983). DSA theory successfully predicts some SEP observations, but fails to prevail as a universal theory in explaining most SEP events, partly because of external drivers that simultaneously affect the observed properties at 1 AU, including species-dependent escape from the IP shock, ambient turbulence, and shock finite size and geometry. For instance, SEP studies have shown that the differential energy spectra of H-Fe nuclei in large SEP events exhibit a distinct form of a broken (i.e., double) power-law (hereafter PL) with a characteristic break-energy (hereafter $E_0$) (A. Tylka et al. 2005). In contrast, DSA theory predicts a single power-law. The location of $E_0$ was found to typically decrease for the heavier ion species as a power-law function of ion's charge-to-mass ratio (Zank, Rice, and Wu 2000; A. J. Tylka, Dietrich, and Atwell 2010; R. A. Mewaldt, Looper, et al. 2012). This systematic Q/M dependence occurs because the energy spectra roll-over or break at the same value of the diffusion coefficient for different species, which depends on ion rigidity or the Q/M ratio (A. J. Tylka et al. 2000; Cohen et al. 2005; R. Mewaldt et al. 2005). Li et al. (2009) generalized an SEP acceleration model by including varying levels of turbulence near shocks of different obliquity and predicted that α could range between ~0.2 for weaker scattering near quasi-perpendicular shocks and ~2 for stronger scattering near quasi-parallel shocks. Finally, Alfvén waves generated by energetic protons streaming upstream of ICME shocks could trap particles locally near the shock (Lario et al. 2005). This indicates that particle scattering and trapping near the shock, in some cases, could be dominated by a dynamic wave spectrum rather than a more universal background Kolmogorov-like wave spectrum (A. Tylka et al. 2005; C. Ng, Reames, and Tylka 2003a).

## 6.2.3 SEP Transport

Effects of interplanetary transport on the temporal evolution of the heavy ion abundances and spectra are also believed to play a critical role in determining SEP observations at 1 AU. Tylka, Reames, and Ng (1999) and Ng, Reames, and Tylka (1999) modeled the energy spectra and systematic temporal evolution of the elemental abundances of ~5–10 MeV/nucleon He, C, O, Ne, Si and Fe ions in two large SEP events in terms of rigidity-dependent trapping and scattering by Alfvén waves generated by streaming energetic protons accelerated at CME-driven shocks. These observations were successfully modeled using self-consistent numerical calculations of wave generation or amplification by shock-accelerated protons escaping or streaming away from the near-Sun CME shock (M. A. Lee 2005b; C. Ng, Reames, and Tylka 2003b; C. K. Ng, Reames, and Tylka 2012). Such self-excited Alfvén waves can scatter and trap particles near the shock and increase its acceleration efficiency. This, in turn, throttles the proton intensities near ~few MeV/nucleon resulting in energy-dependent upper bounds or plateaus, known as streaming limits (Reames 1990). Independently, Mason et al. (2006) pointed out that the dramatic variations in the temporal behavior of Fe/O ratio at all energies between ~0.1 and 60 MeV/nucleon vanish in >70% of the prompt western hemisphere SEP events if the Fe intensities are compared to O intensities at ~twice the Fe kinetic energy-per-nucleon. To explore the physical process involved, Mason et al. (2012) modeled the rise phases in large SEP events and showed that the temporal evolution of Fe/O can be reasonably fitted by a state-of-the-art model where the differences in the transport of Fe versus O are due to the slope of the turbulence spectrum of the IMF. Another effect of turbulence and waves which scatter SEPs can be a significant amount of transport perpendicular to the average magnetic field leading to wider angular particle spreads than the corresponding extent of their acceleration region (Nina Dresing et al. 2012; W Dröge et al. 2014; Wolfgang Dröge et al. 2016). However, the very widespread SEP events observed with the STEREO mission and close to Earth spacecraft (R Gómez-Herrero et al. 2015; Lario et al. 2014; 2016) challenged state-of-the-art transport models based on the longitudinal distribution of electron anisotropies, Dresing et al. (2014) suggested that there exist different types of widespread events, on one hand related to efficient perpendicular diffusion in the IP medium, and on the other hand caused by an extended injection region close to the Sun.

In summary, observations from ACE, Wind, and STEREO during the 1995-2014 epoch have shown that large gradual SEP events are governed by a confluence of multiple processes and effects by the time they are observed at 1 AU. These include: (1) origin and variability of the suprathermal seed populations (G. Mason, Mazur, and Dwyer 1999; G. Mason et al. 2005; Mihir I Desai et al. 2003; 2004; 2006; R. A. Mewaldt, Mason, et al. 2012); (2) the efficiency with which populations from different sources and with distinct distribution functions are injected into the shock acceleration mechanisms; (3) factors that control the efficiency with which particles are accelerated (e.g., CME speed and kinetic energy, shock strength and obliquity (SW Kahler 2001; SW Kahler and Vourlidas 2013; A. Tylka et al. 2005; R. Mewaldt et al. 2008); (4) the presence or absence of multiple, interacting CMEs (N Gopalswamy et al. 2004; G Li et al. 2012) (5) the type, level, and characteristics of the waves and turbulence present near the shock and in the interplanetary medium (A. J. Tylka, Reames, and Ng 1999b; A. Tylka et al. 2005; C. K. Ng, Reames, and Tylka 1999b; Cohen et al. 2003; H. J. Li et al. 2009) and (6) the charge-to-mass (Q/M)-dependence of scattering and transport through the turbulent interplanetary medium (G. M. Mason et al. 2006b; A. J. Tylka et al. 2013). The relative roles of these effects continue to be topics of hot scientific debates, and unravelling their influence remains a major focus of SEP research. In addition, several studies claim a direct flare acceleration of the high-energy (>25 MeV) proton component in large SEP events to augment that produced by coronal/interplanetary shock waves driven by CMEs (Cane and Richardson 2003; Cane et al. 2006; Klein and Posner 2005; Aschwanden 2012).

In the following sections (6.3 and 6.4), we will highlight the progress made during the 2014-2019 timeframe and point out the key areas in which critical observations in the inner heliosphere from Parker Solar Probe and Solar Orbiter will advance our understanding of the physics of SEP events. These efforts are essential for developing models that can reliably forecast and mitigate radiation risks from extreme SEPs and are essential for deep space exploration.

## 6.3. Progress in SEP observations during the VarSITI era (2015 - 2019)

### 6.3.1 SEP origin

Among the more unusual solar phenomena are the long-duration gamma-ray flares (LDGRFs). The prime characteristic of these events is delayed and prolonged γ-ray (>100 MeV) emission after the impulsive phase (Ryan 2000). Recently, the *Fermi* Large Area Telescope (LAT) observed dozens of LDGRFs, with the most intense and longest-duration example the 2012 March 7 event, for which >100 MeV emission was observed for nearly 20 hr (Ajello et al. 2014). Share et al. (2018) characterized and catalogued 30 solar eruptive events observed by Fermi/LAT from 2008-2016, referring to this emission as 'late-phase gamma-ray emission' (LPGRE). These authors produced and presented 'light-bucket' time profiles for all the events, obtaining an estimate of the >100 MeV γ-ray flux from within about 10° of the Sun. Figure 6-2 presents an example of one of these time histories for the 2011 March 7 event, in which the LPGRE lasted ~14 hr. GBM and RHESSI observed impulsive hard X-rays up to only 100-300 keV with no evidence for nuclear-line emission. The >100 MeV fluxes plotted in the inset reveal that the LPGRE started within minutes of the hard X-ray peak. It is clear from the rising LPGRE flux that it is due to a distinct particle acceleration phase and is not just the tail of emission from the impulsive phase of the flare.

In all the events studied, Share et al. (2018) found that the LPGRE is temporally and spectrally distinct from the impulsive phase emission. The spectra are consistent with the decay of pions produced by >300 MeV protons and are not consistent with primary electron bremsstrahlung with synchrotron losses. All but two of the LPGRE events were accompanied by a fast and broad CME. The LPGRE start times range from CME onset to 2 hr later whereas their durations range from ~0.1 to 20 hr and appear to be correlated with the durations of the accompanying >100 MeV SEP proton events. Comparison showed that the number of >500 MeV protons producing the LPGRE is at least a factor of 10 larger than the number producing the impulsive phase >100 MeV γ-ray emission during the associated flare. The number of >500 MeV protons needed to produce the LPGRE ranges in nine events from 0.1% to 50% of the number of protons observed in the accompanying SEP event in interplanetary (IP) space (based on the observations by the GOES/HEPAD experiment and neutron monitors). There are significant systematic uncertainties in the SEP estimates, however.

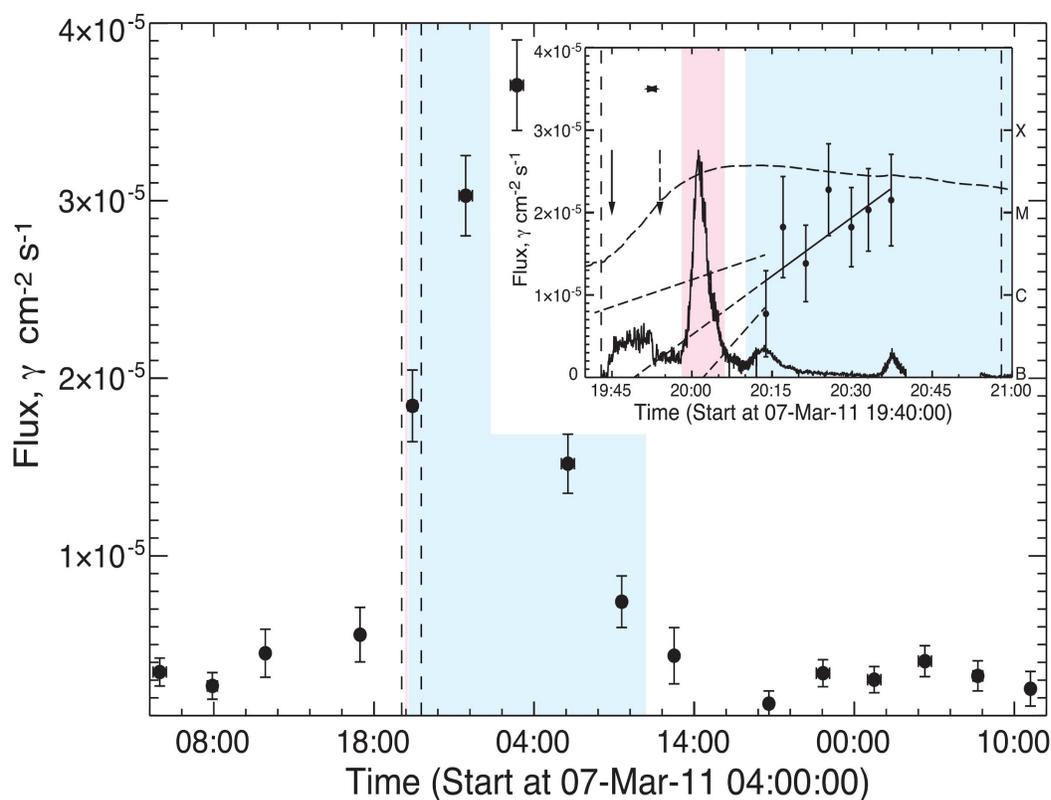

*Figure 6-2. Main plot: Time history of the >100 MeV light-bucket fluxes from <10º of the Sun, revealing LPGRE from the 2011 March 7 solar eruptive event. Vertical dashed lines show the GOES 1-8 Å start and end times. Inset: 4-min*

One of the features of the LPGRE events <u>Share et al. (2018)</u> found is their association with fast CMEs and SEP production, which points to acceleration of particles by the shock produced by a fast CME as a clear candidate for the energy source for the >300 MeV protons that produces LPGRE. Furthermore, (Nat Gopalswamy, Mäkelä, et al. 2018) presented strong quantitative evidence that interplanetary type II radio bursts and such sustained gamma-ray emission (SGRE) events from the Sun are closely related. Out of the 30 SGRE events reported by (Share et al. 2018), they considered 13 events that had a duration exceeding ~5 hr, thus excluding any flare-impulsive phase gamma-rays. These authors found that the SGRE duration has a linear relation with the ending frequency of the bursts (Figure 6-3). The synchronism also found between the ending times of SGRE and the type II emission (Figure 6-3) strongly supports the idea that the same shock accelerates electrons to produce type II bursts and protons (>300 MeV) that propagate from the shock to the solar surface to produce SGRE via pion decay. A CME-shock origin could also explain the wide range of delays observed in LPGRE onset times: short LPGRE onset delays represent shock acceleration low in the corona (N Gopalswamy et al. 2013), while long LPGRE onset delays indicate that the CME had to expand over several solar radii before accelerating >300 MeV protons that could return to the Sun. The smooth time histories of the long duration LPGRE events can be explained by precipitation of particles that are magnetically trapped in a reservoir (Reames 2013) behind the expanding CME.

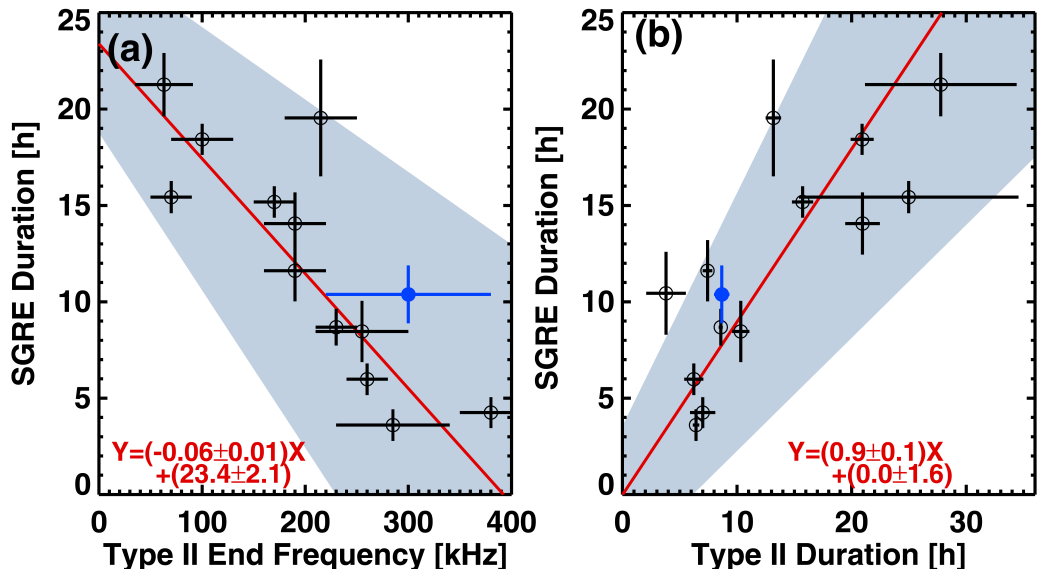

*Figure 6-3. Scatter plots of SGRE duration with type II ending frequency (a) and type II duration (b). The best-fit lines (red) are obtained using the Orthogonal Distance Regression method, which considers errors in both X and Y variables. The shaded area represents 95% confidence interval of the fit (after **Gopalswamy et al. (2018)**)*

Quantitative estimates, however, have indicated that only a very small fraction of accelerated protons can return to interact in the chromosphere (Hugh S. Hudson 2018; Karl-Ludwig Klein et al. 2018) due to the transport of the protons back to the Sun against the magnetic mirror force, well after the flare when the CME is many solar radii above the surface with magnetic field strengths much lower than at the solar surface. <u>Share et al. (2018)</u> go around this difficulty by assuming that there is significant MHD turbulence on the field lines connecting the CME to the Sun, such as required in the model of (Ryan and Lee 1991). Furthermore, Kocharov et al. (2015) presented a shock-wave model and their estimated ratio of the number of protons that return to the Sun and interact to the number that escape into IP space depending on the amount of turbulence is consistent with the range estimated by Share et al. (2018) in their comparison of the number of >500 MeV protons producing LPGRE and those detected as SEPs in space. A variant on the magnetic trap model was proposed by Hudson (2018) and called the 'lasso' model in which the SEP particle accelerator crossed both open (SEP) and closed (LPGRE) field lines, leaving energetic particles on both.

de Nolfo et al. (2019) compared the total number of >500 MeV protons at 1 AU by combining Payload for Matter-Antimatter Exploration and Light Nuclei Astrophysics (PAMELA) and STEREO spacecraft data with the number of high-energy protons at the Sun as deduced from Fermi/LAT (Share et al. 2018). Their analysis showed that the two proton numbers are uncorrelated such that their ratio spans more than 5 orders of magnitude, suggesting that the back precipitation of particles accelerated at CME-driven shocks is unlikely to be the source of the LPGRE emission. They discussed an alternative explanation for LPGREs based on continuous particle acceleration and trapping within large coronal structures that are not causally connected to the CME shock, within the context of new remote observations of these loops available.

## 6.3.2 SEP acceleration

A lively debate has continued in recent years on the question of the principal source of high-energy protons in large SEP events. Some studies have provided new support for a significant contributory or dominant role for flare acceleration of high-energy protons in gradual SEP events, contrary to the generally accepted scenario favoring shock acceleration. This new evidence is mainly based on correlations between the sizes of X-ray and/or microwave bursts and the SEP fluence at different energies, e.g., for 15–40 MeV protons (Trottet et al. 2015), >30 MeV protons (Le and Zhang 2017), >50 MeV protons (Dierckxsens et al. 2015), and >100 MeV protons (Grechnev et al. 2015; Le, Li, and Zhang 2017). In order to assess the above correlations, the technique of partial correlation coefficients (see Trottet et al. (2015) for a detailed description) has been used along with the classical Pearson correlation coefficient, to remove the correlation effects between the solar parameters themselves. In particular, Grechnev et al. (2015) addressed the relation between the >100 MeV proton fluences measured by the Geostationary operational Environmental Satellite (GOES) monitors and the associated flare microwave fluences at 35 GHz, recorded by the Nobeyama Radio Polarimeters, over the 1996–2014 time interval. Grechnev et al. (2015) found a partial correlation coefficient of 0.67, versus 0.001 for a corresponding comparison of the >100 MeV proton fluence and the CME speed, concluding that these SEP events originated in the associated flares. These results were criticized by Cliver (2016), according to which the exclusion in Grechnev et al. (2015) of four outlying "abundant proton" events (black squares in the orange rectangle in Figure 6-4) is not justified if one considers the associated CME speeds and widths, and electron-to-proton ratios which are comparable to those in the main sequence (black circles in Figure 6-4). On the contrary, inclusion of such events in the analysis reverses the conclusion in favor of shock acceleration for the >100 MeV protons. Nevertheless, we point out these results are based on the assumptions that non-DH -associated SEPs are flare generated and those behind the limb are shock generated, which although reasonable are not ultimately proven.

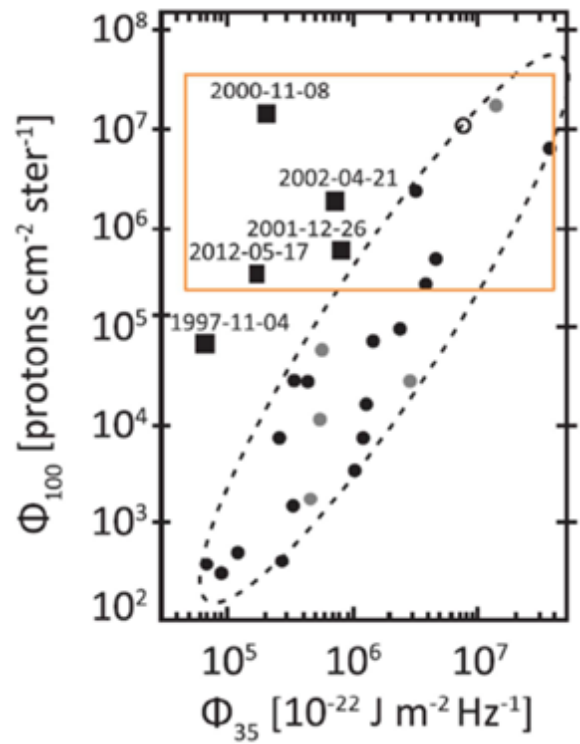

*Figure 6-4. Scatterplot of longitude-corrected >100 MeV proton fluence (Φ100) vs. 35 GHz fluence (Φ35) for solar proton events from 1996 to 2014; black circles and squares (W21–W90); gray circles (E30–W20); open circles (<E30). The orange rectangle isolates events with Φ100 > 2x10⁵ pfu (from Cliver et al., 2016).*

Several other recent studies support the prevailing shock picture for gradual SEP events, such as the observation of a prompt SEP event at widespread locations in conjunction with the longitudinal propagation of a white-light shock (Lario et al. 2016), the SEP source temperatures(Reames 2015) and the hierarchical relationship found between the fluence spectra of gradual SEP events and the kinematics of the CMEs (N Gopalswamy et al. 2017). Gopalswamy et al. (2017) analyzed the SEP fluence spectra of three classes of SEP events (Filament eruption (FE) SEP events, well-connected regular ones and SEP/GLEs) over cycle 23 and 24. They found that: FE SEP events have the softest spectra and lowest initial acceleration; SEP/GLE events have the hardest spectra and the largest initial acceleration; the regular SEP events have intermediate spectral indices and acceleration. It has to be noted that the computed spectral indices by considering a simple power law, without taking into account possible rollovers and breaks, thus mainly representing the spectrum behavior at lower energies. The hierarchical relationship was shown to be present (N Gopalswamy et al. 2017) also in terms of the average starting frequencies of the associated type II bursts and the shock formation height (as obtained by matching the onset time of the type II bursts with the CME leading edge height-time history). Such behavior could be explained by considering that the rapid acceleration of CMEs leads to very high initial speeds and hence a shock formation close to the Sun, where the ambient magnetic field and density are high for an efficient particle acceleration, resulting in harder spectra for GLE events.

Understanding the origin of SEP/GLE events is especially challenging, as particles are accelerated up to GeV energies and can reach the Earth's atmosphere in about 10 minutes, while the various acceleration mechanisms may be expected to exhibit different characteristic timescales. The comparison between the solar particle release (SPR) time and the onset time of the SEP event associated solar phenomena is generally used to pose temporal constraints that any putative acceleration process must meet. Nevertheless, the different methodologies used to perform such studies and underlying assumptions, e.g., the constant SPR at all energies and the particle's scatter-free propagation in the interplanetary space, e.g., questioned by Wang and Qin (2015), can lead to contrasting results about the principal source of particle acceleration. The most recent SEP/GLE event, which occurred on 2017 September 10, has been attributed to both flare and shock acceleration, e.g., by Zhao, Le, and Chi (2018) and Gopalswamy et al. (2018), respectively, due to their different evaluations of the SPR and the Type II bursts onset time. In particular, Gopalswamy et al. (2018) confirmed the parabolic relationship between the eruption longitude and the CME height at SPR, which is considered to be the key to the understanding of particle acceleration by shocks and the magnetic connectivity to the observer. Those authors also showed that the 2017 September 10 GLE did not have an unusually hard spectrum (Schwadron et al. 2018), but a softer-than-average spectrum for a GLE event, having the 10-100 MeV fluence spectral index of 3.17 with respect to the average one of 2.68 for SEP/GLEs (Nat Gopalswamy et al. 2016). They suggested this to be due to the poor longitudinal and latitudinal connectivity of the shock nose with the Earth, possibly compounded by the weak ambient magnetic field reducing the shock acceleration efficiency. Similarly, it is suggested that some gradual SEP events with high initial speeds did not produce a GLE, as only the shock flanks were magnetically connected with the Earth (Gopalswamy et al. 2018). Nevertheless, this has to be reconciled with the hypothesis that high energy protons are accelerated preferentially by quasi-perpendicular shocks which could be located at the shock flanks (Schwadron et al. 2015; Kong et al. 2019), as will be discussed in sub-section 6.3.2.

## 6.3.3 SEP transport

The unprecedented orbits of the two-spacecraft STEREO mission provided well-separated observations at 1 AU and allowed to study the longitudinal distribution of SEPs and especially events with extraordinarily wide particle spreads in great detail (Lario et al. 2014; 2016; R Gómez-Herrero et al. 2015; Nina Dresing et al. 2018). Based on Gaussian functions applied to multi-spacecraft events, Lario et al. (2013) and Richardson et al. (2014) determined their mean widths to be between 36° (27-37 MeV protons) and 49° (71-112 keV electrons). However, a large event to event variation was observed which also limited the determination of the average displacement of the Gaussian center with respect to the longitude of the associated flare site. However, the displacements towards the west of the flare site found in the above studies may be caused by the associated CME-driven shocks that would shift the Gaussian distributions towards the central meridian as viewed from the spacecraft.

In a similar manner, Cohen, Mason, and Mewaldt (2017) have systematically investigated the energy and Q/M dependence of the longitudinal distributions for large ion events using STEREO and close-to-Earth spacecraft reporting comparable values for widths and mean Gaussian-center displacements. While the widths were found to show an energy dependence with distributions narrowing with increasing energy, no Q/M dependence to the widths of Gaussian centers were found for the 41 ion event distributions studied (see Figure 6-5). This suggested that lower energy ions might experience more field line co-rotation, or are accelerated over a larger portion of the CME-driven shock or for longer times as the shock expands. Rigidity-related processes seemed, however, not to be important in terms of longitudinal spreading of the particles.

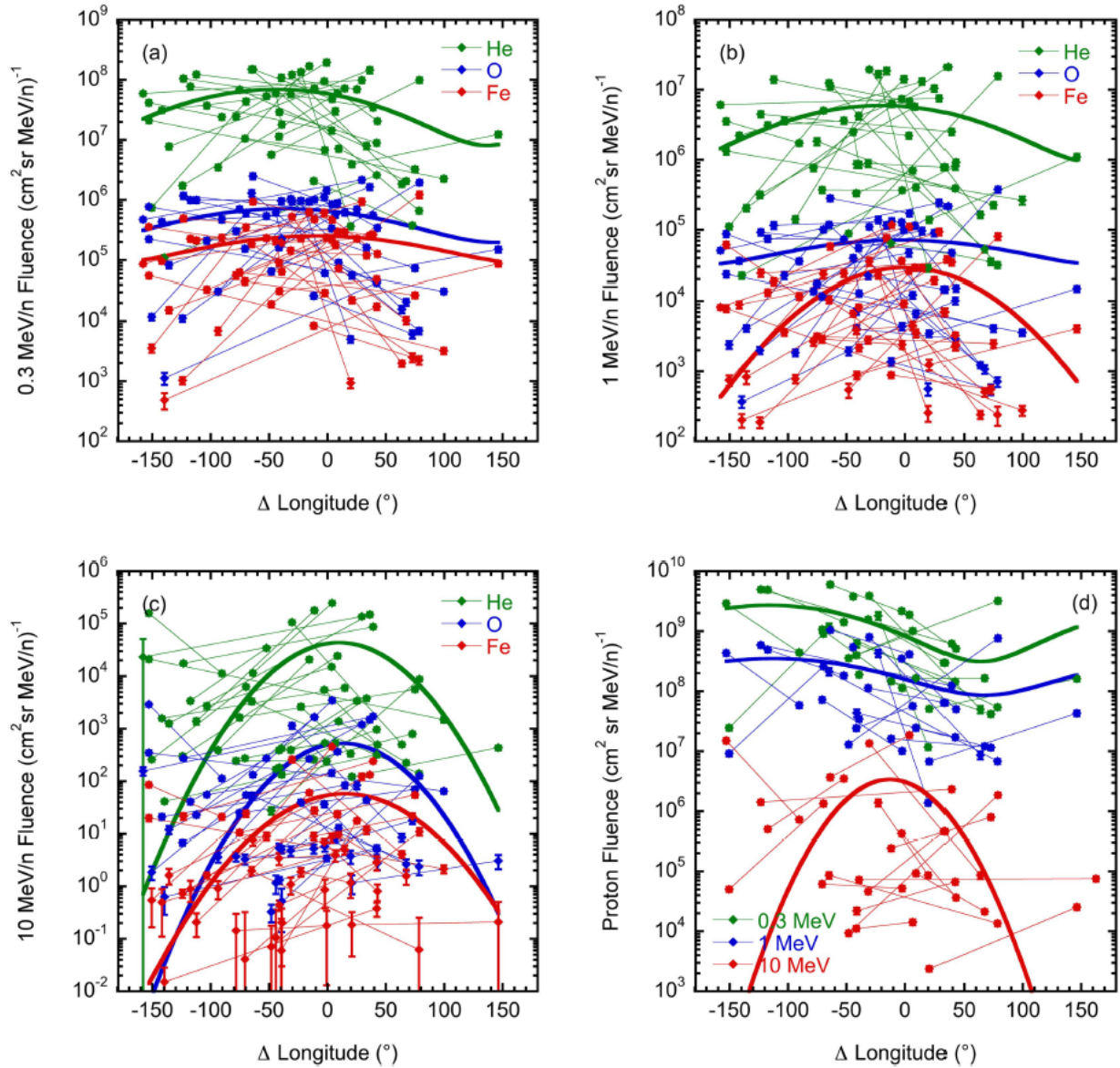

*Figure 6-5. Peak intensities of He, O, and Fe as a function of flare-spacecraft footpoint separations (Δ Longitude) of two-spacecraft events (Cohen et al. 2017). Negative values of Δ Longitude correspond to footpoints west of the flare location, and positive values correspond to locations east of the flare. The different panels show different energies with (a) 0.3, (b) 1, and (c) 10 MeV/n. Panel (d) shows protons at all three energies. Individual events are connected by lines, the thick curves show periodic Gaussian fits.*

The STEREO mission has also enabled the identification and study of some extreme cases of widespread events with distributions up to 360° around the Sun (Nina Dresing et al. 2012; N Dresing et al. 2014; Lario et al. 2014; 2016; R Gómez-Herrero et al. 2015). However, even when the application of interplanetary transport models (see Section 6.3.2) suggests the presence of strong perpendicular transport (Dröge et al. 2014; Dröge et al. 2016; Strauss, Dresing, and Engelbrecht 2017), it seems that all widespread events need at least a somewhat extended injection region of e.g., 25° width (Strauss, Dresing, and Engelbrecht 2017). The nature of this extended injection region may be an extended shock which accelerates and injects the particles over a wide longitudinal range. Modern reconstruction techniques of the coronal and interplanetary shock using EUV and white-light data (see section 2.2) have shown a good agreement between the inferred SEP injection times and the times of the expanding shock intersecting the magnetic footpoint of the spacecraft (A Kouloumvakos et al. 2016; Lario et al. 2016). For well-connected observers, similar correlations were found also for the time when the associated EUV wave intersects the spacecraft magnetic footpoint in the low corona and the role of these waves in terms of particle acceleration and coronal transport were discussed controversially (J. Park et al. 2013; Lario et al. 2014; Miteva et al. 2014). In the case of the farthest separated spacecraft in extreme widespread events, however, the event can usually not solely be explained by the established magnetic connection to the shock or an EUV wave but transport effects are likely to play a role in spreading the SEPs to the farthest observers (Lario et al. 2014; A Kouloumvakos et al. 2016). While shocks are accepted to be the main source of large and gradual solar ion events, their role for efficient electron acceleration is still under debate (Dresing et al. 2020 and references therein). Alternative scenarios providing wide injection regions involved in widespread electron events (N Dresing et al. 2014) are the presence of fan-shaped magnetic field lines (K-L Klein et al. 2008), but non-uniform or non-radial spatial injections (Klassen et al. 2016; 2018) may play a role as well.

The 26 Dec 2013 widespread electron event (Dresing et al. 2018) suggested to be caused by a shock, at a first glance, because of its wide SEP spread and very long-lasting proton and electron anisotropies pointing to a time-extended injection different to a flare. However, other features of the event, like an additional high-energy SEP component, arriving four hours later than the first one, could not be explained solely by the presence of a shock and these authors suggested a trapping scenario to be involved in forming the characteristics of the event. Additionally, to accelerating SEPs, CMEs can also play an important role for SEP transport when they have propagated into interplanetary space. If a previous CME, which is still magnetically anchored at the Sun, is convected over the observer just at the time when another eruptive event occurs at the Sun, new SEPs may be injected into this ICME and propagate through it. The magnetic connection of an observer inside this structure to the Sun is dramatically changed and one loop leg may even provide otherwise unlikely magnetic connections to eastern longitudes so that SEPs from these source regions can be observed even if the source extent is small (I. Richardson, Cane, and Von Rosenvinge 1991). First arriving SEPs may then also arrive in anti-sunward pointing telescopes (Raúl Gómez-Herrero et al. 2017).

Measuring an SEP event inside an ICME confirms not only that the structure is still magnetically anchored at the Sun (O. Malandraki et al. 2002; 2005) but also that solar energetic electron observations constitute a tool to probe the magnetic structure, e.g., the winding of the magnetic field inside the ICME (SW Kahler, Haggerty, and Richardson 2011; L. C. Tan et al. 2012). If combined with reconstruction of the early CME based on coronagraph observations, one can then determine the dimension of the large-scale structure, such as its loop length, at 1 AU, i.e. for times when it propagated far out of the fields of view of coronagraphs (Nina Dresing et al. 2016).

## 6.3.4 SEP compositional results

Measurements of relative abundances of heavy elements, their isotopic and ionic charge state composition in SEP events have been used in a wide variety of ways to infer critical information about the origins of the seed populations and the physical conditions under which these populations are produced, and the manner in which these seed particles are accelerated by CME shocks or in solar jets. In addition, we can infer the conditions that affect their transport through the solar corona, the interplanetary medium, and out into the heliosphere(Reames 2016; M. Desai and Giacalone 2016a). For instance, it is well established that elemental abundances in SEPs exhibit the so-called first ionization potential or FIP effect when compared with corresponding photospheric abundances, as shown in Figure 6-6. Based on well-documented differences between the average SEP abundances of elements such as C, S, and P that have intermediate FIP values (Mihir I Desai et al. 2006; Reames 2018) and those measured in the slow solar wind (Bochsler 2008), Reames (2018) suggested that these differences could be due to the action of ponderomotive forces of Alfvén waves on C, S, and P ions on open field lines and their corresponding absence on neutral C, S, and P atoms on closed field lines in the chromosphere (Laming 2015). According to this hypothesis, coronal material accelerated in SEPs differs from that found in the solar wind (see Figure 6-7), because the chromospheric plasma that later becomes SEPs enters the corona on closed field lines in active regions, while the corresponding plasma that later becomes the solar wind appears on open field lines, thus resulting in the observed differences between the C, S, and P abundances in SEPs and the SW (Reames 2018; 2020).

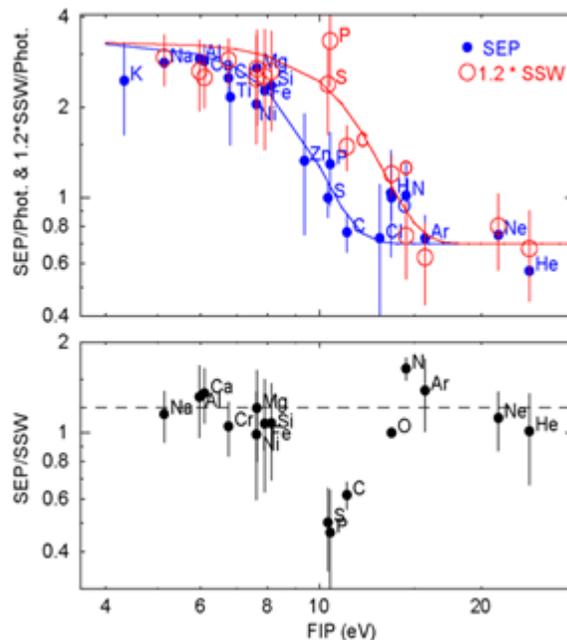

*Figure 6-6. The upper panel shows the SEP/photospheric and 1.2xslow solar wind (SSW)/photospheric abundance ratios as a function of FIP. Curves help show the trends of each data set. The lower panel shows the ratio of the "coronal" abundances from SEPs to those of the slow SW (Bochsler 2009), as a function of FIP. The dashed line suggests the preferred normalization factor of 1.2. Adopted from Reames 2018.*

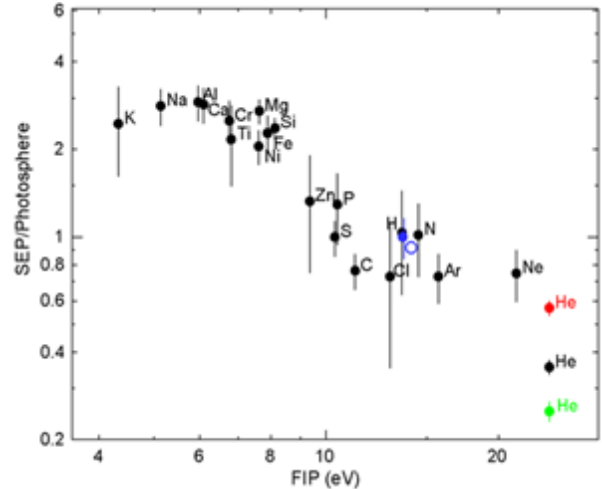

*Figure 6-7. : SEP abundances (from Reames 2014) divided by photospheric abundances (from Caffau et al. 2011, Lodders et al. 2009), normalized at O (blue). Abundances of He in three SEP events are also shown (taken from Reames, 2018).*

Numerous SEP studies have also shown that the differential energy spectra of H-Fe nuclei in large SEP events exhibit a distinct form of a broken (i.e., double) power-law (hereafter PL) with a characteristic break-energy (hereafter $E_o$). (M. I. Desai, Mason, Dayeh, Ebert, Mccomas, et al. 2016; M. I. Desai, Mason, Dayeh, Ebert, McComas, et al. 2016) surveyed the heavy ion spectra in 46 isolated, large gradual SEP events observed in solar cycles 23 and 24, and found that the Fe spectra had lower $E_o$ owing to the lower Q/M ratio or higher rigidity of Fe when compared with O. Figure 6-8, taken from (M. I. Desai, Mason, Dayeh, Ebert, McComas, et al. 2016) shows an example of this relation. This systematic Q/M dependence occurs because the energy spectra roll-over or break at the same value of the diffusion coefficient for different species, which depends on ion rigidity or the Q/M ratio (A. J. Tylka et al. 2000; Cohen et al. 2005; R. Mewaldt et al. 2005; M. I. Desai, Mason, Dayeh, Ebert, Mccomas, et al. 2016; M. I. Desai, Mason, Dayeh, Ebert, McComas, et al. 2016). The authors also found that α varies between ~0.2-3, where extreme SEP events associated with Ground Level Enhancements (GLE) often exceeded the upper limit of 2 (see Figure 6-8), as expected from theoretical predictions (Gang Li et al. 2009).

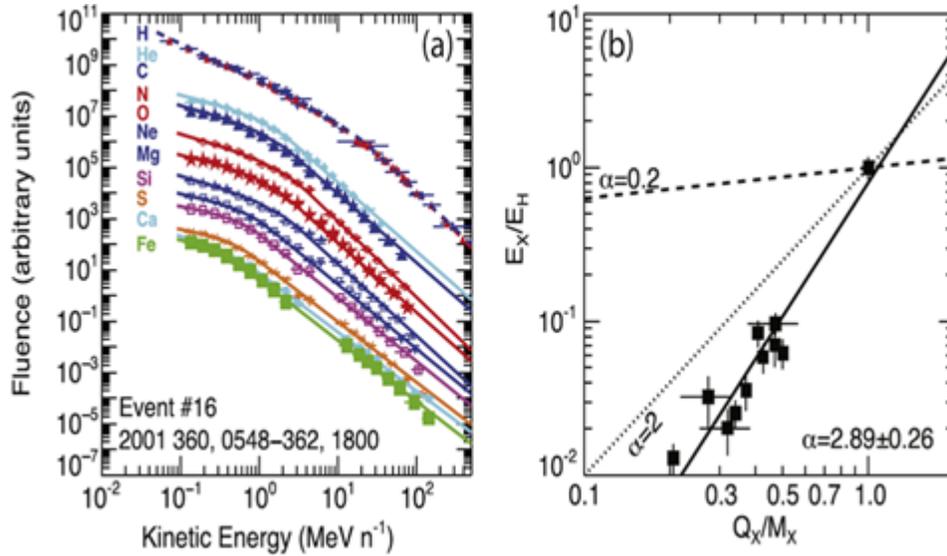

*Figure 6-8. (a) Event-integrated differential fluences vs. energy of ~0.1-500 MeV/nucleon for H-Fe nuclei during a large SEP event, taken from Desai et al. (2016b). The energy spectra for different species are offset for clarity. Solid lines are Band function fits to the spectra. (b) Spectral break energy $E_X$ of species X normalized to $E_H$ -- break energy of H vs. the ion's charge-to-mass (Q/M) ratio. Solid lines are fits to the data $E_X/E_H = no\ (Qx/Mx)^{\alpha}$ ; dashed line: same equation with $\alpha=2$ ; dotted line: same equation with $\alpha=0.2$.*

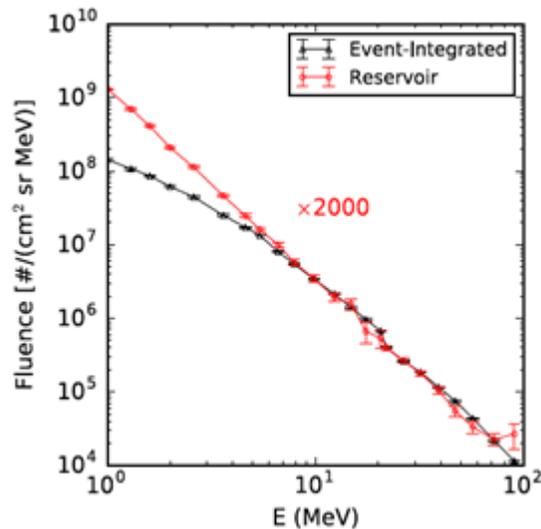

*Figure 6-9. : Event integrated and reservoir spectra during the 11/22/1977 SEP event. The spectrum shows a double power-law feature, while the reservoir spectrum exhibits a single power law (from Zhao et al. 2017). This effect was interpreted in terms of particle streaming and rigidity-dependent scattering effects that modify the spectrum at lower energies.*

Later, (L. Zhao, Zhang, and Rassoul 2016) performed a comprehensive numerical simulation study and found that the single power law spectrum near the Sun transitions into a double power law near 1 AU. The authors found that the spectral indices above and below $E_o$, along with the value of $E_o$, are related to the Kolmogorov-like interplanetary magnetic turbulence spectrum. (L. Zhao, Zhang, and Rassoul 2016) also studied the proton energy spectra in the decay phase (reservoir) of selected SEP events, where transport effects are expected to be minimal. They found that some events (see Figure 6-9) in which the event-integrated spectra were described by double power-laws transitioned into a single power-law during the reservoir or decay phase. This behavior was interpreted in terms of scattering, streaming, and diffusion effects for lower particles below $E_o$

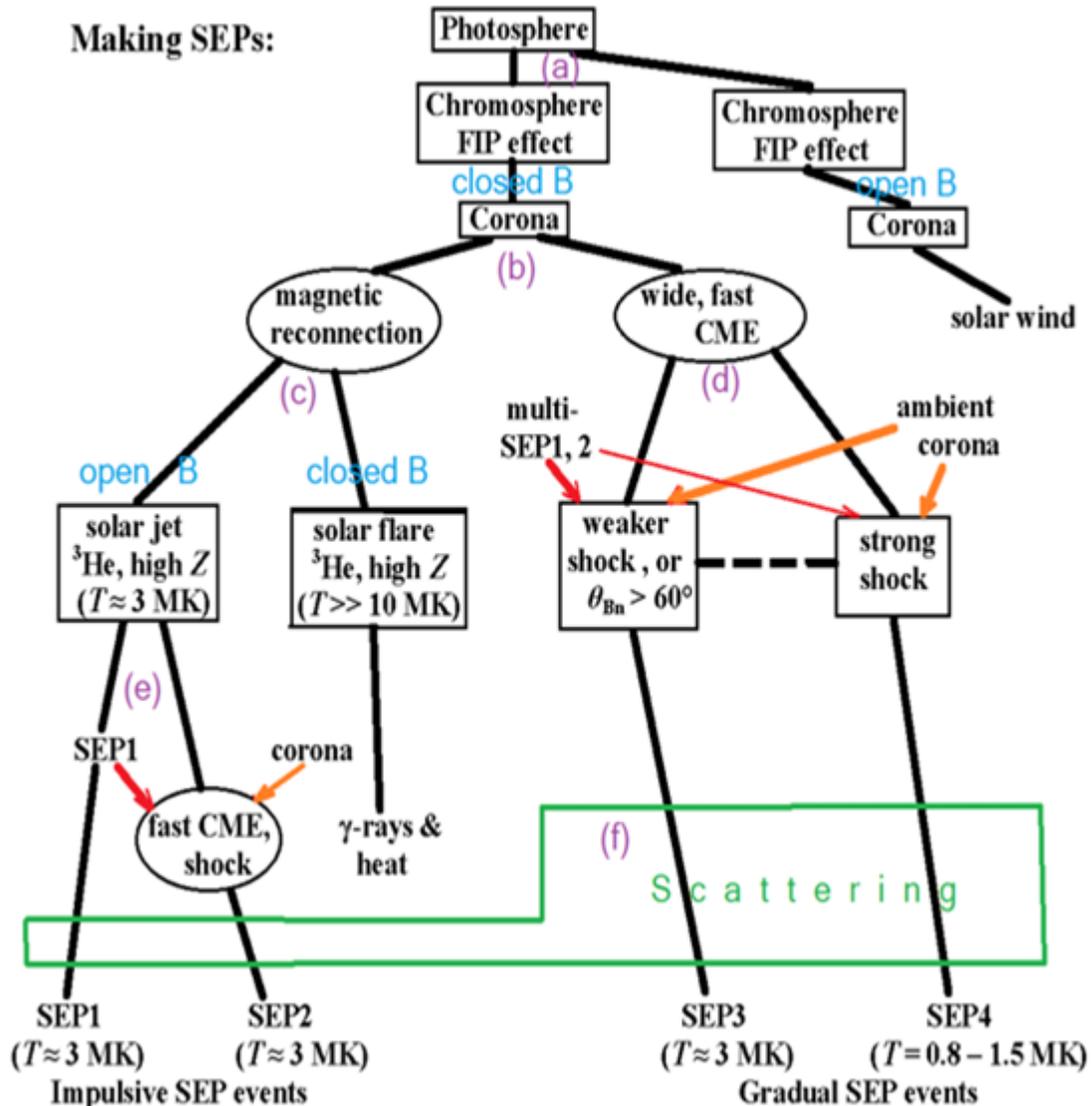

Figure 6-10. : Possible reconciliation between the two-class paradigm of the early 1900's and the puzzling SEP observations from Wind, ACE and STEREO. The SEPs are grouped into four sub-classes according to their acceleration locations and the contributions of relevant physical processes (taken from Reames 2020). (a) Elements experience different FIP processing on open and closed field lines. (b) SEPs are accelerated in magnetic reconnection regions or by CME shocks. (c) SEPs enriched in $^3$He and heavy ions escape along open field lines associated with solar jets, and not from closed field lines associated with solar flares. (d) Weaker CME-driven shocks preferentially accelerate residual suprathermal material left over from multiple jets, while the stronger CME shocks accelerate ambient coronal material. (e) Fast CMEs associated with jets also accelerate suprathermal jet-accelerated material. (f) Rigidity-dependent trapping near CME shocks or scattering in the corona and the IP medium modifies heavy ion abundances and energy spectra in large gradual SEP events. In some large events, self-generated waves can enhance particle trapping and increase the acceleration efficiencies of the CME shocks.

Finally, Zelina et al. (2017) and Doran, Dalla, and Zelina (2019) suggested that the observed temporal evolution of heavy ion elemental abundances and energy spectra during some large SEP events observed at ACE and STEREO could be accounted either by rigidity-dependent scattering (Parker 1965; G. M. Mason et al. 2006b; G. Mason et al. 2012; A. J. Tylka et al. 2013) (or via drift-associated Q/M-dependent deceleration during transport through the interplanetary medium (Kota 1979; Dalla, Marsh, and Laitinen 2015).

To reconcile the puzzling SEP observations from Wind, ACE, and STEREO with the two-class paradigm of the mid-1990's, Reames (2020) has recently proposed that the SEPs observed at 1 AU (see Figure 6-10) can be grouped into four sub-classes based on where and how they are accelerated, namely: 1) Pure impulsive SEP events are produced when the ~3 MK coronal plasma is accelerated by magnetic reconnection islands in solar jets around ~1.5 Rs. These events have large M/Q-dependent enhancements for H through ultra-heavy ions as well as the $^3$He/$^4$He ratio; 2) Impulsive SEP events are produced when CMEs from the same jets are fast enough to drive shocks. These shocks re-accelerate the ~3 MK impulsive suprathermal material mixed with ambient plasma mainly comprising protons; 3) Weak gradual SEP events occur when wide, moderately-fast (>500 km/s) CME-driven shock waves accelerate the ambient coronal plasma, including protons, but also preferentially accelerate the faster impulsive ~3 MK suprathermal heavy ion material left over in the corona from many small jets; and 4) Strong gradual SEP events are produced when wide, fast (>1000 km/s) CME-driven shock waves predominantly accelerate material from the ambient ~1–2 MK coronal plasma at ~2–3 R$_S$. These events have negligible amounts of impulsive suprathermal ions.

## 6.4  Progress in SEP theory and modeling during the VarSITI era (2015 – 2019)

### 6.4.1 SEP acceleration

Valuable insight on particle acceleration at coronal shock waves have been recently obtained by studying the evolution of CME and shocks low in the corona, as well as their interaction with underlying magnetic fields and coronal plasma, both through data-driven or analytical modelling, as well as MHD simulations and combinations of different approaches. New information has been gathered about the relevant parameters for efficient shock acceleration (such as the Mach number, compression ratios, geometry of shock waves), the primary acceleration regions along the shock, the role of coronal magnetic field configuration and how these factors are related to the particle spectra observed in space.

Forward modeling techniques (Alexis P Rouillard et al. 2016; Salas-Matamoros, Klein, and Rouillard 2016; Athanasios Kouloumvakos et al. 2019) have been used to perform the geometrical fitting of shock waves or CMEs with different geometrical models (i.e., the spheroid or the graduated cylindrical model), based on multipoint imaging. (Alexis P Rouillard et al. 2016) performed a triangulation of the three-dimensional (3D) expansion of high-pressure fronts using three simultaneous viewpoints from SOHO, SDO and STEREO observations for the 17 May 2012 GLE event. In conjunction, they inverted remote-sensing observations to derive the background coronal conditions through which the pressure front propagates and modelled the topology of the background magnetic field. To this end, they employed both the PFSS model, based on the line of sight component measured by HMI, and the MHD MAST (Magnetohydrodynamic Around a Sphere Thermodynamic) model with improved thermodynamics including realistic energy equations with thermal conduction parallel to the magnetic field, radiative losses, and coronal heating (Lionello, Linker, and Mikić 2008). They derived the normal speed and the Mach number ($M_{fm}$) over the entire surface of the CME front, as well as the shock geometry and the magnetic connectivity of the near-Earth environment with the shock. A band of high $M_{fm}$ values was found to be co-located with the region of quasi-perpendicular geometry, which evolved within 10 minutes into a quasi-parallel geometry, $M_{fm}$ reaching its highest values near the nose of the CME. In addition, a super-critical shock ($M_{fm}$ values in excess of 3) had formed at the release time of high-energy particles, suggesting that delayed release times of GeV protons could be related with the time needed for the shock to become super-critical. Moreover, the presence of very high $M_{fm}$ values along open field lines crossing the shock in the vicinity of the neutral line, corresponding to the heliospheric plasma sheet, suggests that the neutral line could be a favorable region for particle acceleration, although spatially limited. Thus, it is necessary a good connectivity between the shock regions crossing the vicinity of the tip of streamers and the associated neutral line, e.g. through a large-scale magnetic flux rope or any complex magnetic field structure.

By using the same approach, Plotnikov, Rouillard, and Share (2017) reconstructed the evolving shock front and its properties for three far-side CME events (2013 Oct. 11, 2014 Jan. 06 and Sep. 01) which were associated with LPGREs, as observed by the Fermi/LAT (Large Area Telescope) at energy >100 MeV-300 GeV, and SEP events measured in situ at 1 AU (by SOHO and STEREO A, B). They obtained that for all the three events, the shock became super-magnetosonic ($M_{fm}$>1) and magnetically connected to the visible solar surface within 10-15 min of the start of the flare and just before the onset of the >100 MeV-ray emission, showing a quasi-perpendicular geometry at the flanks during the bulk of the gamma-ray emission. Moreover, by comparing the SEP electron and gamma-ray onset times and the computed electron to proton ratios at the Sun and in space (within a factor 5), they concluded that the same shock processes are responsible for both LPGREs and the production of SEPs. Nevertheless, they found no clear correlation between the shock Mach number levels and the intensity of the gamma-ray flux measured by LAT, suggesting that a more complicated physics might be at work, which requires further investigation.

Furthermore, Kouloumvakos et al. (2019) derived the shock parameters from 3D modelling of the coronal pressure waves and compared them with properties of SEP events over an extended dataset. They analyzed a number of 33 SEP events with energy > 50 MeV clearly observed at least at two interplanetary locations by SOHO and STEREO, and computed the correlation between the peak intensity during the prompt phase and several shock parameters at the shock regions magnetically well-connected with the observers. Correlations with shock speed, compression ratios and Mach numbers were found to be significant for well-connected field lines, having Pearson's correlation coefficients of 65%, 56% and 72%, respectively, supporting previous results (Alexis P Rouillard et al. 2016; Plotnikov, Rouillard, and Share 2017; Afanasiev et al. 2018). On the other hand, no significant correlation was found between the SEP peak intensity and the shock angle as well as no energy dependence on any performed correlation. Finally, shock waves were found to become super-critical at a median distance of 3 RS, while solar particles to be released ~15–20 minutes after the shock waves become super-critical and have connected magnetically with the SEP observing point.

Kozarev et al. (2015) analyzed the initial phase of a CME on 2011 May 11 by combining observations from the SDO/AIA images in the time interval 02:12-02:31 UT and the following models: 1) the Coronal Shock Geometric Surface (CSGS) model which produces a three-dimensional spherical dome surface, propagating through the corona with speed and radius based on the observed time-dependent position of the EUV coronal bright front (describing the shock front) in the radial direction; 2) the Potential Field Source Surface (PFSS) magnetic field model, providing a global coronal vector magnetic field solution for a 3D grid of polar coordinates. They estimated the time-dependent orientation of the wave/shock with respect to the coronal magnetic fields between 1.1 and 1.5 Rs, the part of the coronal surface which shock-accelerated particles during the initial stages as well as heliospheric connectivity during the shock passage in the low corona. They found that the field-shock angle changes significantly throughout the evolution of the shock surface, with higher values preferentially near the flanks, although they reached almost 90° near the shock nose as well in the last phase of the event. Moreover, the open field lines crossed by the shock in the second half of the event were likely both related to high angle values and magnetically connected to the Earth. Thus, the authors concluded that the shock acceleration efficiency and particle release is considerably higher in the second part of the shock evolution in the AIA FOV.

Recent MHD simulations of the CME expansion have shown the formation of shocks or strong compression regions at low coronal heights (< 2 solar radii). As a result of the CME's rapid acceleration, shocks and strong compressions appear on the flank of the CME, showing a large negative velocity divergence and creating the conditions that lead to rapid particle acceleration. By analytically solving the Parker equation in the presence of size-limited acceleration regions as obtained by MAS (Magnetohydrodynamic Algorithm outside a Sphere) simulations, Schwadron et al. 2015) showed that broken power laws can be naturally obtained due to pronounced effects of particle diffusion and particle escape. As a matter of fact, the finite size of the shock or compression limits the maximum energy gain of the particles, because the magnetic field line (or flux bundle), near which these particles are accelerated, moves off of the accelerator, i.e. the shock or compression, leading to the formation of a broken power-law in the particle distribution. Moreover, the break energy and the spectral index of the second power law increase with the shock angle, the size of the shock, the CME driver speed, and with reductions in the rigidity dependence of the scattering mean free path.

The effect of large-scale streamer-like magnetic configuration on particle acceleration at coronal shocks has been investigated by considering a CME-driven shock propagating through a streamer-like magnetic field (Kong et al. 2017) or from its flank (Kong et al. 2019), i.e., when the streamer is rotated with respect to the CME propagation direction. By numerically solving the Parker transport equation with both parallel and perpendicular diffusion in such configuration, Kong et al. (2019) found that the primary sources for particle acceleration are located at different regions and vary significantly as the shock propagates and expands, depending on the particle energy and on time. In particular, the acceleration of particles to more than 100 MeV mainly occurs in the shock-streamer interaction region close to the shock flank, where the shock is quasi-perpendicular and closed magnetic fields are present, favoring particle trapping upstream of the shock, both conditions increasing the acceleration efficiency. They also obtained broken power laws for the particle spectra up to 100 MeV integrated over the simulation domain. Thus, the streamer-like magnetic field can play a critical role in producing gradual solar energetic particle events and they may be a mixture of two distinct populations accelerated in the streamer and open field regions, having different acceleration rates. Nevertheless, these results might be affected by the two-dimensional treatment, leading to non-zero divergence of magnetic field at the shock front and possibly to mismatch parallel diffusion and perpendicular diffusion.

Some previous models attribute the generation of double power-laws and Q/M-dependent spectral breaks to the trapping and subsequent rigidity-dependent escape of ions from the shock during the SEP acceleration processes via DSA at CME-driven shocks in the solar corona (Gang Li et al. 2009; A. J. Tylka and Lee 2006). In the Schwadron et al. (2015) lower coronal SEP acceleration model, the finite size of the ICME shock and stronger Q/M-dependence of the diffusion coefficient facilitates particle escape from the acceleration region, which reduces the $E_o$ and steepens or softens the higher energy spectral slope. Conversely, in the Schwadron et al. (2015) model, the weaker Q/M dependence inhibits particle escape, which increases the break energy and flattens the higher energy spectral slope. Indeed, Desai et al. (2016) were able to infer key properties such as the strength, obliquity, and turbulence conditions associated with the corresponding near-Sun CME shocks using the spectral properties of those 1 AU SEP events where the observed Q/M-dependent spectral breaks were interpreted to be consistent with DSA shock acceleration rather than transport-dominated processes. Recently, Tan, Malandraki, and Shao (2017) also found that the high-energy Ne/O ratio is well correlated with the source plasma temperature of SEPs, and used the variability of Ne/O and Fe/O ratios to investigate the accelerating shock properties.

In contrast, Li and Lee (2015), Zhao et al. (2016) and Zhao, Zhang, and Rassoul (2016) argue that, in some large SEP events, single power-law spectra injected by near-Sun ICME shocks can exhibit spectral breaks at 1 AU due to Q/M-dependent scatter-dominated transport in the IP medium. Specifically, Li and Lee (2015) showed that particle scattering and diffusion from the Sun en route to 1 AU could alter a single proton power law into three distinct power laws by the time the shock arrives at 1 AU, suggesting that scatter-induced particle propagation in the IP medium can also result in spectral breaks at 1 AU even if the ICME shock-accelerated spectra are pure power laws; in this case the Q/M dependence (α parameter) has an upper limit of ~1.4. Thus, in cases where the spectral breaks are interpreted to be caused primarily by interplanetary transport effects, the observed temporal evolution of the heavy ion abundances and spectral breaks can provide insights into the interplanetary turbulence conditions encountered by the SEPs during their transit to 1 AU (Gen Li and Lee 2015; L. Zhao et al. 2016; L. Zhao, Zhang, and Rassoul 2016). On the other hand, another functional form was also found to reproduce the proton energy spectrum during SEP events (Laurenza et al. 2013; 2015), being the Weibull distribution the best fit for the event-integrated spectrum and separately, during the prompt and energetic storm particle phases. A theoretical derivation of the Weibull spectrum was provided (Laurenza et al. 2016; Pallocchia, Laurenza, and Consolini 2017) in the framework of the acceleration by "killed" stochastic processes exhibiting power-law growth in time of the velocity expectation, such as the classical Fermi process, or alternatively, by the shock-surfing acceleration. Thus, those authors suggested that a scenario in which different mechanisms could account for particle acceleration at shocks in different energy ranges, the stochastic or shock surfing acceleration contributing significantly to the acceleration of high energetic particles, in addition to the DSA at lower energies.

## 6.4.2 SEP transport

The well-separated SEP observations with the STEREO mission have also lent themselves to study the transport of SEPs in more detail. The comparison of multi-spacecraft observations with results of 2D or 3D models solving the focused transport equation allowed to study not only transport along the mean magnetic field but also perpendicular to it. It was found that the role of efficient transport perpendicular to the mean magnetic field can eventually be much stronger than expected  (Nina Dresing et al. 2012) and might play an important role, among extended injection and acceleration regions, in the longitudinal spreading of SEPs (Dröge et al. 2014; Dröge et al. 2016)but also in creating asymmetries in their longitudinal distribution at 1 AU (H.-Q. He and Wan 2015; Strauss, Dresing, and Engelbrecht 2017).

The main theories attempting to describe perpendicular transport are diffusion, i.e. by particle scattering at magnetic field irregularities resonant with the particles' gyro radius (Zhang, Qin, and Rassoul 2009; Giacalone and Jokipii 2012; Dröge et al. 2010; Dröge et al. 2014; W Dröge et al. 2016)  and field line meandering. Laitinen et al. ( 2016) and references therein propose a combined scenario with particles remaining on turbulently meandering field lines early in the event which turns over into diffusive transport at a later phase, both leading to particles spreading perpendicular to the mean field.  While a solely diffusive approximation is not able to explain efficient perpendicular transport at the same time like the presence of sharp SEP drop out events (J. E. Mazur et al. 2000), the model of field line meandering is. Furthermore, it is able to explain too early SEP onsets caused by a propagation path length shorter than the nominal Parker spiral length (Laitinen and Dalla 2019). However, both diffusion and field line meandering approaches are struggling to explain the extreme widespread events observed with the STEREO and close-to-Earth spacecraft. This suggests that such events are not only caused by efficient perpendicular transport but by a combination of transport and an extended injection and/or acceleration region. The limited number of well-separated observers during the STEREO era was not fully sufficient to constrain the injection size and the transport conditions at the same time which is expected to improve significantly during the next solar cycle with the presence of new missions like Parker Solar Probe and Solar Orbiter.

An important step taken in SEP transport modelling is the inclusion of a realistic solar wind background. Wijsen et al. (2019) use the data-driven EUropean Heliospheric FORecasting Information Asset (EUHFORIA, Pomoell and Poedts 2018) to generate a background solar wind for their SEP transport code. This allows, on the one hand, the study of the effect of solar wind streams on the SEP propagation and also on adiabatic energy changes (Wijsen et al. 2019b). On the other hand, case studies with a realistic background field will be possible accounting for the effects of transient structures leading to non-Parker field configurations (Leske et al. 2012)

Pacheco et al. (2019) have re-visited 15 relativistic electron events observed with the Helios spacecraft in the 1970's and 80's using the 1D transport code by Agueda et al. (2008), which allows to infer the solar release time profiles and the values of the radial mean free path. Short injection duration (<30min) events were only found in 30% of the cases compared to long-duration events (>30min) otherwise. The radial mean free paths, observed at spacecraft locations between 0.31 and 0.94 AU, vary between 0.02 and 0.27 AU. Agueda and Lario (2016) found indications that the strength of the interplanetary scattering varies with the size of the solar parent event suggesting that energetic particle population itself generates waves. However, Pacheco et al. (2019) did not find such a dependency in their sample. While self-generated waves are an accepted phenomenon for ions, which plays also an important role in proton acceleration at shocks (Bell 1978) (see also section 6.3.1), their presence in relation to solar energetic electrons remains elusive. The four consecutive Helios events studied by Agueda and Lario (2016) are also a famous example used to illustrate the radial effect of SEP transport: the same SEP events were observed by the close-to-Earth spacecraft IMP-8, but the four distinct impulsive increases seemed to have merged into only one gradual event. However, as discussed and modelled by Agueda and Lario (2016) and Strauss, Dresing, and Engelbrecht (2017), not only radial scattering alone but most likely the contribution of perpendicular diffusion is responsible for the loss of the detailed SEP event structure at 1 AU.

## 6.5 Conclusion

In conclusion, for a full understanding of the SEP acceleration, injection, and transport processes which altogether determine the variable particle observations discussed so far, additional systematic, multi-spacecraft studies are needed. SEP observations provided by the Parker Solar Probe (McComas et al. 2016) and Solar Orbiter (Rodríguez-Pacheco et al. 2020) missions in the inner heliosphere in conjunction with modeling efforts will be the basis of upcoming advancements in our understanding.

# 7. Stream Interaction Regions / Co-rotating Interaction Regions (SIRs/CIRs)

## 7.1 Introduction

The solar corona is structured by open and closed magnetic field regions that transition at a certain distance as open field into interplanetary space (see Figure 7-1). The solar atmosphere is permanently in a state of dynamic energy release and renewal, e.g., structures constantly interact with each other. Flares and CMEs are mostly related to active regions and filament eruptions. Other solar features that can cause strong geomagnetic events are stream interaction regions (SIRs) which are related to coronal holes, known as areas on the Sun with predominantly open magnetic field. In that respect, the knowledge about the 360° structuring of interplanetary space is a key input parameter for Space Weather forecasting. As CMEs are affected by the MHD drag which is in quadratic dependence on the ambient solar wind speed, fast streams may strongly influence the predictions of arrival times and impact speeds of CMEs. The structured interplanetary space also actively affects SEP propagation and causes by itself geomagnetic storms. During solar minimum phase, the energy input into Earth's magnetosphere by CIRs is similar to that of CMEs (Richardson, Cliver, and Cane 2001; Tsurutani et al. 2006).

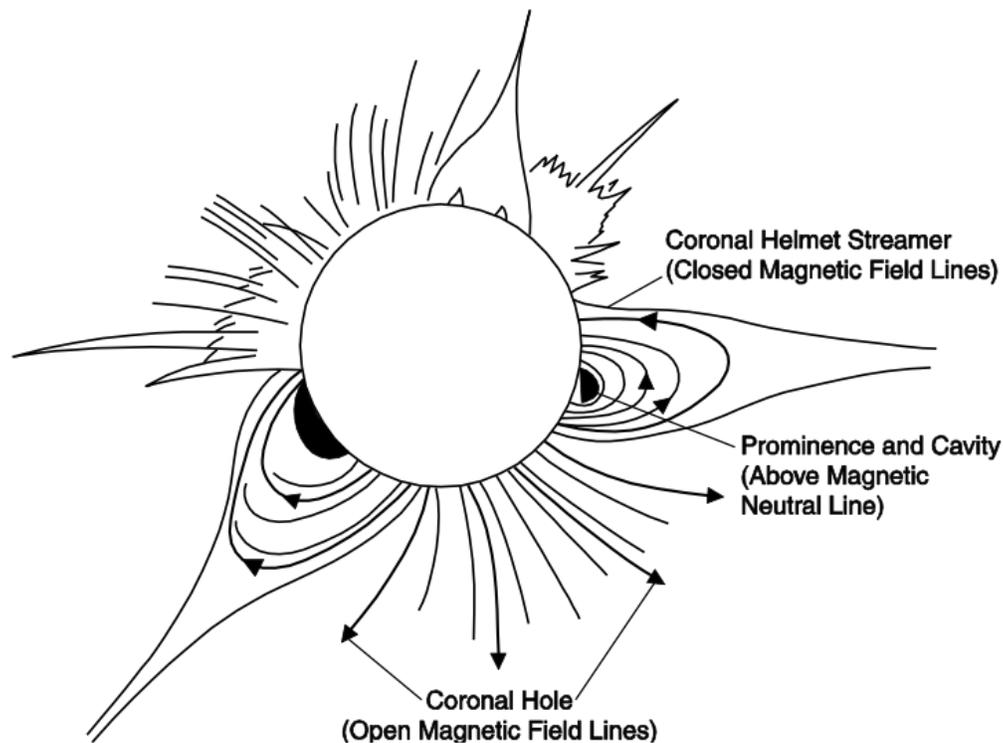

*Figure 7-1. Sketch of the open-closed magnetic field structures on the Sun. Adapted from Hundhausen 1995; Fig 4.8 of the book "Introduction to Space Physics" edited by Kivelson and Russell (1995).*

SIRs are related to the high speed streams originating from regions of open magnetic fields along which plasma is accelerated and may easily escape from the Sun. Primarily open magnetic field (and flux) are observed within coronal holes, dark regions of lower temperature and density as observed in EUV and SXR. Harvey and Recely (2002) describe the evolution of CHs based on observations during solar cycles 22 and 23. Isolated high-latitude CHs evolve into polar CHs, where they appear as stable and rigidly rotating objects, which may exist for several years. Approaching the descending phase of a solar activity cycle, polar CHs extends to lower latitudes that consequently change the solar wind in the ecliptic (i.e., encompassing the planets) quite dramatically. This evolutionary process is revealed in EUV data from the clear change of morphology (area, shape) and location of CHs. Wang and Sheeley (1990) describe the magnetic structure of CHs and find that there is a close relation between the flux tube expansion and the underlying photospheric magnetic field. Recent results show that the evolution of CH area is not correlated to the evolution of the underlying magnetic field, indicating that the magnetic field inside a CH does not drive its evolution (S. G. Heinemann et al. 2020). SIRs have been observed throughout the heliosphere by a large variety of spacecraft, such as Helios in the inner heliosphere, by ACE, Wind, DSCOVR near Earth at 1 au, by Voyager and Pioneer in the outer heliosphere, and also out of the ecliptic by Ulysses.

Due to the quasi-stationary location of low-latitude CHs, the interaction of high and slow speed solar wind streams results in compression of plasma and magnetic field. As the plasma is of enhanced speed compared to neighboring regions, interactions between fast and slow solar wind streams occur at certain distances from the Sun (cf. Figure 7-2). Here the leading edge of the interaction region forms a forward pressure wave that propagates into the slower plasma ahead. Likewise, the trailing edge is a reverse pressure wave (see review by Gosling and Pizzo 1999). Since CHs are slowly evolving but long-lived structures, these interaction regions recur with every solar rotation and are then called co-rotating interaction regions (CIRs). On the large scale, it is assumed that their interplanetary dynamics can be described by ideal MHD equations.

One of the characteristics of SIRs, measured at 1AU distance, is the so-called stream interface, characterized by an abrupt drop in density, simultaneous rise in proton temperature and gradual increase in speed, and an east-west flow deflection, i.e. the region separating the originally slow, dense plasma from originally fast thin plasma back at the Sun (e.g., Wimmer-Schweingruber, von Steiger, and Paerli 1997). Furthermore, the stream interface is usually preceded by a density increase due to the compression (compression region), and is often associated with a sudden change in the magnetic polarity (sector boundary), and a gradually rising solar wind speed profile (L. K. Jian et al. 2009). The change in the magnetic polarity is related to the alternation of magnetic sectors, referring to the neutral line (heliospheric current sheet). The number of magnetic sectors changes with the solar cycle, typically there are 4 sectors, but it can get more complex during solar maximum. Shocks on both sides of the interaction region develop more strongly further out in the heliosphere at 2-3 AU (forward and reverse shock).

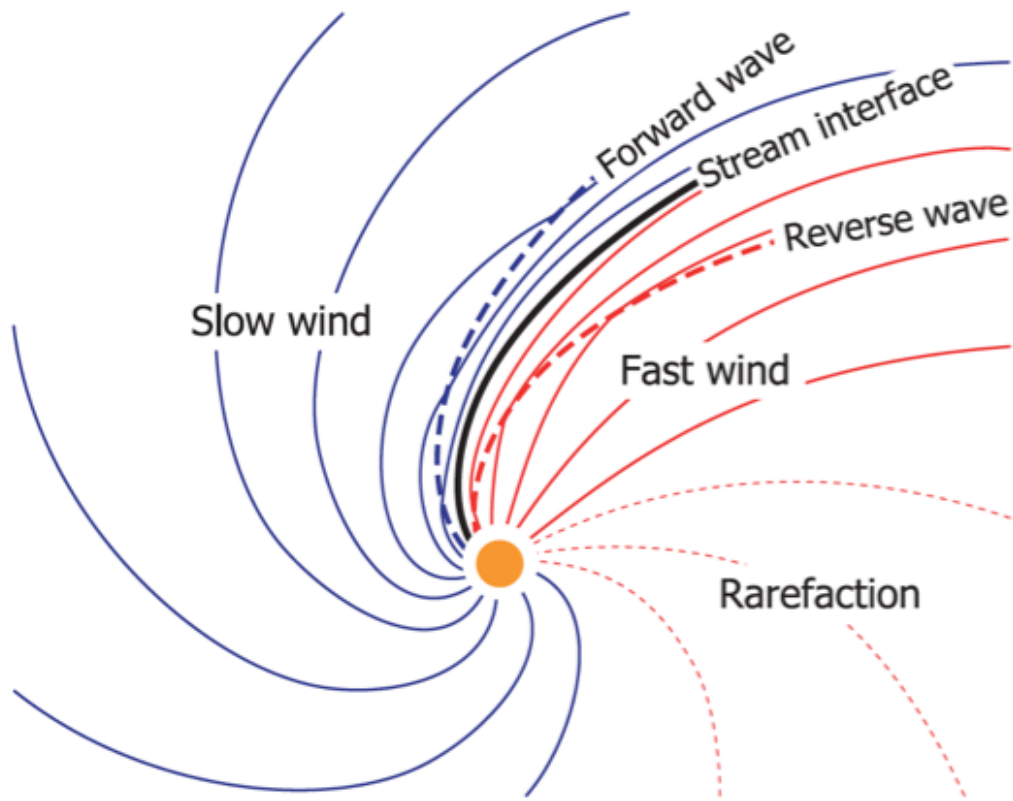

*Figure 7-2. Sketch of an SIR as seen in the ecliptic plane showing the fast (slow) solar wind in red (blue). The frozen-in magnetic field lines become aligned with the stream interface by the reverse (forward) wave. Taken from Owens and Forsyth (2013) adapted from Gosling et al. (1981).*

Because SIRs/CIRs rotate in the direction of planetary motion, the high speed solar wind streams emanating from a centrally located coronal hole arrives at Earth about 4 days later (see Figure 7-3). Typical HSS intervals following CIRs, may drive prolonged geomagnetic activity and cause strong high energy particle enhancements in the Earth's radiation belts (Reeves et al. 2003; Miyoshi et al. 2013; Kilpua et al. 2015). In detail, the strength and impact of geomagnetic storms depends, beside the impact speed, most importantly on the north-south component, $B_z$, of the magnetic field of the solar wind stream (see also Krauss et al. 2015). In the case of CIRs, the cause of the disturbances is often Alfvenic waves in the HSS (B. T. Tsurutani et al. 2018) .

## 7.2 Progress in observations

The evolution of long-lived coronal holes, closely related to SIRs/CIRs, can be studied in detail using multiple views on the Sun from combined STEREO and Earth imagery. Using spacecraft separated by 120°, studies of changes in the large scale EUV structures over time and their relation to the in-situ measured solar wind can be performed. It was found that coronal holes undergo evolutionary patterns revealing a growing and declining phase where area increases and decreases again over several solar rotations (Stephan G. Heinemann, Temmer, et al. 2018; Stephan G. Heinemann, Hofmeister, et al. 2018; S. G. Heinemann et al. 2020). The slow evolution of coronal hole areas and the steady-state of the related solar wind streams is found to be well related to the solar wind speed measured in-situ at 1 AU (Nolte et al. 1976; Bojan Vršnak, Temmer, and Veronig 2007; Abramenko, Yurchyshyn, and Watanabe 2009; Karachik and Pevtsov 2011; Rotter et al. 2012; Tokumaru et al. 2017). With this well-known area-speed relation, empirical forecasting tools for the "pure" background solar wind on the basis of coronal hole area measurements are performed on a regular basis as depicted in Figure 7-4 (Bojan Vršnak, Temmer, and Veronig 2007; Rotter et al. 2012; Reiss et al. 2016a). By understanding the photospheric and coronal evolutionary characteristics of CH, one can aim to gain a better understanding and in turn improve the forecast of CIRs (see Heinemann et al. 2018). Temmer et al. (2017)showed in a case study that the evolutionary trend visible in the CH area is matched by the trend of the peak velocity of the associated HSS over the lifetime of the CH. This is an important indicator that the high-speed peak velocity to CH area relation also persists over a CH's evolution. They also found that the total perpendicular pressure at the stream interface and the in-situ magnetic field at the B- peak and the v- peak do not show the same evolutionary profile. The forecasts are found to be most successful for periods of low solar activity, as during increased solar activity transient events, such as CMEs, strongly disturb the rather stable solar wind outflow for several days (up to 5 days; see Temmer et al. 2017); cf. also Janvier et al., (2019).

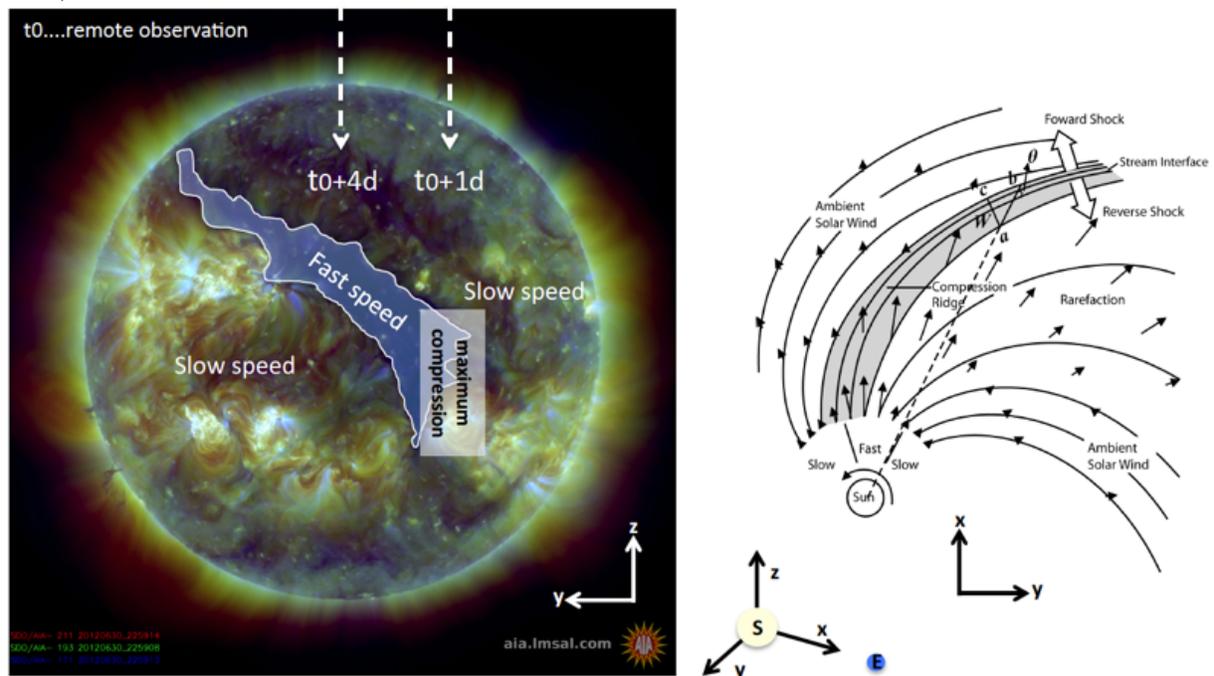

*Figure 7-3. Left: SDO/AIA composite image of the wavelength channels 211-193-171A from June 30, 2012 showing the reduced density region of a coronal hole (shaded area). At the time t0, the central position of the coronal hole is extracted from remote data. At about t0 + 1d the maximum in the density/magnetic field and at about t0 + 4d the maximum in the speed/temperature is observed from in-situ data at 1 AU. Right: Cartoon by Pizzo (1978) to illustrate the fundamental processes involved in the 3D dynamics of stream evolution.*

As the high speed solar wind streams emanate from CHs which are low dynamic structures, CIRs can be forecasted with long lead times. Based on that a variety of persistence models were developed. Under the assumption of persistence, in-situ measurements of the solar wind plasma flow from L1 and varying STEREO vantage point provide a forecast for Earth position with lead times of up to 27 days (depending on the exact STEREO spacecraft position). The closer the measuring spacecraft is ahead or behind Earth, the less the effect of the temporal evolution of the solar wind profile (Opitz et al. 2009). Implementing the actual changes of CH areas (from EUV data) into such simple forecasting tools can improve the forecast quality (Temmer, Hinterreiter, and Reiss 2018). Lead times with about 4.5 days could be achieved when using data from an instrument permanently located at the Lagrangian point L5 (60° behind Earth; ESA preparation for the future L5 mission "Lagrange"). However, it is pointed out that the latitudinal offset between the measuring spacecraft is limiting the accuracy of persistence modeling (Owens et al. 2019) as the streams flow speed profile is rather depending on the latitudinal range of the center of mass (Hofmeister et al. 2018). Recent studies also found latitudinal differences in the geoeffectiveness of events which are caused by variations in the interplanetary magnetic field due to closed flux ropes (CME magnetic structure) and compression regions - CME shock-sheath or SIR ( Huttunen et al. 2008; Yermolaev et al. 2017). Together with the warping of the heliospheric current sheet, the forecasting of high-speed solar wind structures is rather complex, but considering the uncertainties, nevertheless, a good estimation of the flow approaching Earth can be given.

Forecasting solar wind structures in interplanetary space serves also as important information for analytical CME propagation models and Space Weather models. However, SIR/CIRs may not only influence the propagation of CMEs but also the evolution of its internal magnetic structure, as the compression region may represent an obstacle which can hamper the CME expansion (see e.g., Dumbović et al. 2019). Therefore, under the Space Situational Awareness Program of the European Space Agency forecasting services using empirical, and numerical models for the solar wind are available (see http://swe.ssa.esa.int). In that respect, a new four-plasma categorization scheme for the solar wind is given by Xu and Borovsky (2015) that can be used for the automatized detection scheme for solar wind and CMEs (private communication with S. Vennerstroem, 2015). As the forecasting/nowcasting quality is still not sufficient for producing reliable Space Weather alerts, we need to better understand and closely monitor SIRs/CIRs. Main aims for the near future are to 1) verify and evaluate background solar wind models and with that improve the input for CME propagation models (as discussed in Section 4.1 and 4.2) and more accurately predict periodic and recurrent geomagnetic effects from CIRs. In general, single events are easier to forecast compared to multiple events featuring CME-CME interactions, or interactions of CMEs with CIRs. In order to better understand the fast flow plasma in interplanetary space, research on their sources, CHs, is of utmost importance (Wilcox and Howard 1968). By better understanding the physics behind CH evolution we may improve their forecasting capability. With that, we will also gain more insight in the ejection and acceleration processes that define high-speed streams. This is of timeliness as we can exploit data from the NASA mission Parker Solar Probe (PSP) (Fox et al. 2016) measuring the near-Sun space and with that regions where the solar wind actually gets accelerated.

Hofmeister et al. (2017) in a statistical study of 288 low-latitude coronal holes during the time range of January 01, 2011 – December 31, 2013 and Hofmeister et al. (2019) in a statistical study of 98 coronal holes shed light on the magnetic fine structure in photosphere underlying the projected CH boundary. Using SDO/HMI line-of-sight magnetograms, they showed that the magnetic field is made up of a very weak slightly asymmetrically skewed background field ($|B_{BG}|$=0.2-1.2G) and small unipolar magnetic elements. These small unipolar magnetic elements contain most of the signed magnetic flux that arises from coronal holes. It was found that the area that these unipolar, usually long living (lifetimes > 40h), magnetic elements cover determines the total signed flux of a coronal hole (see Figure 7-4). These magnetic elements are important in the context of solar wind acceleration, propagation and forecasting as they are suspected to be the footpoints of flux tubes or magnetic funnels. Flux tubes, or clusters of magnetic fibers open towards interplanetary space, have found to be the small-scale source regions of the plasma outflows within CHs (Hassler et al. 1999; Tu et al. 2005). Wiegelmann et al. (2014) showed, using SUMER data and magnetic field extrapolations, that the regions showing high outflow velocities correspond to strong unipolar flux concentrations within CHs (see Figure 7-9 in the section "Progress in Theory and Simulations"). The substructure of these flux tubes, e.g., a bundle of magnetic fibers, is represented by the abundance of magnetic bright points (MBPs) within magnetic elements as observed in high-resolution HINODE/SOT G-band filtergrams (Hofmeister et al. 2019). This shows the highly structured magnetic field configuration of CHs in comparison to the previously often assumed rather uniform configuration. This structuring also carries out into the solar wind structure in interplanetary space as shown by the first PSP observations of HSS that were found to show a strongly structured and perturbed speed profile (Bale, Badman, and Wygant 2019).

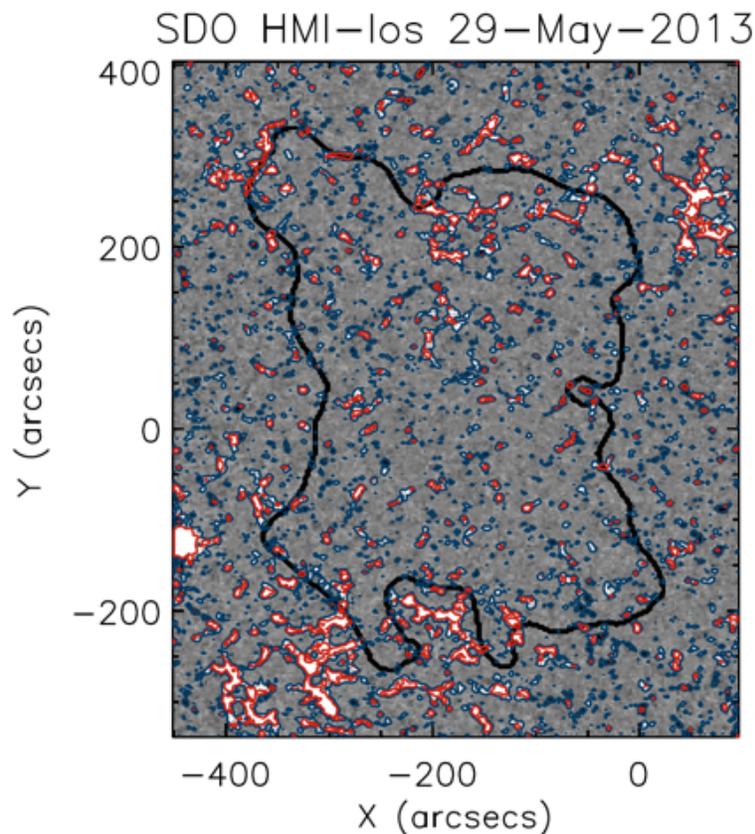

Figure 7-4. Magnetogram (scaled to ±30 G) of a large coronal hole observed on 2013 May 29. The coronal hole boundaries are outlined in black. Regions with an absolute magnetic field density of more than 10 G are outlined in blue and of more than 50 G in red. Taken from Hofmeister et al. (2017).

Though harder to evaluate and forecast, interaction events between SIR/CIR and CMEs are of special interest. Interaction events can lead to significant increase in geomagnetic effects when compared to individual events of similar strength (He et al. 2018; Dumbović et al. 2019). Heinemann et al. (2019) showed in a case study of a CME interacting and propagating within a HSS on June 22, 2011 that the dynamic pressure of the SIR/CIR followed by the CME shock signature within the HSS induces wave-like flaring-motion into the Earth's magnetopause and causes, due to enhanced magnetopause currents, a much stronger Sym-H value than would have been expected. Due to the rather small and weak CME the effects were still only moderate. Also, enhanced substorm activity was recorded.

In a recent paper by Jian et al.(2019), physical properties of a large sample of slow-to-fast SIRs were investigated using STEREO-A and -B data. They identified 518 pristine SIRs, of which more than 50% are associated with crossings of the heliospheric current sheet (HCS) and are of slow speeds but higher densities, and of increased dynamic and total pressure compared to those without HCS (see Figure 7-5). In that respect, HCS related SIRs can be classified as more effective in terms of Space Weather.

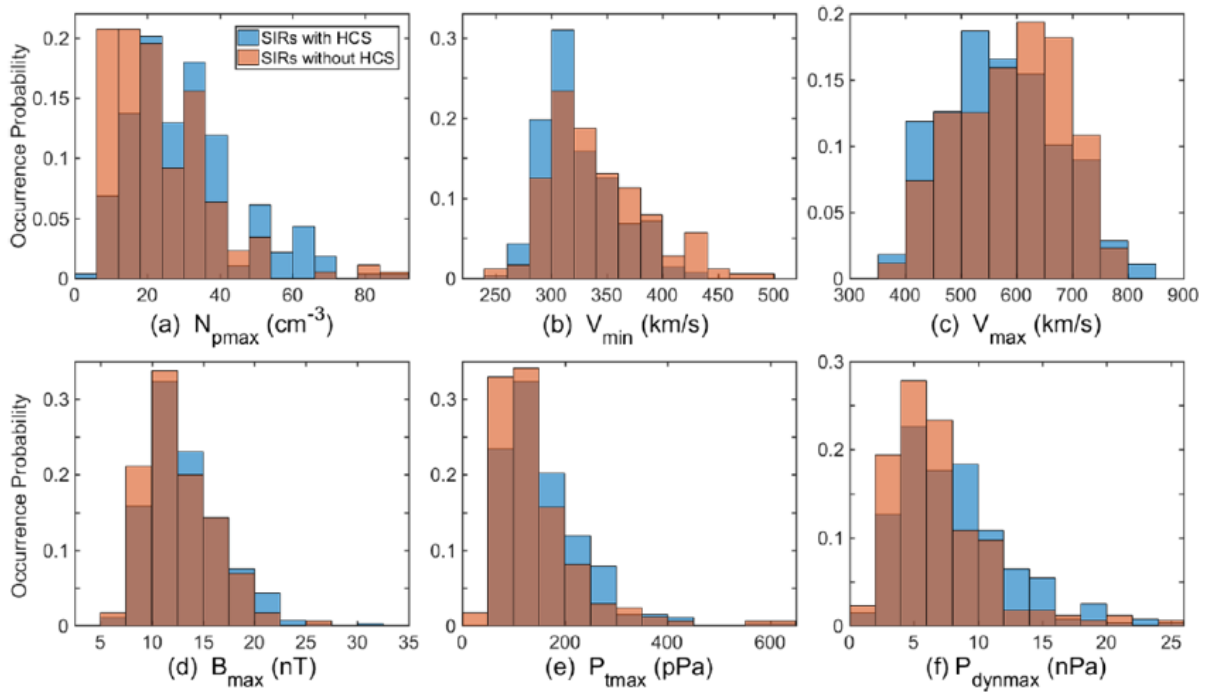

*Figure 7-5.* The comparison of occurrence probabilities between SIRs with an HCS crossing (blue bars) and without any HCS crossing (orange bars) for the following parameters: (a) maximum $N_p$, (b) minimum speed, (c) maximum speed, (d) maximum B, (e) peak $P_t$, and (f) peak dynamic pressure. The brown shaded regions are the overlapping regions between SIRs with an HCS and the SIRs without any HCS crossing. Taken from Jian et al. (2019).

The detection and extraction of reliable CH areas from operational solar observations is extremely important, not only for solar wind forecasting using the area-speed relation, but also for investigating the magnetic open flux on the Sun. At the present, most extraction methods focus on EUV observation taken by SDO/AIA, SOHO/EIT, GOES/SUVI and/or STEREO/EUVI. Due to the optimal filter sensitivity and high contrast, wavelengths of highly ionized iron (e.g., Fe XII : 193/195 Å) are often used. Methodologically, intensity based-methods are the go-to choice, with some form of intensity thresholding being the preferred choice. Intensity threshold methods include: the CHARM algorithm (L. Krista PhD Thesis, 2012), which uses local intensity histograms to determine a fitting threshold; a fixed threshold based on the median solar disk intensity (Rotter et al. 2012; 2015; Reiss et al. 2015; Boucheron, Valluri, and McAteer 2016; Hofmeister et al. 2017; Stephan G. Heinemann, Temmer, et al. 2018); a dual-threshold growing algorithm (Caplan, Downs, and Linker 2016) and a supervised intensity threshold approach modulated by the intensity gradient perpendicular to the CH boundary (CATCH; Heinemann et al. 2019). A multi-wavelength approach was developed by Garton et al. (2018) in the form of the multi-thermal emission recognition algorithm CHIMERA. A spatial possibilistic clustering approach was taken by Verbeeck et al. (2014) which is available as the SPOCA algorithm. Recently, with the dawn of machine learning, new methods, utilizing the increased computational performance have also emerged to provide an additional tool to identify and extract coronal holes (e.g., Illarionov and Tlatov 2018).

Using various techniques, several CH datasets were gathered (especially for CH areas) that are freely available. Automatically created SPOCA boundaries of CHs are available via the Heliophysics Events Knowledgebase (HEK: https://www.lmsal.com/hek/index.html), the automated coronal hole detection and extraction using three SDO/AIA wavelengths (171, 193, 211A) CHIMERA is available via SolarMonitor (https://www.solarmonitor.org/) and an extensive, manually checked, CH catalogue covering the SDO-era (2010-2019) created using CATCH is available via the VizieR catalogue service (http://cdsarc.u-strasbg.fr/viz-bin/cat/J/other/SoPh/294.144).

A recent chain of studies used data from HINODE, SOHO and TRACE to study small-scale changes in the CH boundary. In a first step, Madjarska and Wiegelmann (2009) showed that although CHs maintain their overall shape over short timescales, small loops that are abundant along the boundaries continuously reconnect, changing the small scale magnetic structure (in the order of 1″- 40″). Using XRT observations of coronal bright points within a CH, Subramanian et al. (2010) showed that these small loops could be a source of slow outflowing solar wind. These loops may erupt as X-ray jets ejecting plasma along open field lines into interplanetary space. Madjarska et al. (2012) confirmed these findings using spectrograph data from SUMER/SOHO and EIS/HINDOE. In the last paper of the series, Huang et al. (2012) demonstrated that magnetic flux in CHs undergoes constant reconnection processes. It is suggested that these constant restructuring processes of the small-scale magnetic field within CHs might be largely involved with the overall magnetic flux formation within CHs. The connection between these phenomena and their cause remains an open question.

## 7.3 Progress in theory and simulations

The lack of understanding in solar wind acceleration and solar wind structures is closely related to the problem of coronal heating. In that respect, the properties of the solar corona and its connection to the solar wind are not well understood. As described above, recent studies use high-resolution observations of coronal structures and underlying magnetic field in order to gain a deeper insight into the mechanisms of CH evolution and morphology. Remote observations are coupled with in-situ measurements for investigating the impact of the solar wind structures at larger distances from the Sun. However, due to the scarce in-situ measurements in interplanetary space, we need to rely on improved modeling to make more conclusive interpretations of the physical processes underlying SIRs/CIRs and their Space Weather effects.

As mentioned before, recent studies found that unipolar flux tubes, presumably the main outflow regions of the fast solar wind, cover only a small percentage of the entire area of the CH (see Hofmeister et al. 2017). The open flux problem in that respect is a topic on its own that is tackled by combined observational and modeling efforts.

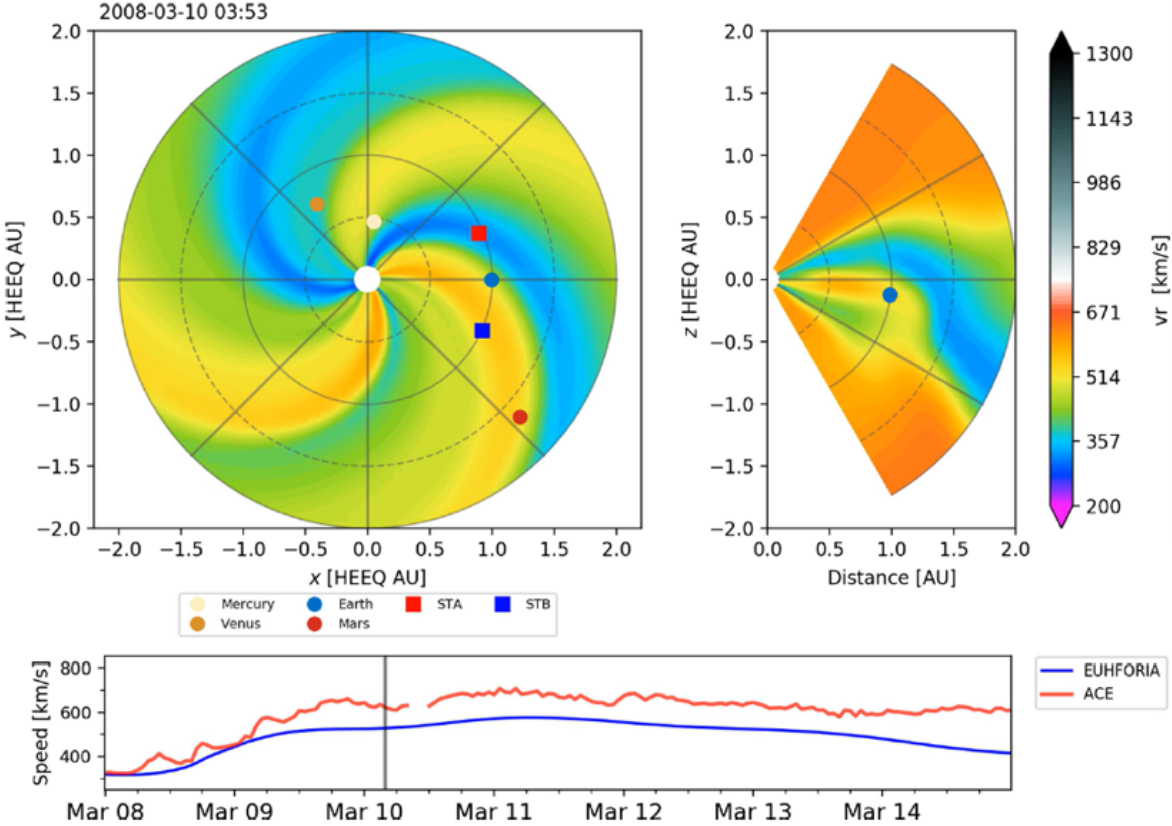

*Figure 7-6. Snapshot of the background solar-wind radial speed modeled by EUHFORIA. The top-left panel shows the MHD solution in the heliographic equatorial plane, and the right panel shows the meridional plane cut that includes the Earth (blue circle). The lower panel shows comparison of the modeled and observed solar wind by EUHFORIA and ACE, respectively. Taken from Hinterreiter et al. (2019b).*

To improve solar wind models and to ensure accurate space weather forecasting the solar wind models have to be tested and validated. The validation of solar wind models is done by comparing the simulation results to in-situ measurements.

The performance assessment of the EUHFORIA solar-wind model was analyzed by Hinterreiter et al. (2019). Within a thorough statistical investigation, a comparison between modeled and in-situ measured solar wind high-speed streams was made to identify possible caveats of the model results. The solar wind was modeled rather well for times of solar minimum (see Figure 7-6). However, during increased solar activity, complex solar-surface situations could be identified that stem from the interplay between evolving and dissipating magnetic field.

In a study by Lee et al. (2009) the heliospheric models ENLIL/MAS and ENLIL/WSA were compared with in-situ measurements from ACE and Wind (time range: 2003-2006). They found that the model results give lower densities for faster solar wind fully agreeing with the solar wind momentum flux conservation. They also derived a general good agreement between the solar wind models and the in-situ measurements for large-scale structures and for time scales of several days. The results are in agreement with findings from Gressl et al. (2014). Jian et al. (2015) performed a comparison of several models installed at CCMC (ENLIL , MAS, WSA, SWMF) with solar wind in-situ measurements and revealed strengths and weaknesses of each model. Common to all studies is the fact that different magnetogram inputs have a huge impact on the model performance. This result should be taken as basis to improve data driven models or at least to add more observational parameters in order to better constrain the models and to identify input magnetograms that match better depending on the specific solar cycle condition. Interesting recent studies for a better understanding of solar wind evolution is given by Jian et al. (2016) who compared ENLIL model results with observations at ACE and at Ulysses for times when ACE and Ulysses were in latitudinal alignment. The alignment made it possible to compare the model results for the same latitude due to the different radial distances (1 AU and 5.4 AU) the evolution of the solar wind could be well observed and interpreted.

For simulating the solar wind magnetic field close to the Sun, EUHFORIA uses an adaption of the semi-empirical Wang-Sheeley-Arge (WSA) model from Arge & Pizzo (2000). WSA is composed of the Potential Field Source Surface (PFSS) (Altschuler and Newkirk 1969; Kenneth H. Schatten, Wilcox, and Ness 1969)  and the Schatten Current Sheet (SCS) (Schatten 1971). The PFSS outer boundary, called "source surface", divides the corona into an inner and outer sector. At the transition from PFSS to SCS modelling domains, artificial kinks appear in the magnetic field lines, which can be improved by putting the inner boundary of the SCS model at a distance below the source surface (McGregor et al. 2008). Presently, no exact heights for those boundaries exist, and actually varying them leads to different results in the computed open magnetic flux (Asvestari et al. 2019) (cf. Figure 7-7). Moreover, as height variations of those boundaries affect the modelled magnetic field topology, e.g., the bending (spatial gradients) of the magnetic field lines, this has further effects on the computation of the expected propagation and extension of high-speed solar wind streams, and how they interact with Earth.

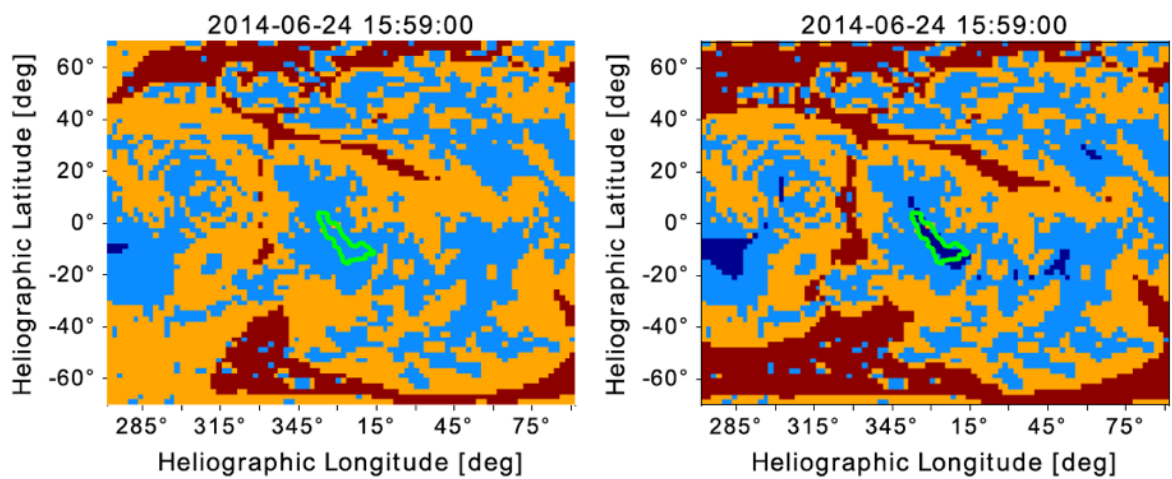

*Figure 7-7. Maps of open-closed flux generated for the same CH by two different model runs. The one on the left resulted by EUHFORIA running for the default pair of heights [2.3, 2.6] Rs, while the one on the right for the pair [1.3, 1.8] Rs. It is clear that the CH is not present in the model result on the left but is present and well captured by the run setup based on lower heights. This example highlights the impact the heights of the source surface and the inner boundary of the SCS model have in the modeling result. Taken from Asvestari et al., 2019.*

The magnetic topology is thought to be key, not only for the shape and morphology of CHs but also for the process of solar wind acceleration. In modeling, the flux tube expansion factor plays an important role in empirically determining the outflowing plasma velocities in CHs. Wiegelmann and Solanki (2004) showed that the magnetic configuration of CHs does not only exist of open field but is dominated by small-scale low-lying loops and few high and long. The small loops that seem to confine the expansion of the flux tubes in the transition region and lower corona were found to be on average flatter than their equivalent in the quiet sun. In that respect, the modeling of flux tubes within a CH is a challenge due to the much weaker magnetic field, but can be achieved as given in Figure 7-8.

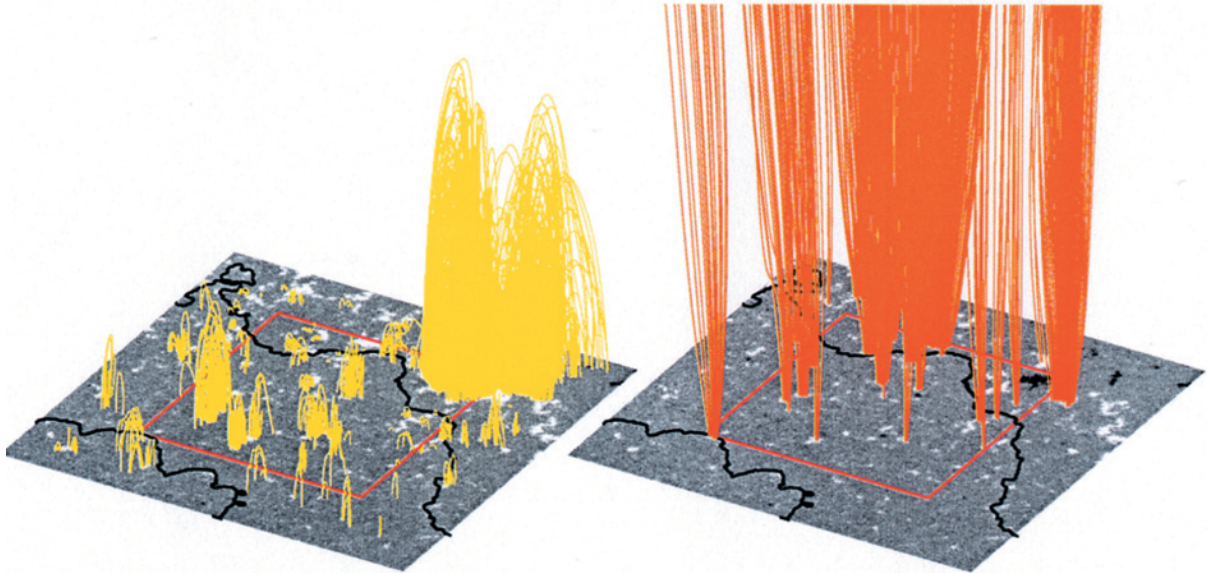

*Figure 7-8. Magnetic structures in a CH. The gray-coding shows the field strength in the photosphere. The black line gives roughly the boundary of the CH. The field of view for SUMER is marked as a red rectangle in both panels. The magnetic field was constructed from a MDI magnetogram. Left figure: mostly closed loops at various scales. Only closed magnetic field lines with B ≥ 30 G are shown. Right figure: only open fields with large photospheric values, B ≥ 100 G. The open flux is bundled in narrow uniform filaments and originates in stronger fields concentrated at small-scale footpoints. The flux tubes expand as they extend into the corona. Taken from Wiegelmann, Xia and Marsch (2005).*

## 7.4 Impact at Earth

The solar wind couples the interplanetary space with the Earth's magnetosphere. Hence, the upper atmosphere reacts to the energy input into the system depending on the speed and magnetic field of the solar wind stream. As such, especially the stream interaction regions and their compressed plasma (sometimes associated to shocks) put energy into the magnetosphere that has consequences for the upper atmosphere (coupled through Poynting flux). With the arrival of CIRs there is a persistent evolution from slow to fast solar wind and the Earth's plasmasphere significantly changes accordingly (Denton and Borovsky 2017). Spatial and temporal variations in the magnetic field are found to be most relevant for the amount of energy input into the system. Moreover, speed and density give the ram-pressure, which is well correlated to the amplitude of sudden storm commencements caused by the rapid compression of the Earth's magnetic field (Gonzalez et al. 1989). In that respect, the faster and stronger the compression, the larger the Space Weather effects. Further, the preconditioning of the magnetosphere plays an important role as pointed out in a well-cited study by Borovsky and Denton (2006). They further concluded that CIR-related storms are more hazardous to space-based assets, particularly at geosynchronous orbit compared to CMEs. The reason is that CIRs are of longer duration and have hotter plasma sheets causing a stronger spacecraft charging. Denton et al. (2016) gave an overview on unsolved problems of the magnetosphere. The atmospheric layers of the Earth all react on CIRs, and effects are measurable down to the neutral atmosphere. Variations in the thermosphere density occur in relation to the arrival of CIRs and CMEs. In that respect, CIRs and CME sheath regions have similar impact on the amount of density increase whereas most strong variations come from the magnetic structure of the CME (e.g., Krauss et al. 2015; Krauss, Temmer, and Vennerstrom 2018).

## 7.5 Conclusions

Studying solar wind streams and their solar sources, CHs, is of utmost importance. The streams highly structure interplanetary space and as such interact with disturbances propagating in the flow.
Consequences for CMEs are strong changes in their propagation behavior (speed, direction) which is the primary cause of large uncertainties in the CME forecast. In view of this, a reliable interpretation of observed changes in the kinematical profiles of CMEs is only possible when we properly understand the variation of the ambient solar wind flow. Besides influencing near-Earth space, high-speed solar wind streams themselves are sources of geomagnetic effects and especially during times of low solar activity put with their recurrent characteristics a comparable amount of energy into the Earth atmosphere as CMEs do over short time scales.

The different atmospheric layers of the Earth are coupled through dynamical, electromagnetic, and photo-chemical processes. With that, geomagnetic effects ue to SIRs/CIRs and CMEs may cause a cascade of impacts down to ground-level enhancements and induced currents. Long and short-term effects were studied during CAWSES that led to a significant improvement in our understanding of the solar influence on our Earth system (see special issue devoted to CAWSES-II (http://progearthplanetsci.org/collection/001.html). SCOSTEP endorsed the continuation of the CAWSES program as CAWSES II during 2009–2013. With the success of the VarSITI program (2014–2018), the path of interdisciplinary studies will be continued in the recently approved SCOSTEP program PRESTO (Predictability of the variable solar-terrestrial coupling) that will act during 2020–2024.

## 8. Forecasting CMEs

Having discussed the observational properties, theory and modeling of CMEs in section 2, 3 and 4, respectively, in this section we review the efforts addressing forecasting CME occurrence (Section 8.1), time-of-arrival at 1 AU (section 8.2), coronal and heliospheric modulations (Section 8.3) and magnetic fields (section 8.4). Recent reviews on these topics can be found in Vourlidas, Patsourakos, and Savani (2019) and Kilpua et al (2019). We will hereby mainly focus on empirical or semi-empirical physics-based approaches, which are more amenable to operational purposes in current stage.

## 8.1 Predicting CME Occurrence

The lack of critical observations (e.g., no routine observations of the magnetic field in the corona), and limitations in theory and models (e.g., idealized initial and boundary conditions), are currently not allowing to predict when a CME would occur. However, thanks to advances in our observational capacity (e.g., new observations from STEREO, Hinode, SDO, IRIS), and in modeling (e.g., increase of realism in models, data-constrained and data-driven models) and in analysis and forecasting techniques (e.g., use of advanced statistical tools and machine learning methods) significant progress in our understanding and eventual prediction of CMEs has been achieved over the last decade.

Identifying and understanding the physical mechanism(s) behind CME onsets would be a key element in developing the capability to predict them on a regular basis. While there exists no doubt about the magnetic origin of CMEs (Forbes 2000; Vourlidas et al. 2000), there is currently no consensus regarding the specifics of the eruption process (Chen 2011; Schmieder, Aulanier, and Vršnak 2015; Xin Cheng, Guo, and Ding 2017; Green et al. 2018; Georgoulis, Nindos, and Zhang 2019) and sections 2 and 3.

A first approach to the prediction of the most powerful CMEs, which are in general the most geoeffective, is to use the observational finding of a rapidly increasing probability of eruption with associated flare magnitude, with flares above X of the GOES classification approaching 100 % (Andrews 2003; Yashiro et al. 2005; Wang and Zhang 2007). Therefore, by assessing conditions/forecasts for major flares, one could also infer whether major CMEs could occur (Anastasiadis et al. 2017). However, the flare magnitude-CME occurrence relationship is *statistical*, and therefore, exceptions should be anticipated (e.g., the super-active AR 12192 which hosted 6 confined X-class (e.g., Thalmann et al. 2019). In addition, this approach excludes CMEs associated with weaker flares as well as CMEs originating from quiet Sun regions.

Eruption predictors based purely on imaging observations include SXR and EUV sigmoids (Canfield, Hudson, and McKenzie 1999; Green and Kliem 2009), EUV and WL cavities (Gibson et al. 2006), EUV hot channels (Zhang, Cheng, and Ding 2012; Patsourakos, Vourlidas, and Stenborg 2013; Xin Cheng, Guo, and Ding 2017). For instance, the statistical analysis of Canfield et al. (1999) found that sigmoidal ARs are more likely to erupt than non-sigmoidal

ARs. These are discussed in the recent reviews by Green et al. (2018) and Xin Cheng, Guo, and Ding (2017).

Given the magnetic nature of CMEs, several magnetic metrics have been considered in the literature as eruption diagnostics. The definition of these metrics is motivated by various proposed eruption models/mechanisms and/or by physical intuition. Pertinent studies calculate a given metric for a set of eruptive and non-eruptive ARs, and then search whether there exist specific thresholds, or region of values of the considered metrics, segregating eruptive and non-eruptive cases. Note here, that since a small fraction of ARs gives rise to eruptive flares, this large class-imbalance between eruptive and non-eruptive cases should be properly accounted for in the corresponding analysis.

A group of CME diagnostics refers to properties calculated within or around polarity inversion lines (PILs), i.e., regions in the photosphere where the vertical magnetic field changes sign. This is motivated by the fact that CMEs, and large-scale eruptive phenomena in general, originate from intense and complex PILs (Webb et al. 1997; Schrijver 2007). The corresponding studies employ line-of-sight or vector photospheric magnetograms. Various metrics are then calculated for each traced PIL and we hereby discuss a sample thereof (see also Figure 8-1). Falconer, Moore, and Gary (2006; 2008) introduced several PIL-related non-potentiality measures related to for example the length of the strong shear PIL(s), the integral of the shear angle along the PIL(s), the magnetic gradient-weighted integral length of the PIL(s). Considering pairs of these measures for a set of ARs, and designating suitable thresholds for each measure, supplied success rates of above 75% for an AR to give rise to a CME in the next few days from the corresponding measurements. Schrijver (2007) calculated the R metric, i.e., the total unsigned magnetic flux within 15 Mm of strong-field and high-gradient PILs. He found that when an R > $2 \times 10^{21}$ Mx is recorded, there is a high probability of a major (>M class) flare, and hence of a CME as well, within 24 hours from the measurement. Georgoulis and Rust (2007) calculated the $B_{eff}$ metric, which essentially calculates the total magnetic field of connected magnetic partitions in strong PILs. They found that the conditional probability for the occurrence of M- and X-class flares within a 12-hr window of the measurement exceeds 0.95 for $B_{eff}$ above 1600 G and 2100 G, respectively.

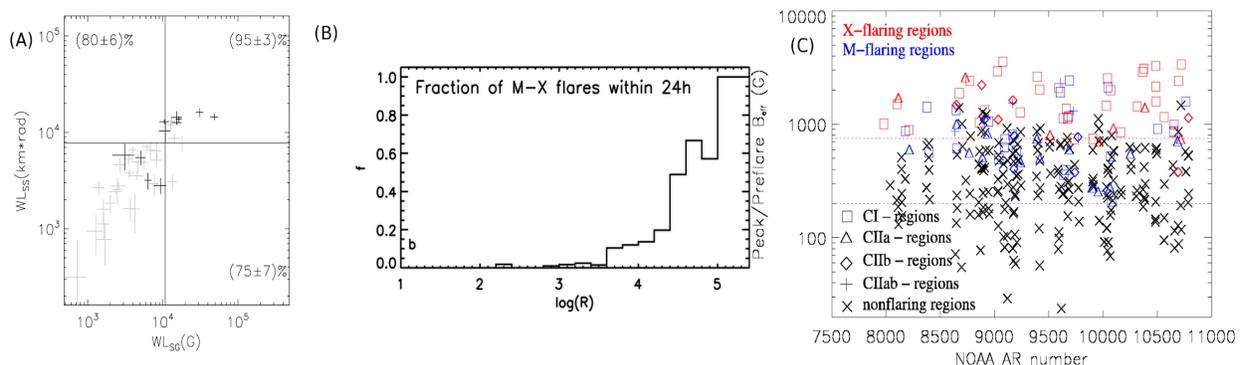

*Figure 8-1. Examples of magnetic metrics applied to major flares/CMEs. Panel (A). The integral of the shear angle along the PILs versus the gradient-weighted integral length of the PILs for 56 bipolar and multipolar ARs. The darker crosses correspond to ARs which gave rise to a CME within the 72 hr search window from the corresponding measurement whereas the lighter crosses correspond to those ARs without a CME with the same search window. From Falconer et al. (2008). Panel (B). Fraction of ARs with an M- or X-class flare as a function of R, within a 24-hour window from the R recording. 2500 ARs were employed. From Schrijver (2007). Panel (C). ARs giving rise to M- or X-*

Magnetic helicity (e.g., review by Pevtsov et al. 2014), a measure of the twist and linkage of magnetic field lines, is an extensively employed quantity given also its conservative nature (e.g., Berger 1984) which allows to draw links between CMEs in the Sun and in the interplanetary medium (Kumar and Rust 1996; Green et al. 2002; Démoulin et al. 2002). In addition, it plays central role in models of filament channel formation such as the helicity condensation model (Antiochos 2013). There exist several methods to calculate the magnetic helicity in the corona or its injection rate in the photosphere. Statistical surveys supplied important clues on the importance of magnetic helicity in CMEs. Nindos and Andrews (2004) studied the pre-flare helicity of active regions which gave rise to big flares, and showed, that statistically speaking active regions with eruptive flares are associated with larger magnetic helicity compared to active regions with confined (i.e., non-eruptive) flares. LaBonte, Georgoulis, and Rust (2007) found that active regions giving rise to X-class flares, hence with a high probability of being eruptive, exhibit peak helicity flux injection rates above $6x10^{36}$ Mx$^2$/s in 24-hour windows prior to the considered flares. Tziotziou et al. (2012) found that helicity and magnetic free energy thresholds of $\sim 2x10^{42}$ Mx$^2$ and $\sim 4x10^{31}$ erg separate eruptive with non-eruptive ARs. Using MHD simulations, Pariat et al. (2017) suggested that the ratio between the current-carrying part of magnetic helicity to the total (volume-integrated) helicity could be used as a discriminator between eruptive and non-eruptive cases. This has been recently tested in a handful of cases (Moraitis et al. 2019; Thalmann et al. 2019), where it was found that this ratio increases significantly prior to eruptive flares while it does not significantly change prior to confined flares. The recent study of Pagano, Mackay, and Yardley (2019) showed that a metric based on the calculation of the Lorentz force from data-driven nonlinear force-free field models could help into discriminating between eruptive and non-eruptive active regions.

Another important physical parameter pertinent to eruptivity is the decay index ($n$) of the overlying horizontal magnetic field. This essentially measures how fast the strapping magnetic field declines with height above the erupting flux. Its value assumes a key role in certain models such as the torus instability (Kliem and Török 2006). Depending on the properties of the magnetic set-up (e.g., bipolar, multipolar, aspect ratio and shape of flux rope, etc.), $n$ should be at least as steep as ~[0.5,2] for an eruption to take place (Kliem and Török 2006; Fan and Gibson 2007; Démoulin and Aulanier 2010; Oscar Olmedo and Zhang 2010; Zuccarello, Aulanier, and Gilchrist 2016; Syntelis, Archontis, and Tsinganos 2017). Note that stability against the torus instability could become more involved for complex systems like multiple flux ropes (Inoue, Hayashi, and Kusano 2016). Magnetic field extrapolations are normally employed in the calculation of the coronal magnetic field in regions that erupted (or not). In some cases, the starting height of the eruption, as inferred from height-time single or multi-viewpoint measurements, is used as the bottom boundary for the decay-index analysis.

The statistical study by Liu (2008) showed that the difference of $n$ between eruptive (>1.74) and non-eruptive cases (<1.71) is statistically significant. Analysis of $n$, for a set of confined and eruptive flares which took place in the same active region, showed that the eruptive flares were associated with steeper $n$ in the low corona compared to the confined ones (Cheng et al. 2011).

The statistical study of Wang et al. (2017) showed that the critical $n$ (i.e., 1.5) as per Kliem & Török (2006), is achieved at somehow larger heights above active regions which gave rise to confined flares as compared to cases associated with eruptive flares. Surveys of eruptive prominences by McCauley et al. (2015) and Vasantharaju et al. (2019) found $n$ in the range [0.8, 1.3] above the starting heights of the eruptions. Duan et al. (2019) calculated $n$ along *inclined* paths, reflecting non-radial CME propagation rather along the local vertical, and found that all eruptive flares they considered had $n>1.3$. Recently, Cheng et al. (2020) found that the average $n$ at the onset heights of the main acceleration was close to the torus-instability threshold for the 12 CMEs they analyzed. The segregation between eruptive and non-eruptive cases in terms of $n$ is also reported in several case studies as discussed in the review of Cheng, Guo, and Ding(2017). However, taking into account the temporal evolution of $n$ seems that a steep decay index of the overlying magnetic field is a necessary but not a sufficient condition for eruptions to take place (Suzuki, Welsch, and Li 2012; Chintzoglou, Patsourakos, and Vourlidas 2015). In addition, cases of high-lying QS filaments with overlying magnetic fields with $n>1.5$ associated with confined eruptions could be also found (Z. Zhou et al. 2019). Strong overlying fields seem to prevent CMEs as vividly illustrated for the case of super-active AR 12192 which exhibited a multitude of confined flares, including X-class (Chen et al. 2015; Sun et al. 2015).

The twist number (Tw) of magnetic field lines is another parameter that is extensively used in CME onset studies. Magnetic twist is an integral part of magnetic helicity discussed above, and comparison of its properties/distribution in the pre-eruptive/eruptive configurations in the solar atmosphere and at 1 AU supplies important physical clues about CMEs (Yuming Wang, Zhuang, et al. 2016a). In addition, magnetic twist plays a key role in the triggering of the helical kink instability, and related Tw thresholds in the range ~ [1.25, 2.0], depending on the specifics of the implementation (e.g., twist profile, employed geometry, etc.), were derived (Hood and Priest 1981; Fan and Gibson 2003; Török and Kliem 2003; Török, Kliem, and Titov 2004; Hassanin and Kliem 2016). Jing et al. (2018) calculated the spatial average of Tw for a sample of 38 eruptive and confined flares. They found that Tw was not playing a role in separating between confined and eruptive flares, by indeed all eruptive cases having Tw smaller than the lower bound of the Tw kink instability threshold (i.e., 1.25) discussed above. On the other hand, a similar survey of 45 eruptive and confined flares by Duan et al. (2019), calculating this time the maximum Tw per flux rope, showed that Tw above (below) 2, i.e., close to the upper bound of the kink-instability thresholds discussed above, were relevant to the majority of the considered eruptive (confined) flares, therefore allowing to segregate between confined and eruptive cases. The differences between these two studies could be possibly attributed to the different approaches used to calculate the coronal magnetic field and Tw, i.e., NLFFF extrapolations and average Tw in Jing et al. (2018) and magnetic relaxation and maximum Tw in Duan et al. (2019). Irrespectively of whether kink instability assumes the main role in setting CMEs, it may still, in cases of confined eruptions triggered by this instability, lift flux ropes to heights where the torus instability could take over.

The existence of coronal null points (i.e., points of vanishing magnetic field) has central role in CME models such as the breakout model (Antiochos, DeVore, and Klimchuk 1999). Therefore, magnetic field extrapolations are used to investigate whether coronal null points exist above erupting ARs, e.g., Aulanier et al. (2000) for the first such application. Searches of coronal null points and their association with CMEs have been extended to larger statistical samples by Ugarte-Urra et al. (2007) and Barnes (2007). These studies found that a significant fraction of the analyzed ARs with coronal null points were eruptive, 73% and 26%, respectively. Both studies reported that for the majority of the considered eruptive cases (~ 75%), no pre-eruptive coronal null points were found.

As discussed in the introduction of this section, the problem of predicting major CMEs could be mitigated to predicting major flares. Flare forecasting is a field that has really boomed over the last decade because of the quantum increase of vector magnetic field data, thanks to HMI on SDO, and the advent of machine-learning (ML) in heliophysics. ML schemes allow to digest in autonomous or semi-autonomous means large volumes of data, explore multi-dimensional parameter spaces, and are particularly suited for identification and classification tasks (e.g., review of Camporeale (2019) of ML applications in heliophysics. Recent reviews on flare forecasting, including ML-schemes could be found in Leka et al. (2019) and Park et al. (2020). Major conclusions from their extensive benchmarking of a large number of methods currently used in flare forecasting, are that numerous such methods do better than climatology, no method clearly outperforms the others, and consideration of prior flare history improves the corresponding skill scores.

ML has been used directly in CME predictions. Bobra & Ilonidis (2016) applied a Support Vector Machine classification scheme to 18 parameters derived from HMI vector magnetograms for more than 3000 ARs and found that only a handful (i.e., 6) amongst these parameters is sufficient to separate erupting and non-erupting ARs within 24 hours from the corresponding measurements. These parameters (e.g., mean gradient of the horizontal magnetic field, mean current helicity, mean twist parameter) are *intensive* (i.e., do not depend on the AR size but are spatial averages) and not *extensive* (i.e., depend on the AR size and correspond to spatial sums). Interestingly, extensive measures seem more appropriate for the prediction of *any flare,* irrespectively of its eruptivity (e.g., Bobra and Couvidat 2015).

MHD simulations are invaluable in evaluating existing metrics as well as for supplying physical insight into new metrics. For example, Guennou et al. (2017) analyzed a set of eruptive and non-eruptive MHD simulations and found, in agreement with the ML work discussed in the previous paragraph, that intensive parameters, are more relevant to eruptivity. Another example is the study of Moraitis et al. (2014) who validated the existence of particular magnetic helicity-free magnetic energy regimes pertinent to eruptivity as reported in the observational study of Tziotziou et al. (2012). Routine observations of the AR vector magnetic field at several layers above the photosphere may be instrumental into predicting CMEs (e.g., Patsourakos et al. 2020).

## 8.2 Predicting CME Time of Arrival

The essential questions regarding the CME forecast are if and when it will hit Earth. Therefore, various models and methods have been proposed in the past decades to give (a reliable) answer to these questions. Recently, Vourlidas, Patsourakos, and Savani (2019) gave an extensive overview of the time-of-arrival (ToA) forecast models, whereas earlier reviews include that of Zhao and Dryer (2014) and Siscoe and Schwenn (2006). In this review, ToA models are noted and described in Sections 3 (analytical and semi-empirical models) and Section 4 (numerical models), whereas in this Section we will primarily focus on their performance.

Vourlidas, Patsourakos, and Savani (2019) made a comprehensive summary of the mean absolute error (MAE) reported by numerous studies and found that the unweighted mean of all MAE is 9.8 ± 2 h, representing the value for the current state of accuracy of ToA studies. However, as noted by Vourlidas, Patsourakos, and Savani (2019) and Verbeke et al. (2019), most of the studies on the ToA prediction does not report their performance validation consistently and moreover a comparison of the ToA performance between different methods/models is generally missing. Not only should different methods be compared on the same sample and under the same conditions, but also a community-agreed metrics and validation methods should be used in order to assess the current state of CME modeling capabilities unbiasedly. An effort in that direction was recently made by Riley et al. (2018), who compared the performance of different methods and models that performed predictions on the CME scoreboard (https://kauai.ccmc.gsfc.nasa.gov/CMEscoreboard/). The CME scoreboard, facilitated by Community Coordinated Modeling Center (CCMC), is one aspect of the efforts of the CME Arrival Time and Impact Working Team that started in 2017 in the scope of the community-wide International Forum for Space Weather Capabilities Assessment. Riley et al. (2018) explored the accuracy and precision of the predictions made by 32 teams 2013-2017 and found that the models on average predict arrival times to within ±10 hours with the precision around the average of ±20 hours. In addition, they found that the "Average of all Methods" forecasts generally performs as well as, or outperforms the other models, thus acting as a simple super-ensemble approach. It should be noted that the ensemble approach was implemented in several ToA models/methods in the past years (Mays, Taktakishvili, et al. 2015a; Dumbović, Čalogović, et al. 2018; Amerstorfer et al. 2018; Kay and Gopalswamy 2018b; Napoletano et al. 2018), as it can provide a probabilistic forecast of CME arrival time as well as an estimation of arrival-time uncertainty from the spread.

The current state of accuracy of ToA prediction, regardless of the method used, seems to revolve around 10 hours. This resemblance in the performance of very different propagation models indicates that the major drawback lies in the lack of reliable observation-based input. This includes the CME input parameters as well as the input of the heliospheric background in which it propagates, which on their own contain errors. Namely, to obtain CME input parameters, such as the CME velocity and angular width, different methods and models have to be used to go past the problems related to the projection effects (for observational properties of CMEs see Section 2). Different methods can present a rather wide spread in the obtained CME parameters on a case by case basis (Mierla et al. 2010). Therefore, it is important to keep track of the CME parameter measurement metadata and to test the CME input errors. On the other hand, the model performance also depends on the input of the heliospheric background, which hugely relies on our current capabilities of the solar wind modelling (see section 4). Recently, Kay, Mays, and Verbeke (2020) performed an analysis of the ToA sensitivity of arrival times to various input parameters for drag-based models. They found that the ToA tends to be more sensitive to CME parameters than solar wind parameters, and that different precisions on the input parameters are needed for different "strength" CMEs. We expect more such studies will be performed in the near future and especially in the scope of the new SCOSTEP program PRESTO.

## 8.3 Predicting the magnetic field of CMEs

Predicting the magnetic field distribution within CMEs/ICMEs represents a holy grail in heliophysics, given the fact that extended intervals of intense southward magnetic fields, typically associated with ICMEs, spawn the stronger geomagnetic storms (e.g., Gonzalez, Tsurutani, and Clúa de Gonzalez 1999). A thorough account of the state-of-the art in this important problem was recently given in (A. Vourlidas, Patsourakos, and Savani 2019). Currently, we are able to routinely observe the magnetic field of CMEs only at the "end of the road" to geospace, via in-situ observations at 1 AU by the WIND, ACE etc spacecraft (e.g., Chi et al. 2016; Nieves-Chinchilla et al. 2019 and Section 3). Occasionally we can observe it also in the inner heliosphere beyond ~ 0.3 AU, with HELIOS (e.g., Bothmer and Schwenn 1998), and more recently with Messenger and VEX (Miho Janvier et al. 2019b; Good et al. 2019). Much closer to the Sun, in the corona, we have only a few reported cases of direct observations of the magnetic field of CMEs in the radio domain. These observations exploit gyroresonance emissions from mildly relativistic electrons spiraling in CMEs (Bastian et al. 2001; Tun and Vourlidas 2013; Carley et al. 2017) and Faraday rotation of the electromagnetic radiation of either natural (e.g., pulsars) or artificial sources going through the CME body (Jensen and Russell 2008; T. A. Howard et al. 2016; Kooi et al. 2017). Lack of continuous monitoring of the Sun with solar-dedicated instruments and low sensitivity of the existing instrumentation is behind the scarcity of radio diagnostics of CME magnetic fields.

Therefore, we have to mainly rely on modeling, either MHD, or empirical/semi-empirical to remedy this critical deficiency. MHD simulations of CMEs covering the domain spanning the lower solar atmosphere, the outer corona, and the inner heliosphere out to 1 AU, could in principle deal with the problem in a self-consistent manner, since they simultaneously treat CME initiation, evolution and propagation in a realistic corona and solar wind (e.g., Jin et al. 2017; Török et al. 2018 and review by Manchester et al. 2017). However, such simulations, given the huge resources they require, are currently used almost exclusively for research purposes and not for forecasting.

A more tractable approach and recently developed capability is to use heliospheric models of *magnetized* CMEs; see Section 4.2 for details. These models represent a major step over heliospheric CME models which treat CMEs as purely hydrodynamic disturbances such as the widely-used ENLIL model (Odstrcil, Riley, and Zhao 2004), which is indeed the standard operational space weather model (Mays, Taktakishvili, et al. 2015b). Heliospheric CME models launch CMEs in their inner boundary, typically in the range 10-20 Rs, and follow their evolution in the inner heliosphere. The prescribed CMEs are empirically constrained by STEREO observations supplying their speed, size and orientation. Magnetized CME models require also inputs for the CME magnetic field in the inner boundary of their computational domain such as the axial magnetic field. The less computing resources that these models require allows to run them in almost real-time, with ensemble studies testing the influence of uncertainties of the input parameters on the ICME properties upon impact at 1 AU.

To deal with the lack of direct routine observations of the near-Sun magnetic field of CMEs, an ever-growing number of empirical or semi-empirical models to infer this vital parameter has appeared over the last decade (A. Vourlidas et al. 2000; Kunkel and Chen 2010; Savani et al. 2015b; 2017; Isavnin 2016b; S. Patsourakos, Georgoulis, Vourlidas, Nindos, Sarris,

Anagnostopoulos, Anastasiadis, Chintzoglou, Daglis, Gontikakis, Hatzigeorgiu, Iliopoulos, Katsavrias, Kouloumvakos, Moraitis, Nieves-Chinchilla, Pavlos, Sarafopoulos, Syntelis, Tsironis, Tziotziou, Vogiatzis, Balasis, Georgiou, Karakatsanis, Malandraki, Papadimitriou, Odstrčil, Pavlos, Podlachikova, Sandberg, et al. 2016; S. Patsourakos and Georgoulis 2017; C. Möstl et al. 2018; Sarkar, Gopalswamy, and Srivastava 2020). These models can be used to supply inputs to the heliospheric magnetized CMEs' models, and to predict on their own CME magnetic profiles at 1 AU. Various empirical inputs (e.g. CME positional, orientation and size information, reconnected magnetic flux and SXR light-curve of the associated flare, eruption-related magnetic helicity, height-time measurements of the CME, CME energetics, etc.) and underlying assumptions (conservation of energy, magnetic flux and magnetic helicity; self-similar expansion; force-free/non-force-free magnetic fields, etc.) are used. The handedness of the near-Sun CME flux ropes is derived from empirical schemes based on for example the CME source region hemisphere, the sense of winding of observed features etc (Bothmer and Schwenn 1998; Palmerio et al. 2017). Given that the inputs to these models could be retrieved from easy-to-obtain observations from magnetographs, imagers and coronagraphs, and their analytical or semi-analytical nature, they could be used to routinely infer the near-Sun magnetic field of CMEs and to predict it upon impact at 1 AU. In addition, ensemble studies to account for uncertainties in the inputs are feasible at minimal computational cost. Meaningful and rigorous comparisons between forecasts and observations could benefit from tools/concepts developed for terrestrial weather (Austin and Savani 2018). A major conclusion was that while existing models supply encouraging results, they nevertheless require further development and validation. For instance, there exist models that lack the ability to derive beforehand the near-Sun CME magnetic field and have to rely on the in-situ observations upon the corresponding ICME arrival at 1 AU in order to properly scale the predicted CME magnetic field vectors at 1 AU. In addition, there is no benchmarking of these models. A zero-order comparison of the near-Sun CME magnetic fields from two of these models seems encouraging (see Figure 8-2).

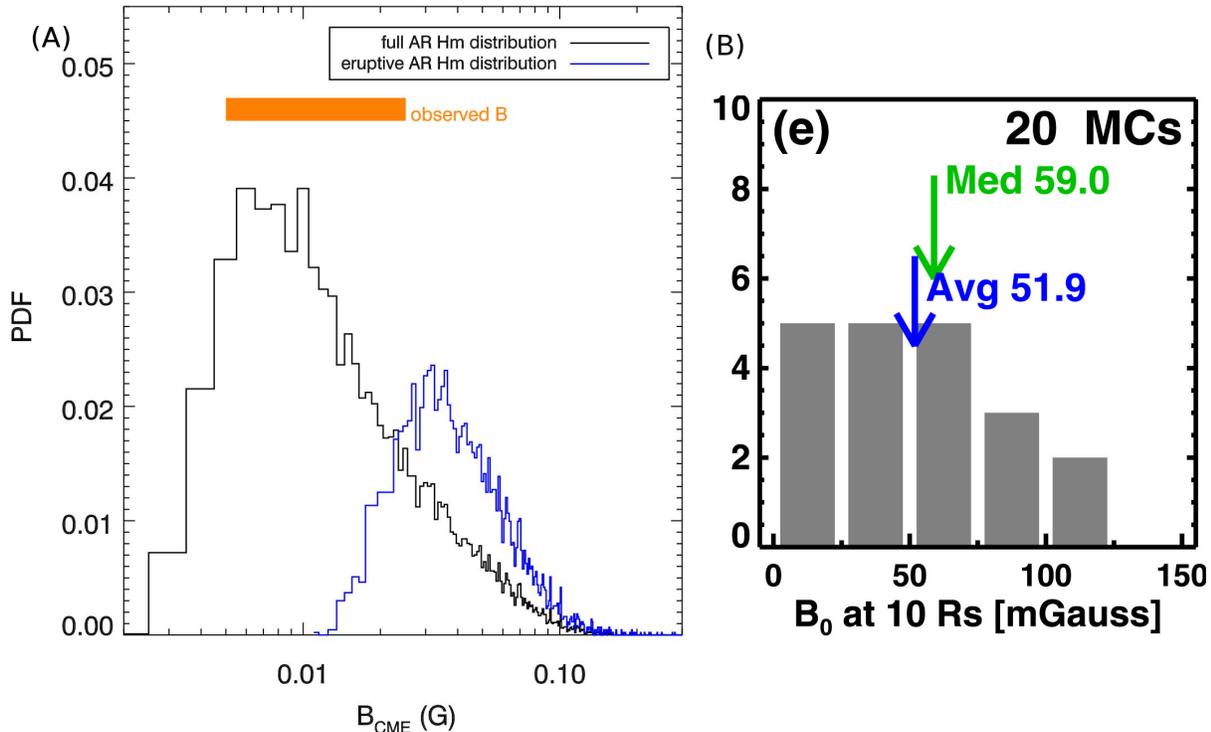

*Figure 8-2. Histograms of near-Sun CME axial magnetic fields at 10 Rs from two different methods. Panel A. From Patsourakos & Georgoulis (2017) (blue histogram corresponds to eruptive cases). Panel B. From Gopalswamy et al. (2018) (results from CMEs corresponding to magnetic clouds upon impact at 1 AU).*

Another major limiting factor in our ability to forecast CME magnetic fields upon impact at 1 AU is related to the uncertainties of the input parameters of the employed models. This includes uncertainty related to the deflection and rotation that CMEs undergo, in the determination of the initial (i.e., source region) and coronal CME location and orientation. In addition, since models also rely upon on the properties of the background corona and solar wind, which are derived from different models/approaches, which constitutes another source of uncertainty. A common outcome of these studies, whether based on Sun-to-Earth MHD simulations (e.g., Török et al. 2018), heliospheric MHD models (e.g., Verbeke, Pomoell, and Poedts 2019; Scolini et al. 2019), or semi-analytical physics-based models (e.g., Kay et al. 2017) is that rather small changes in the CME positional and orientation parameters in the range 2-20 degrees, could have significant impact on the predicted CME magnetic field profiles at 1 AU, and particularly on the field components (e.g., Figure 8-3). Pattern recognition applied to in-situ observations of incoming ICMEs at 1 AU could be also used to predict CME magnetic fields with however much shorter lead times of a few hours only (James Chen, Cargill, and Palmadesso 1997; Riley et al. 2017; Salman et al. 2018; Camporeale 2019).

Very recently, important results regarding the nature of young CMEs started to emerge from the first observations of the recently launched PSP mission (Teresa Nieves-Chinchilla et al. 2020; Hess et al. 2020). As more CMEs are observed by PSP, as well as by the SoLO mission, it should be possible to validate and to eventually increase the physical realism of models of CME magnetic field forecasting. In-situ monitoring of ICMEs at Venus orbit tied with either empirical scaling laws and/or propagation models could supply predictions of Earth-bound CMEs with a lead-in of ~ 1 day prior to impact (Kubicka et al. 2016). Recent or upcoming facilities observing in radio like LoFAR, MWA and SKA will finally supply more systematic observations of CME magnetic fields in the corona (Nindos, Kontar, and Oberoi 2019).

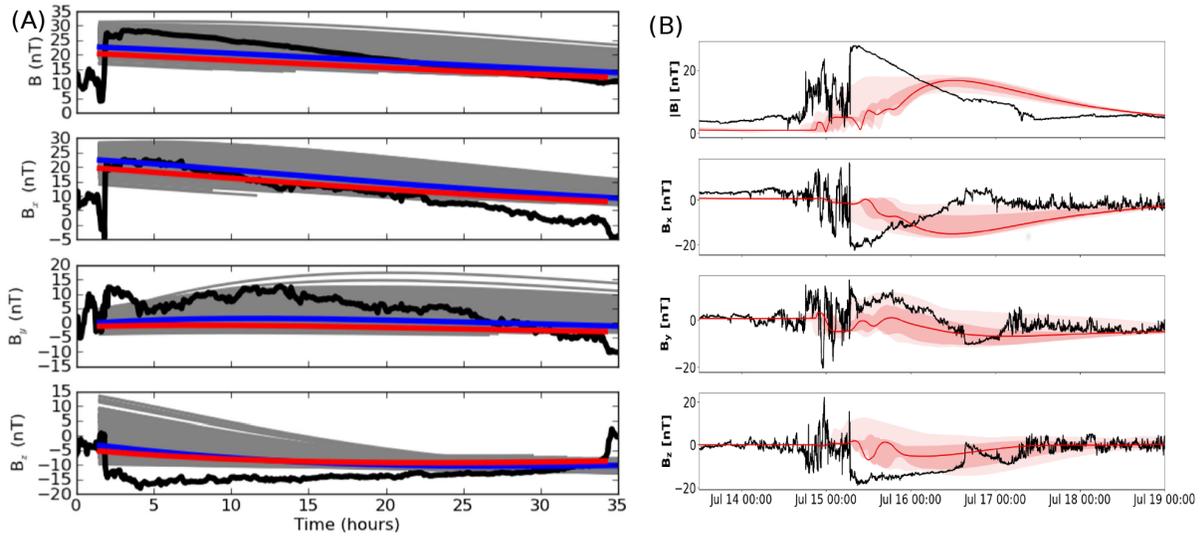

*Figure 8-3. Predicted (non-black curves and shaded areas) and observed (black curves) CME magnetic field profiles at 1 AU from two different models. Panel (A): from Kay et al. 2018; panel (B): from Scolini et al. 2019. The predicted profiles consider uncertainties in input parameters. In both panels from top to bottom the magnetic field magnitude, and its Bx, By, and Bz components in the GSE system are plotted.*

# 9. Minimax24 project

The ISEST/MiniMax24 non-flare target is an email alert service about non-flare related, but possibly geo-effective phenomena provided daily by the observer on duty. The observation overview is shown in Figure 9-1. The service was first established in the scope of the SCOSTEP/CAWSES "MiniMax24 Campaign" in 2013, which was declared as the year of "MiniMax24" to note that, even though the Sun is going through activity maximum conditions, the activity is rather low. The goal of the action was to understand and explain the current behavior of the Sun and its potential impact on human society and Earth's space environment through yearlong scientific and outreach activities. The campaign team counted 37 institutions from 17 countries focused on the solar-terrestrial observations of solar eruptive events through the MaxMillenium program of solar flare research (http://solar.physics.montana.edu/max_millennium/) as well as CHs, filaments and CIRs (i.e. non-flare related phenomena) through the newly established email alert service. By the end of the year of "MiniMax24", MiniMax24 email list reached more than 140 participants from more than 30 countries. As the MiniMax24 has shown to be a very useful and successful hub for the scientific community, the action transcended from its original 1-year-campaign scheme and was included in the new SCOSTEP program VarSITI, as one of the working groups of the ISEST project.

The aim of the ISEST/MiniMax24 non-flare target is to monitor and warn against potentially geoeffective phenomena that are not related to solar flares. This includes CMEs that are not accompanied by solar flares, but rather eruptive filaments, as well as stream interaction regions associated with coronal holes.

## 9.1 The scientific outline

CMEs may either originate from strong active regions, large-scale relatively weak magnetic fields or filaments (Green et al. 2018). The production of CMEs is highly likely when associated with eruptions of filaments (Schmieder, Démoulin, and Aulanier 2013) and eruptive filaments are often used as on-disk signatures of the non-flare related CMEs. These do not show significantly different properties from the flare-related CMEs (Vršnak, Sudar, and Ruždjak 2005a; Chmielewska et al. 2016), unlike e.g. 'stealth CMEs', i.e. CMEs without obvious on-disk signatures, which are slower and therefore potentially less geoeffective (Robbrecht, Patsourakos, and Vourlidas 2009; Kilpua et al. 2014; Nitta and Mulligan 2017). Therefore, eruptive filaments can be regarded as potential sources of significant geoeffective events. Since filaments are regarded as cool plasma suspended in the magnetic dips of the flux rope (Sarah E. Gibson 2018), the scale of the filament is directly related to the scale of the flux rope and therefore by that logic large and dark filaments are indicative of large flux ropes, i.e. more massive/energetic CMEs, which are then more likely to be significantly geoeffective (Gopalswamy, Yashiro, and Akiyama 2007). Thus, the ISEST/MiniMax24 focuses on detecting and monitoring only large and dark filaments. Since the CMEs are largely propagating radially, those originating from sources close to the center of the solar disc are more likely to arrive at Earth and, therefore, more likely to be geoeffective (Srivastava and Venkatakrishnan 2004; Gopalswamy, Yashiro, and Akiyama 2007; Zhang et al. 2007). Therefore, the ISEST/MiniMax24 focuses on filaments located close around the central meridian. It should be noted though, that not all filaments erupt and that filament eruptions are not necessarily related to CMEs and might be triggered by e.g. magnetic flux emergence or local and large scale photospheric motions (Parenti 2014b). From that perspective, ISEST/MiniMax24 only alerts on the *possibly* geoeffective filament targets.

Another significant source of non-flare related geoeffectiveness, as pointed out in Section 7, are SIRs formed by the interaction of the high-speed solar wind originating from a CH with the preceding slower solar wind (Richardson 2018). CHs are the darkest and least active regions of the Sun, associated with rapidly expanding open magnetic fields and acceleration of the solar wind (Cranmer 2009) and can be easily outlined automatically in EUV images using the threshold technique (Rotter et al. 2012). Moreover, the area of the CH was found to be highly correlated with the speed of the corresponding HSS at Earth (Nolte et al. 1976; Tokumaru et al. 2017) and was found to typically need about 4 days to arrive at Earth (Vršnak, Temmer, and Veronig 2007; Manuela Temmer, Vršnak, and Veronig 2007). Since generally geoeffectiveness is related to the dawn-to-dusk electric field and therefore solar wind flow speed (Richardson and Cane 2011), the ISEST/MiniMax24 relies on the premise that the potentially geoeffective HSS emanate from large CHs close to the central meridian. Empirical relations between coronal holes and HSS have been utilized to produce a tool for automatic detection of the CH area in a meridional slice around the center of the solar disc to predict solar wind speed near Earth 4 days in advance (Vršnak, Temmer, and Veronig 2007; Rotter et al. 2012; Reiss et al. 2016b). This HSS forecast algorithm is called 'Empirical Solar Wind Forecasting' (ESWF) tool and it operates automatically using near real-time SDO/AIA 193Å images to forecast solar wind speed 4 days in advance (http://cesar.kso.ac.at/programme/minimax.php).

## 9.2 Non-flare Target alert description

The non-flare target alerts are sent on a daily basis, with a 'NO non-flare targets' descriptor for quiet times and brief notification and description of the activity for non-quiet times. The non-quiet times are defined based on the following criteria:

***Criterion #1.*** There is a CH located within +/-7.5 degrees in longitude exceeding a ratio area of 0.2, or predicted SW speed at Earth exceeding 500km/s as observed/forecasted by the ESWF tool.

***Criterion #2.*** There is a prominent (i.e. dark and wide) filament located in longitude within +/-30 degrees around the central meridian as detected in H-alpha images provided by Kanzelhöhe Observatory or GONG H-alpha network (http://halpha.nso.edu). If one or more CHs are identified as non-flare targets based on criterion 1, the observer notifies in the alert the total ratio area (calculated cumulatively for all CHs across the whole meridional slice), approximate position of each CH and the arrival time and speed of the corresponding HSS. If one or more filaments are identified as non-flare targets based on criterion 2, the observer notifies in the alert the position, E-W and N-S spread of each filament. In both cases, the reference to the observing image is provided. The email alerts were not systematically stored until 2018, although some of the observers archived a significant part of the alerts. Since the beginning of 2018 the alerts are systematically archived and together with previously stored alerts compiled into a single non-flare target catalogue maintained by the ISEST/MiniMax 24 team and available upon request (contact email: mateja.dumbovic@geof.unizg.hr).

## OBSERVATION OVERVIEW

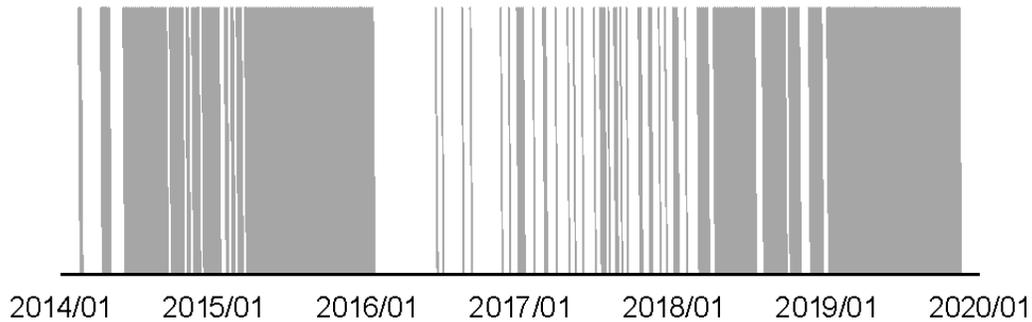

### OBSERVING TIME

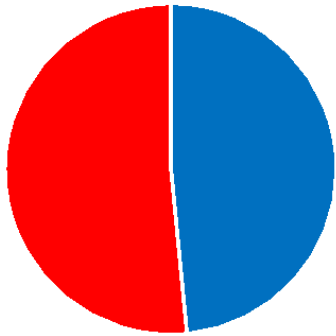

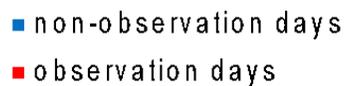

### ACTIVITY IN OBSERVED DAYS

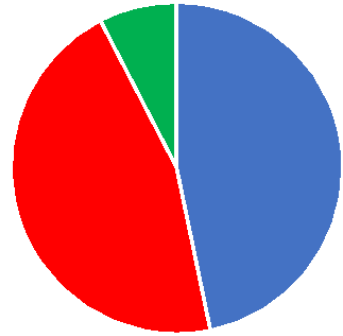

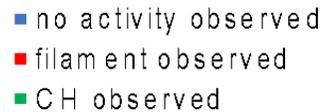

*Figure 9-1. On the top: archived non-flare target alerts 2014-2020; Bottom left: number of archived alerts (observation days) vs number of non-archived alerts (observation days); bottom right: number of quite days, filament and CH observations in the archived alerts*

## 9.3 Summary

The ISEST/MiniMax24 non-flare Target activity started as a 1-year SCOSTEP/CAWSES activity and continued throughout the SCOSTEP/VarSITI program in the scope of the ISEST project. It has been shown that it is a useful scientific community service that provides a daily overview of the non-flare solar activity. Moreover, as an added value we highlight that this activity has significantly helped to improve the visibility of the young scientists acting as the daily observers. Therefore, the MiniMax24 activity is planned to continue within the new SCOSTEP program PRESTO, in scope of which it is in addition planned to archive and compile all the alerts into a non-flare target catalogue to be used for future statistical analysis.

# 10.    Conclusion and Outlook

The aim of the ISEST project is to understand the origin, propagation and evolution of solar transients through the space between the Sun and the Earth and develop the prediction capability for space weather with particular emphasis on the weak solar activity prevailing in Solar Cycle 24. The ISEST project dealt with short-term solar variability in the form of flares, CMEs, and SIRs and the associated phenomena such as interplanetary shocks and SEP events by forming several working groups to focus on specific problems. The MiniMax24 program, started during the previous SCOSTEP program, has proved to be extremely useful. This program will be continued and become a permanent feature. The program has developed extensive data base on Earth-affecting transients that is available to the scientific community (Figure 9-1). The ISEST program also brought together hundreds of scientists from all over the world to focus on specific problems that resulted in rapid progress. In following, we discuss the implications of the current status and suggestions for future directions.

CMEs are the most recently (1971) discovered phenomena compared to other disturbances. The discovery of solar wind and interplanetary shocks preceded the CME discovery by a decade, while the discovery of SEPs was even earlier (1940s). Observationally, our understanding of CMEs has progressed significantly over the next four decades: a well-defined magnetic structure ejected along with the coronal plasma. While the vast majority of CMEs near the Sun do not drive a shock, the small number of shock-driving CMEs have the most intense consequences in the heliosphere. Although we treated SEP events as a transient, they are closely related to CMEs via their shocks. SEPs not related to shocks are generally the weak and short-lived impulsive events thought to be accelerated in the flare reconnection region. Once SEPs are released from the shock or flare site, their further propagation and evolution depends on the wave/plasma/magnetic properties of the background solar wind through which they propagate before reaching the observer. Shocks are also readily inferred from type II radio bursts, providing information on shocks from their origin close to the surface to 1 AU and beyond. Interplanetary scintillation is another technique that can be used to track the turbulent sheath region of interplanetary shocks. SIRs/CIRs also drive shocks and accelerate particles, but generally at large distances from the Sun. Only about a fifth of IP shocks detected at Earth are due to CIRs. Particles acceleration in CIRs is significant beyond 1 AU. Geospace phenomena such as energetic storm particle events, ultra relativistic electron events, storm sudden commencement, and the onset of geomagnetically induced currents. The magnetic structure of the shock sheath and the driving CME determine the onset of a geomagnetic storm following shock arrival. The primary requirement for a storm is the presence of intense and prolonged southward pointing magnetic field in the sheath and/or the CME. This is the motivation behind the attempts to assess the internal magnetic field of CMEs when they are still near the Sun. For example, if the magnetic structure of the CME can be determined near the Sun, it should be possible to predict the structure in the heliosphere taking into account of the environmental conditions.  Both CIRs and shock sheaths are compressed heliospheric plasmas and hence have similar impact on the magnetosphere in causing geomagnetic storms.

A wealth of observational information on CMEs and SIRs has accumulated over the past two decades, thanks to the fleet of space missions observing the Sun-Earth system. These data have contributed greatly to the current understanding of CMEs and SIRs. SOHO has provided extensive data over two solar cycles (23 and 24). In cycle 24, STEREO has provided multiview observations, enabling the determination of three-dimensional morphology of CMEs. The extended STEREO field of view allowed CMEs to be tracked from the coronal base to beyond Earth orbit, significantly enhancing observational knowledge of CME propagation. SOHO observations have shown that a magnetograph to measure the photospheric magnetic field, EUV imager to observe eruption signatures, white-light coronagraph to image CMEs, heliospheric imager for tracking CMEs to 1 AU, and a low frequency radio telescope to detect shock signatures from close to the Sun to 1 AU. In addition to these, we also have instruments for in-situ measurements of plasma, magnetic field, and energetic particles to complete the data set needed for investigating earth-affecting solar transients. The STEREO mission's twin spacecraft transited through the Sun-Earth Lagrange points L4 and L5 and demonstrated that these are ideal locations for placing these instruments to better observe Earth-directed CMEs. STEREO did not have a magnetograph to observe the photospheric magnetic field; SOHO did not have a magnetometer for detect CMEs and CIRs in the solar wind. Multiview magnetograms are important not only to track potentially eruptive active regions from behind the east limb before they rotate on to the disk, but also build global magnetic field distribution used as input to background solar wind models. While both L4 and L5 observations help characterize CMEs near the Sun, L5 vantage is useful in identifying active regions before they rotate into Earth view. On the other hand, L4 can observe Earth-directed CMEs without being affected by "snowstorm" of secondary particles created by SEPs hitting the spacecraft. Earth-directed energetic CMEs are magnetically well-connected to L5, so coronagraph images are vulnerable to such "snowstorms". Another advantage of the L5 location is that CIRs arrive at L5 a few days before they arrive at Earth, so one can predict the nature of CIRs arriving at Earth.  Placing similar instruments at L4 and L5 will be ideal for a better characterization of CMEs near the Sun. Ideally, one should have multiple spacecraft at various locations in Earth orbit to provide space-weather relevant information on transients as well as the global magnetic field. Future efforts should also be directed toward using other techniques such as Faraday rotation to measure the magnetic content of CME flux ropes.

There has also been a rapid development of several MHD models with sophisticated simulation techniques to describe most or all stages of CMEs, i.e., pre-eruptive stage, destabilization and eruption, and propagation. In particular, models involving flux rope are growing in number so that the currently used hydrodynamic pulse representing a CME can be eventually replaced by a flux rope, which is more realistic and consistent with in-situ observations. Such a transition would have the potential to predict the magnetic field vectors in the heliosphere rather than predicting just the CME arrival time. MHD models have also started considering the simulation boundaries closer to the Sun to account for forces that significantly affect the propagation of CMEs. The ultimate goal is to predict the magnetic field vectors at any point in the inner heliosphere, soon after the eruption at the Sun. Recent work on deriving the magnetic properties of CMEs near the Sun using source properties and eruption data will help test and improve global MHD modeling of CME propagation. However, we still have a long way to go in understanding when a magnetic region on the Sun hosts an eruption, but significant progress over the recent years in both terms of our observational and modeling capabilities is seamlessly contributing towards an eventual resolution of this cornerstone challenge. We still do not have a reliable set of active region parameters that would indicate an eruption. This is a common problem to both flares and CMEs because they are manifestations of a common energy release in the source magnetic region. Observations of the magnetic field at several layers above the photosphere and all around the Sun tied with advances in our capability to model magnetic fields from emergence to eruption are lending reasonable optimism.

The VarSITI program was launched at the peak phase of solar cycle 24, which turned out to be only half as strong as cycle 23. VarSITI investigations dealt with Earth-affecting solar transients in the background of the diminished solar activity and the related changes in the heliosphere into which solar disturbances propagated. The space weather consequences of CMEs and CIRs have proved to be mild in solar cycle 24. This is the combined effect of the diminished number of energetic CMEs and the weakened heliospheric state. This was the smallest cycle in the space age, so we are able to expand our knowledge of extended parameter space of Earth-affecting phenomena. Current predictions of the strength of solar cycle 25 point to a weak cycle as well, and one can expect another cycle with mild space weather. The extended and uniform data set from SOHO have helped us characterize the solar cycles from the point of view of CME evolution and particle acceleration. The Parker Solar Probe and the Solar Orbiter are sampling the weak heliosphere and we expect to learn a lot on the behavior of Earth-affecting transients in the heliosphere.

# Abbreviations

ACE: Advance Composition Explorer

ADAPT: Air Force Data Assimilate Photospheric Flux Transport

AWSoM: Alfvén Wave Solar Model

CAWSES: Climate And Weather of the Sun-Earth System

CME: Coronal Mass Ejection

CIR: Corotating Interaction Region

CORHEL: Coronal-Heliosphere

CH: Coronal Hole

EVE: EUV Variability Experiment

DSA: Diffusive Shock Acceleration

EEGGL: Eruptive Event Generator Gibson-Low

EUHFORIA: European Heliospheric Forecasting Information Asset

ESP: Energetic Storm Particle

FIP: First Ionization Potential

ICME: Interplanetary Coronal Mass Ejection

IP: Interplanetary

LASCO: Large Angle and Spectrometric Coronagraph Experiment

LAT: Large Area Telescope

LDGRF: Long-duration Gamma-Ray Flare

PFSS: Potential Field Source Surface

PSP: Parker Solar Probe

ISEST: International Study of Earth-affecting Solar Transients

LFM: Lyon-Fedder-Mobarry

MAS: Magnetohydrodynamic Algorithm Outside a Sphere

MESSENGER: Mercury Surface, Space Environment, Geochemistry and Ranging

MHD: Magnetohydrodnamics

PCA: Polar Cap Absorption

SCOSTEP: Scientific Committee on Solar-Terrestrial Physics

SDO: Solar Dynamic Observatory

SECCHI: Sun Earth Connection Coronal and Heliospheric Investigation

SEP: Solar Energetic Particle

SIR: Stream Interaction Region

SO: Solar Orbiter

SOHO: Solar and Heliospheric Observatory

SPR: Solar Particle Release

SSN: Sunspot Number

STEREO: Solar Terrestrial Relations Observatory

STEREO/A: STEREO Ahead

STEREO/B: STEREO Behind

SWMF: Space Weather Modeling Framework

VarSITI: Variability of the Sun and Its Terrestrial Impact

## Declarations

## Availability of Data and Material

Data sharing not applicable to this article as no datasets were generated or analyzed during the current study.

## Competing interests

The authors declare that they have no competing interest.

## Funding

The ISEST project, which results in this review, is supported by SCOSTEP/VARSITI program. JZ was funded by NASA grant NNH17ZDA001N-HSWO2R, NNH19ZDA001N-TMS. SF is supported by the Strategic Priority Research Program of Chinese Academy of Sciences Grant No. XDB 41000000, the National Natural Science Foundation of China (41774184 and 41974202) and the Specialized Research Fund for State Key Laboratories. NVN's work was supported by NASA contract NNX17AB73G. MID acknowledges support of NASA Grants 80NSSC18K0520, 80NSSC19K0079, 80NSSC20K1255, and 80NSSC18K1366. NL was supported by NASA grant 80NSSC20K0700. XF is supported by the Natural Science Foundation of China under grants (41531073,41861164026, 42030204).

## Authors' contributions

JZ provided the overall management of writing this review paper, which was a group effort 19 individual authors. JZ, NG and MT conceived the structure of the paper.  JZ (leading author) contributed to S1 (Section 1). JZ (leading author), NG and KD contributed to S2. BV (leading author), YW (leading author), BZ and MD contributed to S3. FS (leading author), NL and XF contributed to S4. DW (leading author) and NVN (leading author) contributed to S5. OM (leading author), NG, MID, ND and ML contributed to S6.  MT (leading author), KD, MD and SGH contributed to S7. SP (leading author), MD and JZ contributed to S8. MT (leading author), MD and KD contributed to S9. NG (leading author) contributed to S10.

## Authors' Information

No

# Acknowledgement

JZ, NG, SF and MD gratefully acknowledge the travel grant from Nagoya University to attend the workshop held in Nagoya, Japan in November 2019, for initiating this review work. JZ thanks Prof. Kazuo Shiokawa for his assistance in guiding the process of this review work.

637X/731/2/109.